\documentclass[floats,floatfix,amssymb,prd,aps,twocolumn,nofootinbib,nolongbibliography,reprint,groupedaddress]{revtex4-2}
\usepackage{amssymb,amsmath,verbatim,mathtools,needspace,enumitem,etoolbox,graphicx,physics,microtype,afterpage,xspace,tabularx,lmodern,multirow,booktabs,fontawesome}
\usepackage{gensymb}
\usepackage[dvipsnames, usenames]{xcolor}
\usepackage[unicode, colorlinks=true, linkcolor=linkcolor, citecolor=linkcolor, filecolor=linkcolor, urlcolor=linkcolor, linktocpage, breaklinks]{hyperref}
\usepackage{orcidlink}
\usepackage[nolist,nohyperlinks]{acronym}
\usepackage{tikz}
\usetikzlibrary{decorations.markings}
\usetikzlibrary{decorations.pathmorphing}
\usepackage{algorithm}
\usepackage[noend]{algpseudocode}

\newcommand{\be}{\begin{equation}}
\newcommand{\ee}{\end{equation}}

\definecolor{linkcolor}{rgb}{0.0,0.0,1.0}
\definecolor{olive}{rgb}{0.4,0.7,0.2}
\newcommand{\Eeffmrg}{\hat{E}_{\mathrm{eff}}^{\mathrm{mrg}}}
\newcommand{\bmrg}{\hat{b}_{\mathrm{mrg}}}
\newcommand{\jmrg}{j_{\mathrm{mrg}}}

\def\i{{\rm i}}
\def\e{{\rm e}}

\allowdisplaybreaks

\begin{document}

\pagenumbering{arabic}


\title{Highly eccentric non-spinning binary black hole mergers:\\ quadrupolar post-merger waveforms}

\author{Nishkal Rao \orcidlink{0009-0006-4551-7312}}\email{nishkal.rao@students.iiserpune.ac.in}
\affiliation{Department of Physics, Indian Institute of Science Education and Research, Pashan, Pune - 411 008, India}
\author{Gregorio Carullo \orcidlink{0000-0001-9090-1862}}\email{gregorio.carullo@ligo.org}
\affiliation{School of Physics and Astronomy and Institute for Gravitational Wave Astronomy, University of Birmingham, Edgbaston, Birmingham, B15 2TT, United Kingdom}

\begin{abstract}
    We present numerically-informed closed-form expressions for the dominant $(\ell,m)=(2,2)$ waveform harmonic of the post-merger emission from mergers of non-spinning binary black holes with comparable masses on highly eccentric orbits. Using 233 non-spinning eccentric simulations from the RIT catalog, we construct time-dependent complex quasinormal mode amplitudes via a Bayesian procedure. We build multivariate polynomial models, represented as functions of the symmetric mass ratio $\nu$ and two dynamics parameters evaluated at merger: the mass-rescaled effective energy $\Eeffmrg$ and angular momentum $\jmrg$. We further validate the post-merger non-circular waveform model by comparing it against simulations from the SXS catalog. Our models achieve mismatches around $\sim10^{-3}$, including for near-extreme eccentricities. The model can be directly combined with effective-one-body and phenomenological inspiral waveforms to produce accurate inspiral-merger-ringdown waveforms, essential for parameter estimation of both astrophysical and fundamental physics properties of the signals' sources.
\end{abstract}

\maketitle

\section{Introduction}

Black-hole (BH) coalescences are nowadays routinely detected through laser interferometric techniques, with more than 200 events reported in the fourth Gravitational-wave Transient Catalog (GWTC-4)~\cite{LIGOScientific:2025slb}, constructed through the LIGO-Virgo-Kagra (LVK)~\cite{LIGOScientific:2014pky, VIRGO:2014yos, KAGRA:2018plz} detector network. 
With the upgraded sensitivity of gravitational-wave (GW) detectors during the fourth observational run (O4)~\cite{LIGOScientific:2025slb}, the rate of detection of binary BH mergers has nearly doubled, enabling new probes on the astrophysics of binary compact objects~\cite{LIGOScientific:2025pvj}, cosmology~\cite{LIGOScientific:2025jau} and the strong-field dynamics of General Relativity (GR)~\cite{LIGOScientific:2026qni, LIGOScientific:2026fcf, LIGOScientific:2026wpt}.

BH coalescences are characterized by a long \textit{inspiral}, during which the BHs orbit gradually shrinks, losing energy through GW emission. 
This is followed by a rapid plunge, resulting in the \textit{merger} phase forming a common horizon. 
Subsequently, during the \textit{ringdown}~\cite{Vishveshwara:1970cc, Press:1971wr, Teukolsky:1973ha, Press:1973zz, Chandrasekhar:1975zza, Chandrasekhar:1975zz, Chandrasekhar:1976zz} phase, the newly formed remnant settles down to a perturbed BH which dynamically approaches equilibrium by mainly emitting a spectrum of damped sinusoidal GW signals. 
As predicted by linear BH perturbation theory~\cite{Regge:1957td, Zerilli:1970wzz, Zerilli:1970se, Teukolsky:1972my, Teukolsky:1973ha, Press:1973zz, Teukolsky:1974yv}, the ringdown is mostly characterized by QNMs~\cite{Nollert:1999ji, Ferrari:2007dd, Berti:2009kk, Berti:2025hly} with (complex) frequencies completely described by the BH mass $M$ and angular momentum $J$, due to the no-hair conjecture (for electromagnetically neutral BHs, consistent with LVK data~\cite{Carullo:2021oxn, Gu:2023eaa}). 
This simplicity has sparked interest in utilizing BHs not only as astrophysical tracers, but also as testing grounds for GR and the BH paradigm, a program known as Black Hole Spectroscopy~\cite{Berti:2025hly, Echeverria:1989hg, Dreyer:2003bv, Berti:2005ys, Gossan:2011ha, Berti:2015itd, Berti:2016lat, Baibhav:2017jhs, Baibhav:2018rfk, Carullo:2022qak, Carullo:2025oms}.

Current analyses carried out by the LVK collaboration mostly use waveform models that describe binary BH coalescences on quasi-circular orbits, the exception being Ref.~\cite{LIGOScientific:2025rid}.
This is motivated by the fact that GW radiation is very efficient in circularizing binaries during a long inspiral~\cite{Peters:1963ux, Peters:1964zz}.
This is especially the case in the \textit{isolated} formation channel~\cite{Mandel:2018hfr, Mapelli:2021taw}, where the binary has little interaction with the surrounding environment.
Consequently, in this scenario, one expects to have negligible eccentricities shortly before the merger of stellar mass compact binaries, soon after entering the frequency window of ground-based detectors~\cite{Stevenson:2017tfq}.
However, dynamical formation and interactions in dense stellar environments can imprint residual eccentricity even close to merger~\cite{Samsing:2013kua, Zevin:2021rtf, Chattopadhyay:2023pil, Antonini:2013tea, Samsing:2017xmd, Rodriguez:2018pss, Zevin:2018kzq, DallAmico:2023neb}.
Although a priori one expects that such large residual eccentricities close to merger are unlikely, to obtain an unbiased inference of astrophysical and fundamental scenarios, it is necessary to confirm this expectation using observational data.
Importantly, several massive events (e.g., GW190521~\cite{LIGOScientific:2020iuh, LIGOScientific:2020ufj} and GW231123~\cite{LIGOScientific:2025rsn}) present peculiar features, such as unexpectedly misaligned or even near-extremal spin magnitudes, which are challenging to obtain in standard formation scenarios and gave rise to a variety of interpretations~\cite{Liu:2025yok, Islam:2026yxx, Lai:2026yvm, Tenorio:2026dcc, Croon:2025gol, Cuceu:2025fzi, Shan:2025dcd, Chakraborty:2025pxt, DeLuca:2025fln, Stegmann:2025cja, Hu:2025lhv, Goyal:2025eqo, Chan:2025kyu, LIGOScientific:2025cwb, Delfavero:2025lup, Ray:2025rtt, Passenger:2025acb, Chatterjee:2025avc, Li:2025pyo, Baumgarte:2025syh, Popa:2025dpz}.
This motivates the exploration of alternative formation models, including those with non-negligible eccentricity at merger.
Indeed, several studies pointed to the need to include non-circular corrections to accurately analyze GW190521~\cite{CalderonBustillo:2020xms, Gayathri:2020coq, Gamba:2021gap}.
More broadly, precise modeling of eccentric GW waveforms is key to disentangle formation channels~\cite{Romero-Shaw:2021ual, Zevin:2021rtf, Favata:2021vhw, Romero-Shaw:2022fbf, Carullo:2023kvj}.

Even before starting to infer population-level binary properties, ignoring eccentricity could lead to a significant reduction in the detection rates of eccentric binaries by current GW detectors~\cite{Huerta:2013qb, LIGOScientific:2019dag, Phukon:2024amh} and introduce systematic biases in parameter estimation~\cite{OShea:2021faf, Ramos-Buades:2019uvh, Divyajyoti:2023rht, Stegmann:2025shr}, which could manifest as spurious deviations from GR~\cite{Bhat:2022amc, Narayan:2023vhm, Bhat:2024hyb, Gupta:2024gun, Shaikh:2024wyn}, or give false evidence for spin precession in sufficiently short signals~\cite{CalderonBustillo:2020xms, Divyajyoti:2023rht, Romero-Shaw:2022fbf}.
For this reason, full Inspiral-Merger-Ringdown (IMR) waveform models that are routinely used for data analysis by the LVK collaboration have been extended to include signatures of orbital eccentricity in the inspiral~\cite{LIGOScientific:2019dag, Islam:2021mha, Romero-Shaw:2022fbf, Liu:2023dgl, Gamboa:2024hli, Planas:2025feq}, and applied to measure orbital eccentricity in detected events~\cite{Iglesias:2022xfc, Bonino:2022hkj, Clarke:2022fma, Ramos-Buades:2023yhy, Gupte:2024jfe, Morras:2025xfu, Planas:2025jny, Tiwari:2025fua, LIGOScientific:2025rid, Jan:2025zcm}.

There is a rich history in the development of waveform models for binary mergers, evolving from systems with no intrinsic angular momentum (``non-spinning''), to the inclusion of spin components aligned with the orbital angular momentum, and finally misaligned components giving rise to spin-induced precession~\cite{Chatziioannou:2024hju}.
One of the semi-analytical approaches that has been proven to be accurate, flexible, and efficient is the Effective-One-Body (EOB) framework~\cite{Buonanno:1998gg, Buonanno:2000ef, Damour:2000we, Damour:2001tu, Damour:2014yha, Damour:2015isa, Nagar:2019wds, Nagar:2020pcj, Albanesi:2023bgi}.
EOB is a perturbative framework in the symmetric mass ratio $\nu=m_1m_2/(m_1+m_2)^2$ of the binary components, describing the inspiral of the coalescing system by translating the problem of two comparable-mass merging objects into the evolution of a test mass orbiting a deformed Kerr metric.
This description was only carried out until the merger phase, leaving the post-merger waveform portion to be described through phenomenological ans\"atze~\cite{Buonanno:2000ef}, later informed to numerical simulations.
Because of the smooth transition from inspiral to plunge~\cite{Buonanno:2000ef, Blanchet:2013haa}, EOB has proven to be highly effective in allowing for the integration of Numerical Relativity (NR) calibration~\cite{Buonanno:2006ui, Damour:2007xr} onto waveform models.
NR-informed strategies proved to be faithful, accurate, and sufficiently fast for analyses of GW waveforms.
Thereby, contemporary GW analyses often rely on waveform models derived from a combination of multiple analytical approaches to the coalescing two-body problem in GR, which are later calibrated against NR simulations for the late stages of binary mergers.

Specifically to the ringdown, Ref.~\cite{Damour:2014yha} provided a phenomenological, NR-informed, accurate representation of the ringdown for non-precessing, equal-mass, and equal-spin quasi-circular binaries.
The effectiveness of this type of post-merger ans\"atze has been extensively verified in Ref.~\cite{DelPozzo:2016kmd}, and full waveforms generated by stitching it together with EOB inspiral waveforms including higher harmonics~\cite{Cotesta:2018fcv, Nagar:2019wds}, aligned-spin effects~\cite{Damour:2014sva, Bohe:2016gbl, Nagar:2018zoe, Nagar:2020pcj}, orbital precession~\cite{Estelles:2020osj, Estelles:2020twz, Gamba:2021ydi, Pompili:2023tna}, tidal effects~\cite{Akcay:2018yyh, Nagar:2020xsk, Akcay:2020qrj}.
Due to efficient circularisation, the same family of quasi-circular merger-ringdown templates has also been faithfully used to complement inspiral waveforms that include sufficiently small eccentricity~\cite{Chiaramello:2020ehz, Albanesi:2021rby, Nagar:2021gss, Nagar:2021xnh, Riemenschneider:2021ppj, Nagar:2024dzj}.

In view of extending these closed-form post-merger models to arbitrary eccentricities in bound orbits, Ref.~\cite{Carullo:2024smg} constructed a model for the late-time QNM-driven regime.
In the context of the EOB formalism, recent efforts in Refs.~\cite{Faggioli:2026alx, Albanesi:2026qtx} have also been undertaken in the test-mass regime to characterize and phenomenologically model the GW merger-ringdown modes emitted by a small mass plunging and merging into a Kerr BH on eccentric equatorial orbits, extending the Schwarzschild analysis presented in Ref.~\cite{Albanesi:2023bgi}.
Alternative approaches to model non-circular waveforms of comparable-mass binaries include phenomenological or NR-informed surrogates for the merger-ringdown phase, albeit currently these models cover configurations in which most of the eccentricity has been radiated before the merger, and non-circular corrections are small~\cite{Cao:2017ndf, Liu:2019jpg, Liu:2021pkr, Ramos-Buades:2021adz, Islam:2021mha, Gonzalez:2025xba, Planas:2025plq, Nee:2025nmh, Islam:2025bhf}.
In this paper, within the context of modeling comparable-mass astrophysical binaries, we extend the late-time model presented in Ref.~\cite{Carullo:2024smg} for eccentric, non-spinning ringdown waveforms to describe the early post-merger emission.

Eccentricity has been classically defined~\cite{Peters:1963ux, Peters:1964zz, Mora:2002gf} by generalizing the Newtonian definitions of deviations from circularity, such as the eccentricity and mean anomaly, measuring fractions of the orbital period through an angle.
Recent progress~\cite{Islam:2021mha, Ramos-Buades:2019uvh, Ramos-Buades:2022lgf, Romero-Shaw:2022fbf, Knee:2022hth, Clarke:2022fma, Bonino:2022hkj, Shaikh:2023ypz, Bonino:2024xrv, Shaikh:2025tae} is based on redefining eccentricity through quantities constructed directly from the waveforms observed by a far-away observer (e.g. by interpolating the waveform frequency evolution).
While these parameterizations are (nearly, see below) gauge-invariant, they are constructed from the waveform frequencies at apastron and periastron. 
In practice, estimating these frequencies for highly eccentric configurations, often with only a few waveform cycles available from finite-length simulations, is challenging.
Further, eccentricity is intrinsically ill-defined at merger~\cite{Carullo:2023kvj}, preventing the application of this approach to dynamical captures that do not complete several orbits before merging.
Thereby, to develop non-circular ringdown models applicable to short-duration or dynamical capture waveforms, we use the dynamics-based parameterization introduced in Ref.~\cite{Carullo:2023kvj}, valid for arbitrary orbits (including highly eccentric, or dynamically bound).
In particular, we provide parametric fits of the waveform using the (dimensionless) effective energy $\Eeffmrg$ and angular momentum $\jmrg$, which we at times also combine to arrive at the effective impact parameter at merger $\bmrg$ (see Ref.~\cite{Carullo:2023kvj} for details). 
This approach, coupled to the non-circular ans\"atz presented in Ref.~\cite{Albanesi:2023bgi} for extreme mass-ratio inspirals, and here extended to the comparable-mass case, provides all the necessary ingredients to build a non-circular model for arbitrary eccentricities.

The paper is structured as follows.
We define the conventions used in the manuscript in Sec.~\ref{sec:conventions}, and the employed model is detailed in Sec.~\ref{sec:model}. 
Extraction of the relevant quantities, including a description of the numerical dataset, is presented in Sec.~\ref{sec:extraction}.
Global fits are constructed in Sec.~\ref{sec:global_fit}. 
Finally, Sec.~\ref{sec:conclusions} contains discussion, future work, and conclusions.

\section{Conventions and numerical dataset}\label{sec:conventions}

We use geometric units $c=G=1$ throughout the paper.
A spin-weighted spherical harmonics $_{-2}Y_{\ell m}(\theta,\varphi)$ decomposition~\cite{Buonanno:2006ui, Blanchet:2013haa, London:2014cma, London:2018gaq} of the outgoing GW radiation is considered

\be
    h_+-{\rm i}h_\times =  \sum_{\ell=2}^{\infty}\sum_{m=-\ell}^{\ell} h_{\ell m}(t) {}_{-2}Y_{\ell m}(\theta,\varphi) \, .
\ee

where $h_{+}$ ($h_{\times}$) is the plus (cross) polarization of the GW waveform, and we refer to $h_{\ell m}$ as ``modes''.

Individual initial Christodoulou masses of the two BHs are denoted as $m_{1,2}$, the total initial rest mass $M = m_1 + m_2$ (set to unity)~\cite{Boyle:2009vi}, and the mass ratio $q \equiv m_1 /m_2 \geq 1$.

\subsection{Quasi-circular simulations}

As the algorithm validation dataset, we use GW waveforms simulated using the Spectral Einstein Code (\texttt{SpEC}) developed by the \textit{Simulating eXtreme Spacetime} (\texttt{SXS}) collaboration~\cite{Boyle:2019kee} interfaced through the \texttt{sxs} package~\cite{sxs}. 
We consider the dominant mode $(\ell, m)=(2,2)$ from the highest resolution waveforms extrapolated with a third-order polynomial to future null infinity.
\paragraph*{Non-spinning:}
As a baseline for our non-circular model extension of non-spinning progenitors, we use the quasi-circular dataset present in Table~II, Ref.~\cite{Nagar:2019wds}, pertaining to symmetric mass ratio, $0.08\lesssim\nu\lesssim0.25$ (corresponding to $1\leq q\leq 10$). 
\paragraph*{Equal-mass, equal spins:}
To validate the \emph{RatExp} template's effectiveness against previous quasi-circular results, we use the equal-mass ($m_1=m_2$) and equal-spin waveforms with the individual spins either both aligned or anti-aligned with the orbital angular momentum, for the set of simulations considered in Ref.~\cite{Damour:2014yha}.

\subsection{Eccentric simulations}

We consider 233 non-spinning, non-circular bounded simulations in highly eccentric orbits, available in the fourth public release of the RIT catalog of strain waveforms extrapolated to future null infinity~\cite{Healy:2022wdn}. 
The RIT catalog is used to probe higher orbital eccentricities close to merger compared to the SXS catalog. 
The convergence of the waveforms has been extensively studied and evaluated for different mass ratios and spins of the binaries~\cite{Lovelace:2016uwp}.
Future work will include the newer release of the SXS catalog~\cite{Scheel:2025jct}.

We employ the non-spinning dataset in Table~I, Ref.~\cite{Carullo:2023kvj}, which spans the range $0.18\lesssim\nu\lesssim0.25$ (corresponding to $1\leq q\leq 3$). 
The dataset is constructed keeping into account data quality considerations, by checking for balance laws through computations of fluxes~\cite{Carullo:2023kvj}, limiting the available dataset to binaries with $e_0 \leq 0.9$.
When studying the dependence of the template coefficients on the parameter space, for qualitative purposes we will make use of the gauge-dependent initial eccentricity $e_0$, but for quantitative purposes we always use the notions of eccentricity given by the dynamical quantities $\{\bmrg,\,\Eeffmrg,\,\jmrg\}$ accessed through the simulation metadata and appropriately evolved to merger via the GW fluxes~\cite{Carullo:2023kvj}.
These quantities are computed at infinity, and are thus gauge-invariant~\cite{Damour:2011fu, Hopper:2022rwo}, modulo Bondi-van der Burg-Metzner-Sachs transformations which act non-trivially at infinity~\cite{Mitman:2022qdl}.
In this work, we won't consider the latter effect, which is expected to be negligible for quadrupolar waveforms.
In terms of this dynamics-based parameterization, we correspondingly analyze the range $1.91\lesssim\bmrg\lesssim3.41$, $0.87\lesssim\Eeffmrg\lesssim0.96$, and $1.85\lesssim\jmrg\lesssim3.22$.
This range sets the region of validity of the constrained global fits developed below.

Due to simulation inaccuracies, the strain does not always converge towards zero at late times~\cite{Cheung:2023vki, Carullo:2024smg}.
Hence, we pre-process the simulations by subtracting a constant complex value extracted at late times.

\section{Waveform model}\label{sec:model}

Our model is constructed as a modification of the non-precessing \texttt{TEOBPM} model~\cite{Gennari:2023gmx}, falling into the class of \texttt{KerrPostmerger} templates, as defined in Chapter~6, Ref.~\cite{Berti:2025hly}.
The template is implemented within the \texttt{pyRing} package~\cite{Carullo:2019flw, pyRing}.
Via phenomenological fits, this template aims to phenomenologically effectively capture transients~\cite{Berti:2006wq, Lagos:2022otp, Albanesi:2023bgi}, non-linearities~\cite{Gleiser:1996yc, London:2014cma, Sberna:2021eui, Cheung:2022rbm, Mitman:2022qdl, Lagos:2022otp, Baibhav:2023clw, Cheung:2022rbm, Bucciotti:2023ets, Perrone:2023jzq, Redondo-Yuste:2023seq, Ma:2024qcv, Bucciotti:2024zyp, Bourg:2024jme, Zhu:2024rej} and mass-spin variations~\cite{Sberna:2021eui, Redondo-Yuste:2023ipg, Zhu:2024dyl, May:2024rrg, Capuano:2024qhv} in the comparable mass regime, where no first-principles model exists yet.

We start by factoring out the contribution of the stationary fundamental mode from the $h_{\ell m}$ modes, and introduce the QNM-rescaled waveform~\cite{Damour:2014yha, Nagar:2019wds, Nagar:2020pcj, Albanesi:2023bgi}

\be
    \bar{h}_{\ell m}(\tau) = \e^{\i(\omega_{\ell m}^+\tau+\phi_{\ell m}^{\rm ref})}h_{\ell m}(\tau) \, .
\ee
where $\phi_{\ell m}^{\rm ref}$ is the value of the phase $\phi_{\ell m}^+$ at the reference time $t_{\rm ref}$ (see below), $\omega^+_{\ell m}$ indicates the prograde fundamental QNM (complex) frequency, fixed by the final Kerr BH mass and spin and read through the \texttt{qnm} package~\cite{Stein:2019mop}, and $\tau=t-t_{\rm ref}$ a relative time coordinate.
Negative-$m$ modes are constructed through $h_{\ell,-m} = (-1)^{\ell} h^{*}_{\ell m}$, as a consequence of the symmetries inherent in the Kerr metric and equatorial symmetry of the initial data~\cite{Blanchet:2013haa, London:2018gaq}.

The model only considers co-rotating contributions (see Sec.~5.1.6, Ref.~\cite{Berti:2025hly}), since counter-rotating contributions are excited only for high anti-aligned intrinsic binary spins, precessing systems and extreme mass ratios~\cite{Berti:2005ys, Apte:2019txp, Lim:2019xrb, Li:2021wgz, Zhu:2023fnf}. 
We restrict our model to the dominant quadropolar mode and ignore higher modes for the current study.
Their inclusion can be achieved with analogous methods and will be considered in future work.
The model also neglects spherical-spheroidal mixing~\cite{Berti:2005gp, Buonanno:2006ui, Kelly:2012nd, Berti:2014fga, London:2018nxs} (although this is not a fundamental limitation~\cite{Pompili:2023tna}) and beyond-linear QNM couplings~\cite{Cheung:2022rbm, Mitman:2022qdl, Lagos:2022otp, Baibhav:2023clw, Cheung:2022rbm, Bucciotti:2023ets, Perrone:2023jzq, Redondo-Yuste:2023seq, Ma:2024qcv, Bucciotti:2024zyp, Bourg:2024jme, Zhu:2024rej}.

For each mode, the QNM rescaled complex amplitude $\bar{h}(\tau)$ is in turn factored in an amplitude and a phase $\bar{h}(\tau)=A_{\bar{h}}(\tau) \e^{\i\phi_{\bar{h}}(\tau)}$.
We use different ans\"atze for the quasi-circular and eccentric cases, introduced below.\\

\paragraph*{HypTan:} 
The \emph{HypTan} template for the amplitude and phase is based on Ref.~\cite{Damour:2014yha}, and given by

\begin{eqnarray}
    A_{\bar{h}}(\tau) &=& c_1^A \tanh\big(c_2^A \tau + c_3^A\big) + c_4^A \, , \label{eq:qc_amp} \\
    \phi_{\bar{h}}(\tau) &=& -c_1^{\phi}\ln\!\left(\frac{1 + c_3^{\phi} \e^{-c_2^{\phi}\tau} + c_4^{\phi} \e^{-2c_2^{\phi}\tau}}{1 + c_3^{\phi} + c_4^{\phi}}\right) \, . \label{eq:qc_phase}
\end{eqnarray}

where $\{c_3^A, \, c_3^{\phi}, \, c_4^{\phi}\}$ are the free parameters that we aim to extract from numerical simulations.

The rest of the coefficients are constrained by the NR amplitude at merger, and its derivative, along with the phase evolution~\cite{Damour:2014yha}, as detailed in Appendix~\ref{app:constraints}.
To determine them, we use the parametric fits of the amplitude and frequency at merger obtained in the quasi-circular case in Ref.~\cite{Nagar:2019wds}. 
In quasi-circular models, $t_{\rm ref}$ is usually set to correspond to the peak of the $h_{22}$ mode~\cite{Bhagwat:2017tkm}.
Hence, these quantities are defined as $A_{22}^{\rm mrg} \equiv A_{22}(t=t_{\rm mrg}) = \underset{t}{\max} \, |A_{22}(t)|$ and $f_{22}^{\rm mrg} \equiv 2\pi \dot{\phi}_{22}(t_{\rm mrg})$, where $h_{22}(t) = A_{22}(t) \cdot \e^{i\phi_{22}(t)}$ and $t_{\rm ref} = t_{\rm mrg}$.
To geometrically scale the amplitudes, frequencies, and time axis, we use the NR simulation metadata. 
We use the symmetric mass ratio $\nu$ in the metadata to construct global fits for the free coefficients using a non-eccentric parameterization.\\

\paragraph*{RatExp:}
In this template, we use a modified factorization for the amplitude as a rational exponential, introduced in Ref.~\cite{Albanesi:2023bgi}, to account for the more complex structure observed in eccentric orbits

\begin{eqnarray}
    A_{\bar{h}}(\tau) &=& \left(\frac{c_1^A}{1 + \exp(-c_2^A\tau + c_3^A)} + c_4^A\right)^{\frac{1}{c_5^A}} \, , \label{eq:nc_amp}
\end{eqnarray}

while the phase template is instead the same as Eq.~\eqref{eq:qc_phase}, so that $\{c_2^A, \, c_3^A, \, c_2^{\phi}, \, c_3^{\phi}, \, c_4^{\phi}\}$ are free parameters for this template.
Note that in the \emph{HypTan} template $c_2^{\phi}$ was constrained by the pre-merger emission, while here we leave it as a free parameter to improve the phase description of $\ell=m$ modes~\cite{Albanesi:2023bgi}.
Continuity relations at $\tau=0$ for the amplitude, its first two time derivatives, and the frequency, fix the other coefficients~\cite{Albanesi:2023bgi}, as elaborated in Appendix~\ref{app:constraints}.

In highly eccentric scenarios, the peak of the waveform does not generally provide an accurate estimate of the merger time~\cite{Carullo:2023kvj, Carullo:2024smg} due to a large burst of GW radiation on the closest approach before merger~\cite{Carullo:2023kvj, Nagar:2020xsk, Gamba:2024cvy, Albanesi:2024xus, Kankani:2024may}, resulting in non-monotonicity of the amplitude peak against eccentricity.
Thereby, in the \emph{RatExp} template, we define the merger time ($t_{\rm ref} = t_{\rm mrg}$) as the inflection point of the gradient of the waveform phase (i.e., waveform frequency) right before the QNM-dominated phase (see Fig.~\ref{fig:single_sim_waveform}).
We found that this latter choice provided slightly smoother relationships than setting this time as the last amplitude peak before merger, as in Ref.~\cite{Carullo:2024smg}.
Due to the simulation's finite resolution, before determining the inflection point, we apply a filter to eliminate high-frequency noise.
We use the merger-remnant quantities as previously described to constrain the coefficients for the amplitude and phase, and proceed with constructing parametric fits for the free coefficients against notions of eccentricity.
The merger time (inflection point) then determines the NR merger frequency ($f_{22}^{\rm mrg}$) amplitude $A_{22}^{\rm mrg}$ and its derivative at merger $\ddot{A}_{22}^{\rm mrg}$ as defined above.
These quantities are then used to fix the remaining template coefficients as detailed in Ref.~\cite{Albanesi:2023bgi}.

The improvements observed in the non-circular case, showcased below, can be attributed not only to the new functional form of the \emph{RatExp} template, but also the inclusion of eccentricity into the merger data to constrain the template coefficients, and to the incorporation of the non-circular parameters into the parameter-space fit.

\section{Fitting algorithm workflow and validation}\label{sec:extraction}

\subsection{Coefficients extraction procedure from single simulations}\label{sec:single_extraction}

We perform a Bayesian fit to determine the amplitude and phase coefficients defined in Sec.~\ref{sec:model} and reported in Table~\ref{tab:priors}, treated as the free parameters to be inferred from the numerical data.
We employ the Bayesian algorithm implemented in the \texttt{bayRing} package~\cite{bayRing, Redondo-Yuste:2023seq}, relying on the \texttt{cpnest} nested sampler~\cite{cpnest}. 
We determined the appropriate sampler settings as a balance between computational speed and accuracy of extraction after testing for convergence (by increasing sampler settings).
We employ $256$ live points and $512$ maximum Markov Chain Monte Carlo (MCMC) steps, an evidence tolerance of $0.1$, with three independent chains set up at different random initial seeds. 
For reporting, we selected the run with the largest evidence and quoted the posterior median as the point estimate with the $90\%$ credible interval as the uncertainty.
We verified that the three chains produced consistent posterior medians and evidence estimates, ensuring the selected run is representative.
The priors were wide enough to capture the variation of the amplitude and phase in their domains of existence, and to ensure accurate recovery, and are also outlined in Table~\ref{tab:priors}.
In all cases, the phase $\phi_{22}^{\rm ref}$ is varied uniformly in $\mathcal{U}[0,\,2\pi]$.

\begin{table}[htp!]
    \centering
    \begin{tabular}{c c}
        \toprule
        \textbf{Coefficient} & \textbf{Prior} \\
        \midrule
        \multicolumn{2}{c}{\textit{Quasi-Circular}} \\
        \midrule
        $c_3^A$ & $\mathcal{U}[-5,0.1]$ \\
        $c_3^{\phi}$ & $\mathcal{U}[2,7]$ \\
        $c_4^{\phi}$ & $\mathcal{U}[0,10]$ \\
        \midrule
        \multicolumn{2}{c}{\textit{Eccentric}} \\
        \midrule
        $c_2^A$ & $\mathcal{U}[-5,5]$ \\
        $c_3^A$ & $\mathcal{U}[-5,1)$ \\
        $c_2^{\phi}$ & $\mathcal{U}[0,10]$ \\
        $c_3^{\phi}$ & $\mathcal{U}[0,10]$ \\
        $c_4^{\phi}$ & $\mathcal{U}[0,10]$ \\
        \bottomrule
        \end{tabular}
        \caption{List of parameters and priors for the Bayesian extraction of the amplitude and the phase coefficients. 
        We have $\{c_3^A, \, c_3^{\phi}, \, c_4^{\phi}\}$ as free parameters for the quasi-circular template, and $\{c_2^A, \, c_3^A, \, c_2^{\phi}, \, c_3^{\phi}, \, c_4^{\phi}\}$ for the eccentric one.}
    \label{tab:priors}
\end{table}   

Since RIT waveforms have a single resolution and extraction radius, a Gaussian error estimate on the GW complex strain is assumed, thereby setting the likelihood function.
The estimate of the noise floor is obtained by extracting the late-time constant amplitude from the simulations, which sets the precision threshold required for the fit.
We use a consistent estimate for the errors in the strain obtained from the SXS catalog.
Data beyond $t_{\rm end} = t_{\rm peak} + 80 \, M$ are excised from the fit.
Although the choice of the end-time only slightly reduces signal power for lower eccentricities, it eliminates contamination from simulation artifacts at late times across all scenarios, including those with lower data quality~\cite{Carullo:2023kvj, Carullo:2024smg}.

To quantify the goodness of the template waveform, we further compute the mismatch~\cite{Owen:1995tm, Ajith:2007kx} between the NR simulation, $h_{\ell m}$, and the template model, $\mathsf{h}_{\ell m}$ as the effective norm of the difference of the normalized strain, or equivalently, as the complement of the match (overlap) of the waveforms, given by

\begin{align}
    \mathcal{M}[h_{\ell m}, \mathsf{h}_{\ell m}] &= \frac{1}{2}\left\Vert \frac{\mathsf{h}_{\ell m}}{\sqrt{\left\Vert \mathsf{h}_{\ell m}\right\Vert^2}} - \frac{h_{\ell m}}{\sqrt{\left\Vert h_{\ell m}\right\Vert^2}} \right\Vert^2 \nonumber\\
    &= 1-\frac{\left\langle\mathsf{h}_{\ell m}|h_{\ell m} \right\rangle}{\sqrt{\left\Vert h_{\ell m}\right\Vert^2\left\Vert\mathsf{h}_{\ell m}\right\Vert^2}} \, ,
\end{align}

where $\left\Vert a \right\Vert^2=\left\langle a | a \right\rangle$, and the overlap is computed after time and phase alignment, with the inner product defined in the time-domain~\cite{Berti:2025hly, Isi:2021iql,  Crescimbeni:2025ytx} by 

\be
    \left\langle a| b \right\rangle = \sum_{i,\,j=N_i}^{N_{\rm trunc}} a_i \cdot \mathsf{C}_{ij}^{-1} \cdot b_j \, ,
\ee

where $\mathsf{C}$ is the noise-covariance matrix, and we define the scalar product in a finite region $[t_{\rm start},t_{\rm end}]$, setting $N_i,N_{\rm trunc}$.
To compute the mismatches, we utilize the advanced LIGO design sensitivity curve: \texttt{aLIGODesignSensitivityT1800044}~\cite{LIGOT1800044} for the noise-covariance, and use GW150914-like values~\cite{LIGOScientific:2016lio} for the total mass, $M=60~{\rm M}_{\odot}$, to rescale the waveform phasing from geometric to physical units.
Finally, residuals are defined as $\delta h_{\ell m}=h_{\ell m}-\mathsf{h}_{\ell m}$, where $h_{\ell m}$ is the NR simulation, and $\mathsf{h}_{\ell m}$ the template model.

\begin{figure*}[ht!]
    \centering
    \includegraphics[width=0.37\textwidth]{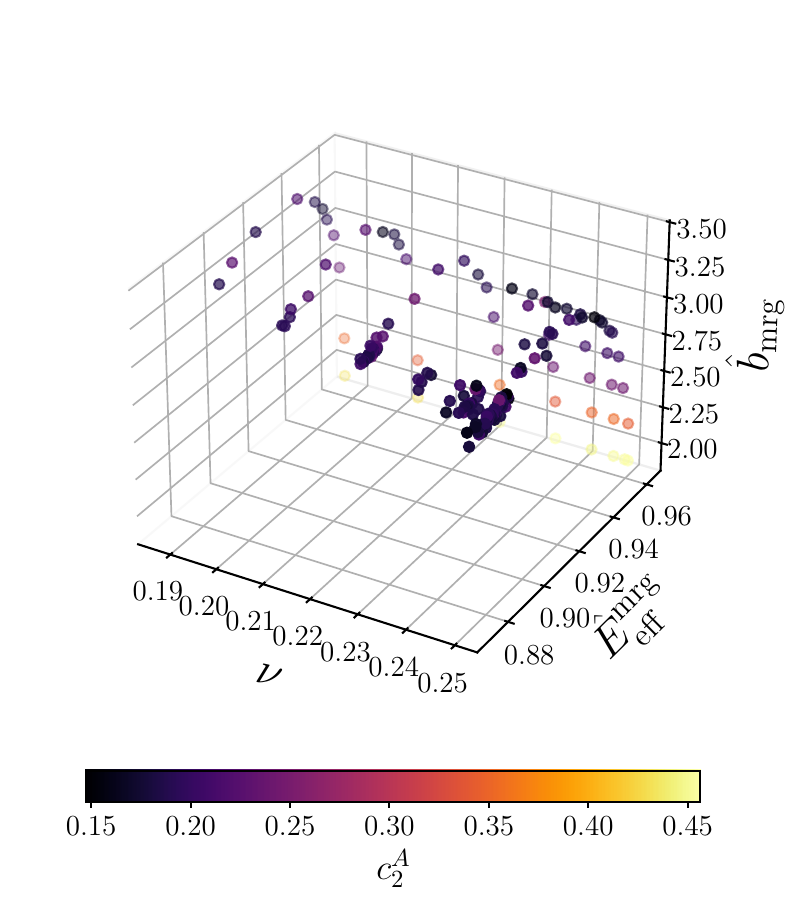}   
    \includegraphics[width=0.37\textwidth]{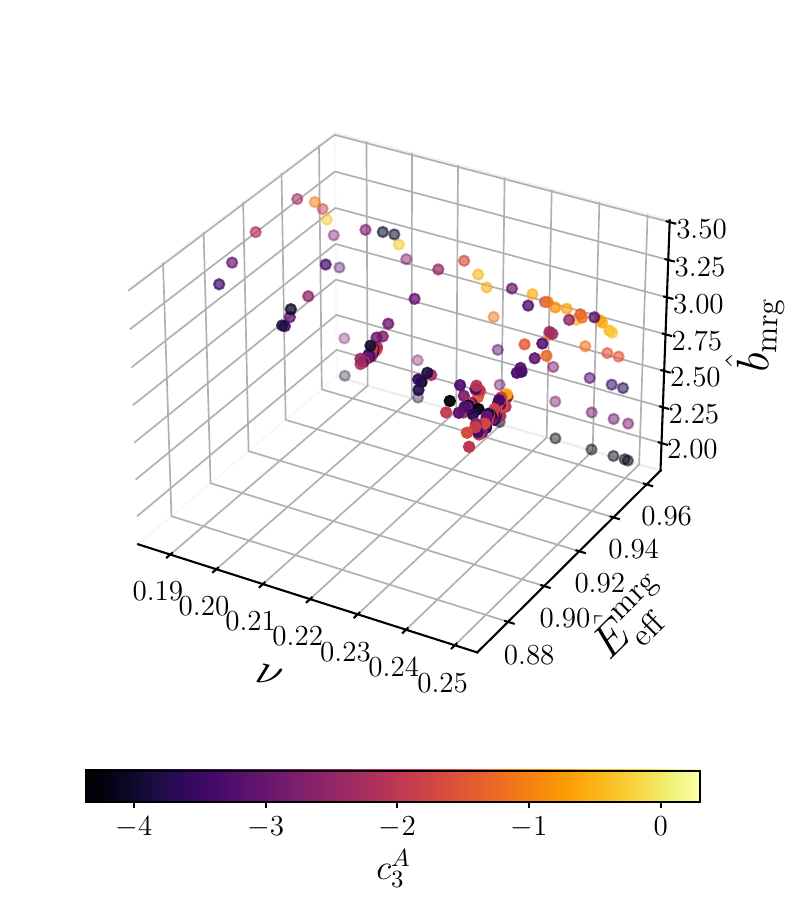}   
    \includegraphics[width=0.37\textwidth]{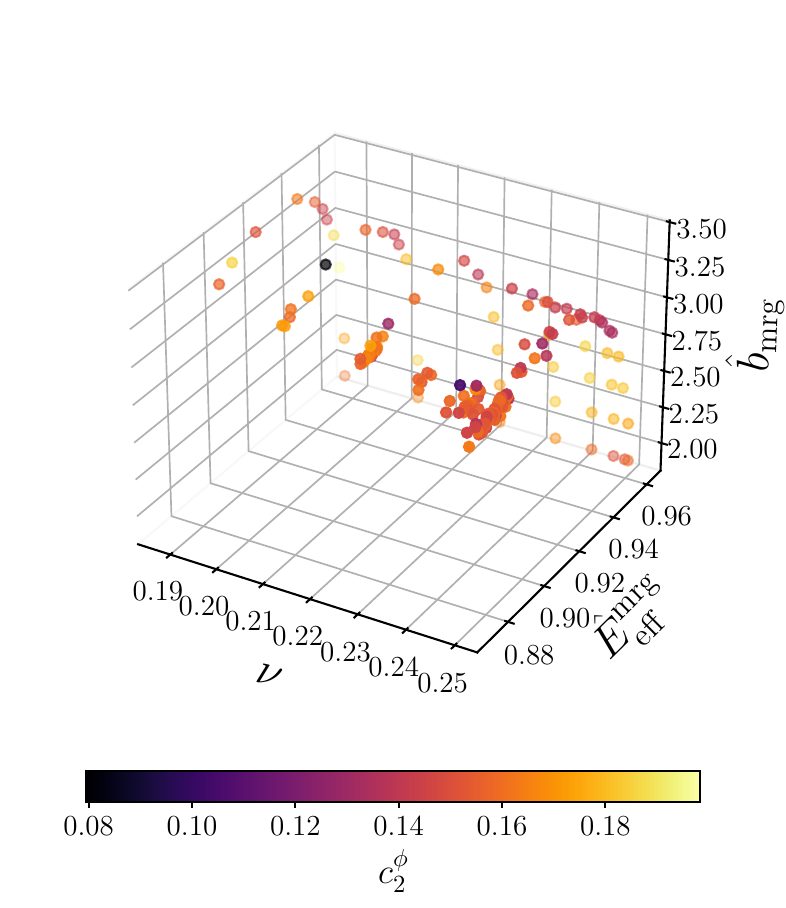}   
    \includegraphics[width=0.37\textwidth]{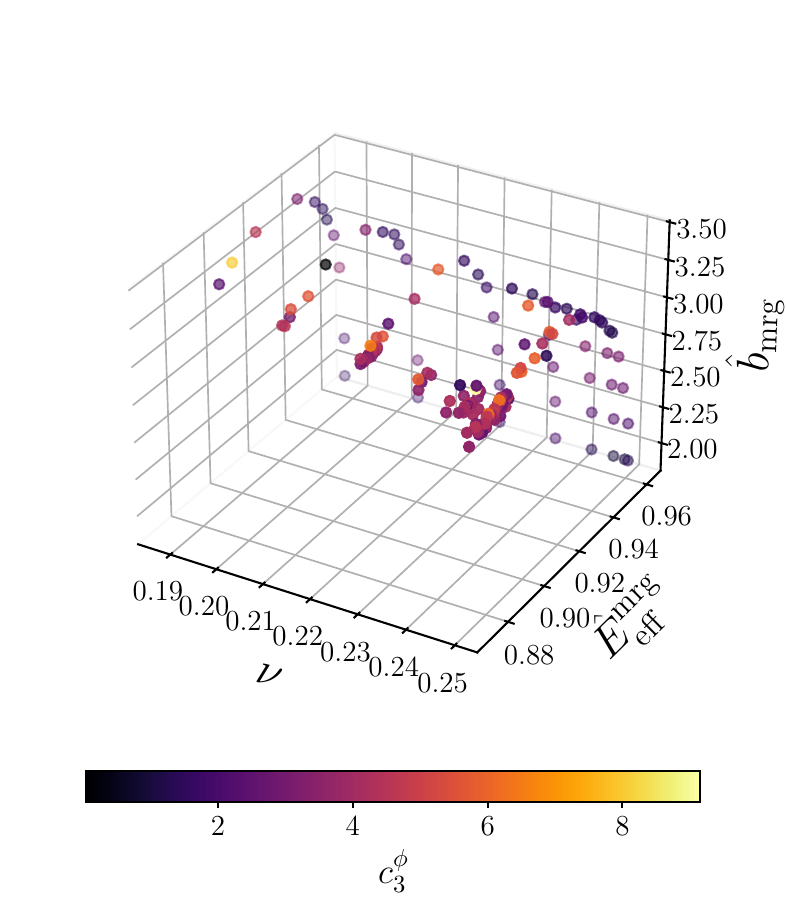}   
    \includegraphics[width=0.37\textwidth]{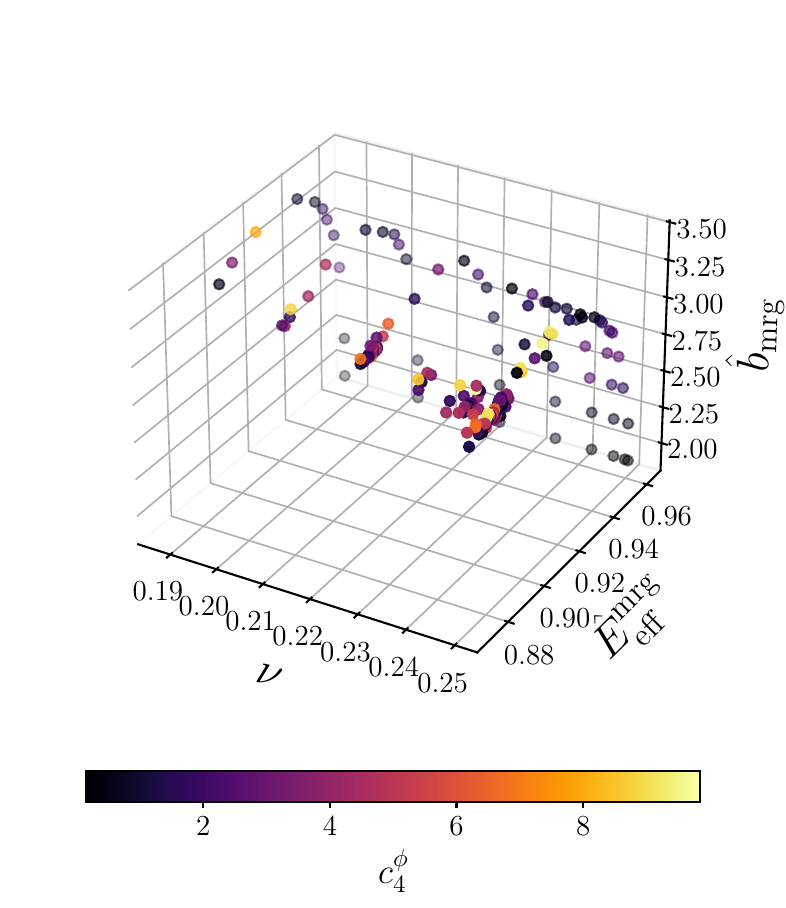}
    \includegraphics[width=0.39\textwidth]{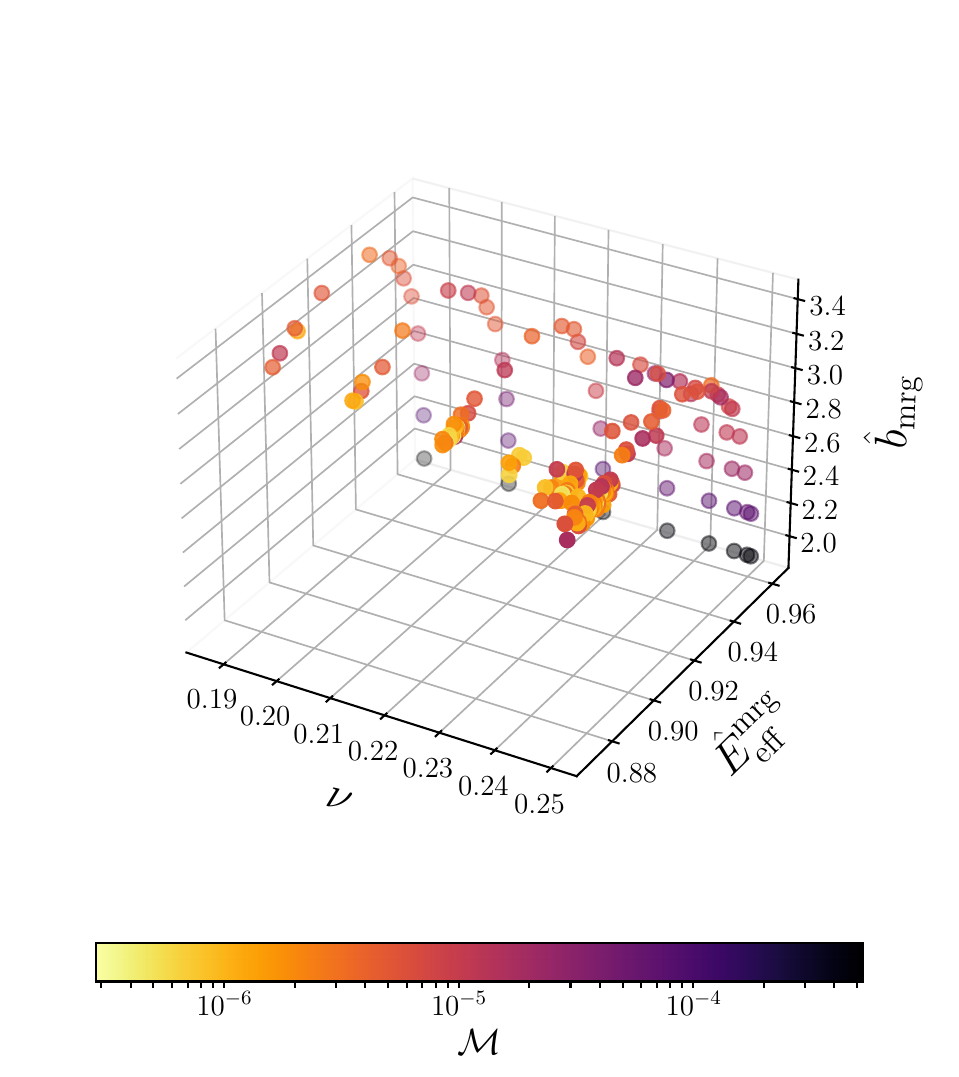}
    \caption{
    \textit{Top, Middle, Bottom Left:} Extracted median amplitude and phase coefficients from the Bayesian fits for the eccentric non-spinning RIT catalog in terms of two non-circular parameters $\{\bmrg,\, \Eeffmrg\}$ and mass ratio $\nu$.
    These results constitute the dataset entering the global fits constructed below, to describe the coefficient variation as a function of the binary parameters.
    \textit{Bottom Right:} Mismatch of the template model with the NR simulation, obtained from extracting the coefficients through the Bayesian fits independently for each simulation (local fits), as a function of the two non-circular parameters and mass ratio.
    }
    \label{fig:nc_fits}
\end{figure*}

\subsection{Global fit construction}\label{subsec:bayesian_global_fit}

After fitting the amplitude and phase coefficients $\{c_2^A, \, c_3^A, \, c_2^{\phi}, \, c_3^{\phi}, \, c_4^{\phi}\}$ for each NR simulation (``local fits''), through the Bayesian procedure just described, we proceed to construct fits of the coefficients variation across the binary parameter space.
We parameterize the coefficients using the symmetric mass ratio $\nu$ and effective spin $\chi$, and combinations of non-circular parameters $\{\bmrg, \, \Eeffmrg, \, \jmrg\}$~\cite{Carullo:2023kvj, Carullo:2024smg}.

The procedure is similar to what is described in Refs.~\cite{Carullo:2023kvj, Carullo:2024smg}, but now using a simpler polynomial fitting template for the parameters of interest

\be\label{eq:template_2D}
    \tilde{Y} = Y_0 + \sum_{k=1}^{\mathsf{d}}\sum_{|\boldsymbol{\alpha}|=k}p_{\boldsymbol{\alpha}}\prod_{i=1}^m Q_i^{\alpha_i} \, ,
\ee

where $\boldsymbol{Q}=\{Q_1, \, \dots, \, Q_m\}$ are $m$ binary parameters, $\{p_{\boldsymbol{\alpha}}\}$ with $p_{\boldsymbol{\alpha}} \in \mathbb{R}$ the global coefficients to be determined, and $\mathsf{d}$ being the maximum polynomial degree considered in the fit, such that the multi-index $\boldsymbol{\alpha}=\{\alpha_1\, \dots, \, \alpha_m\}$ where $\alpha_i \in \mathbb{N}$, with total degree $|\boldsymbol{\alpha}|=\sum_i\alpha_i\leq\mathsf{d}$.
The target quantities are $\tilde{Y}  \in \{c_2^A, \, c_3^A, \, c_2^{\phi}, \, c_3^{\phi}, \, c_4^{\phi}\}$.
We consider single, dual and triple combinations of parameters $\boldsymbol{Q} = \{\nu, \, \bmrg, \, \Eeffmrg, \, \jmrg\}$ with $m=1, \, 2, \, 3$ respectively, and we investigate $\mathsf{d}\in\{1, \, 2, \, 3\}$ as linear, quadratic and cubic fits.
To gain intuition on the fits' behaviour, we also consider fits against the gauge-dependent parameter $e_0$, which is not used in any of the final fits construction.

From the obtained posterior distributions in the local fits, we compute the medians $Y$ and their respective errors $\sigma$ defined as the $90\%$ confidence intervals. 
We then proceed with Bayesian sampling, relying again on the nested sampling algorithm implemented in \texttt{cpnest}~\cite{cpnest}, to extract the global dependence of the coefficients against the relevant parameter space of symmetric mass ratio and non-circular parameters $\boldsymbol{Q}$.
We verify the accuracy and convergence of the Bayesian sampling by comparing with a constrained multivariate global linear fit, as described in Appendix~\ref{app:lsq_linear}. 

We infer the posterior distribution of the polynomial coefficients $p_{\boldsymbol{\alpha}}$, describing the amplitude and phase coefficients $\tilde{Y}$ across the parameter space.
We use the (log)-likelihood function defined by the residuals, assuming that errors are Gaussianly distributed

\be
    \ln\mathcal{L}\propto-\frac{1}{2}\sum_i \frac{1}{(\sigma_i)^2}\left(\tilde{Y}_i-Y_i\right)^2 \, .
\ee

It is essential to account for errors by incorporating them into the inferred parameters, so that we not only obtain a point estimate of the quantities of interest but also present the complete information available at a given resolution, including associated uncertainties.

To prevent fits instabilities and overfitting, we (linearly) normalize the parameter space of $\boldsymbol{Q} = \{\nu, \, \bmrg, \, \Eeffmrg, \, \jmrg\}$ to $[-1,\,1]$ and use broad priors for the polynomial coefficients, $p_{\boldsymbol{\alpha}}\in(-500,\,500)$.
The physical bounds (constraints) on the parameters $\boldsymbol{Q}$ are imposed by setting a \texttt{NaN} likelihood to the cases excluding the values in Table~\ref{tab:priors}.
We use $256\times2^{\mathsf{d}-1}$ live points where $\mathsf{d}=\{1,\,2,\,3\}$ is the order of the fit, and use $4$ parallel processes to check for convergence and combine them in a cumulative posterior distribution by weighting them through their Bayesian evidence, with $5000$ maximum MCMC steps and an evidence tolerance of $0.1$.
We further checked for convergence by increasing the live points and reducing the evidence tolerance, and by assessing the stability of the posteriors and the inferred values.

The reduced chi-squared statistic $\tilde{\chi}^2$ is employed to assess the performance of the Bayesian nested sampling runs, and further compare with the least square residual minimization.
The obtained coefficients and a \texttt{python} implementation of the fits will publicly available, at the time of journal submission, on the Github repository \texttt{nc\_ringdown}~\href{https://github.com/nishkalrao20/nc_ringdown}{\textcolor{blue!50}{\faGithubSquare}}.

In Appendix~\ref{sec:quasi_circular_testing}, we validate our fitting algorithm by showing that in the quasi-circular case, we improve by an order of magnitude in mismatch upon previous results in the literature that relied on simpler least-squares fits.

\section{Eccentric Global fit}\label{sec:global_fit}

In this section, we first present the results of fitting the single RIT catalog simulations, and then present a closed-form global fit in terms of the binary parameters for the coefficients extracted above, using the dynamics-based parameterization of Ref.~\cite{Carullo:2023kvj}.

\subsection{Single-simulations results}

\begin{figure*}
    \centering
    \includegraphics[width=0.9\textwidth]{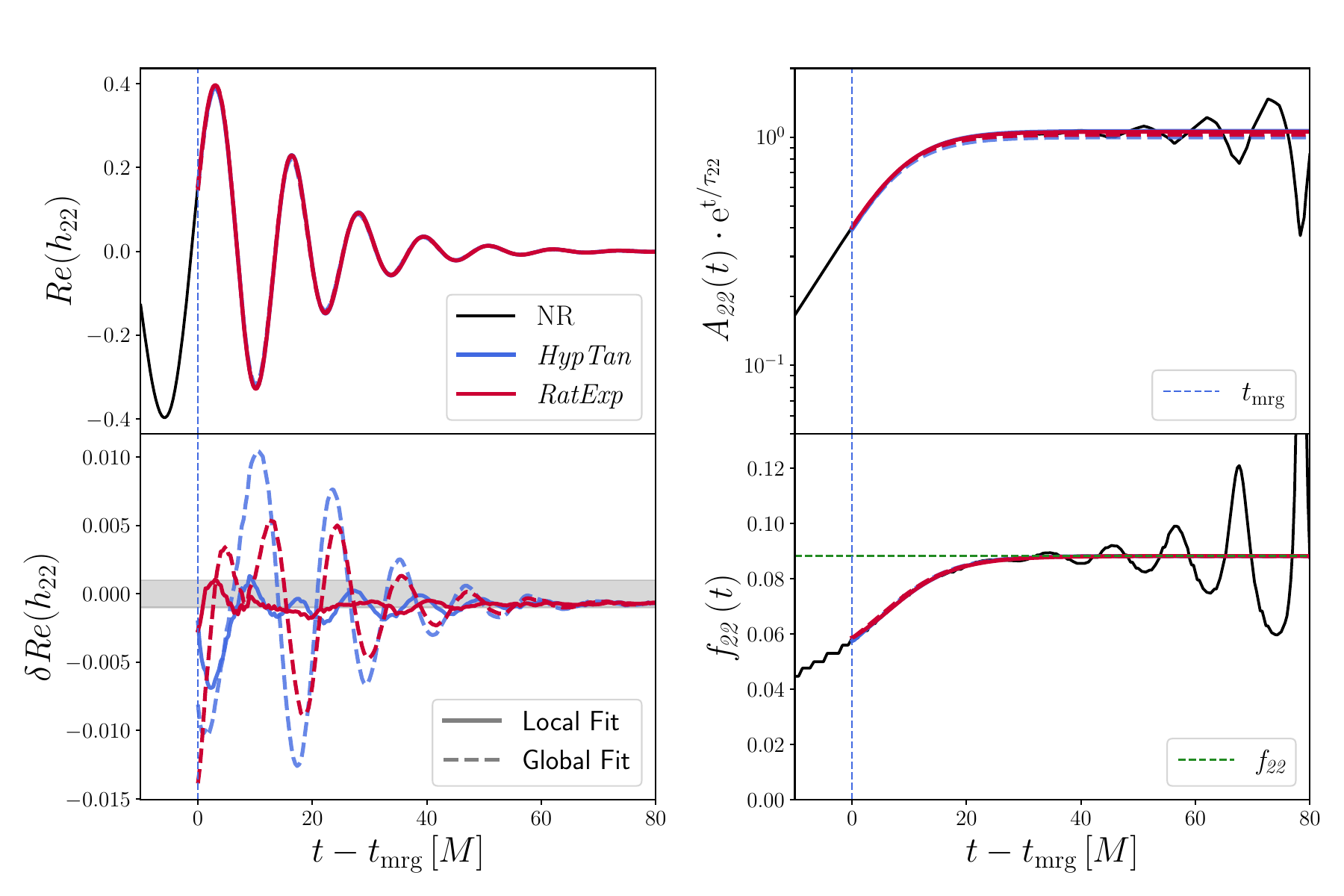}     
    \caption{
    Quadrupolar strain $h_{22}$ (black) for the illustrative case \texttt{RIT:1334}, with $e_0 \simeq 0.5$ and $q \simeq 1.1$.
    The red (blue) line represents the fit using the extracted coefficients for the \emph{RatExp} (\emph{HypTan}) template defined in Eq.~\eqref{eq:nc_amp} (Eqs.~\eqref{eq:qc_amp}, and~\eqref{eq:qc_phase}), fitting the numerical data only after $t_{\rm mrg}$.
    The \emph{RatExp} (\emph{HypTan}) templates respectively use eccentric (quasi-circular) merger data (parametric fits), as outlined in Sec.~\ref{sec:model}.
    The red dashed curve represents the global-fit coefficients used for the \emph{RatExp} template using a cubic parameterization involving $\{\nu,\,\bmrg,\,\Eeffmrg\}$, determining the parametric dependence of the coefficients in Eq.~\eqref{eq:nc_amp}.
    The blue dashed curve uses the past global-fit coefficients for the \emph{HypTan} template in Ref.~\cite{Nagar:2019wds} pertaining to Eqs.~\eqref{eq:qc_amp}, and~\eqref{eq:qc_phase}, implemented by default in \texttt{bayRing}.
    The corresponding solid curves instead represent the case where the same template is used, its coefficients are left free to vary, and inferred via nested sampling (local fit).
    The gray bar in the residual plot represents the NR error used for the Bayesian fit, and the green horizontal line represents the value of the fundamental QNM frequency $f_{22}$.    
    The residuals (bottom) and to a lesser extent, amplitude (top right) panels allow to distinguish the templates, unlike in the strain (top left) and frequency (bottom left) panels, where they overlap.
    The mismatches obtained for the local fits are $\mathcal{O}(10^{-5})$ for the \emph{HypTan} template and $\mathcal{O}(10^{-6})$ for the \emph{RatExp} template, while we obtain $\mathcal{O}(10^{-4})$ for both global-fit templates.
    }
    \label{fig:single_sim_waveform}
\end{figure*}

\begin{figure*}
    \centering
    \includegraphics[width=0.9\textwidth]{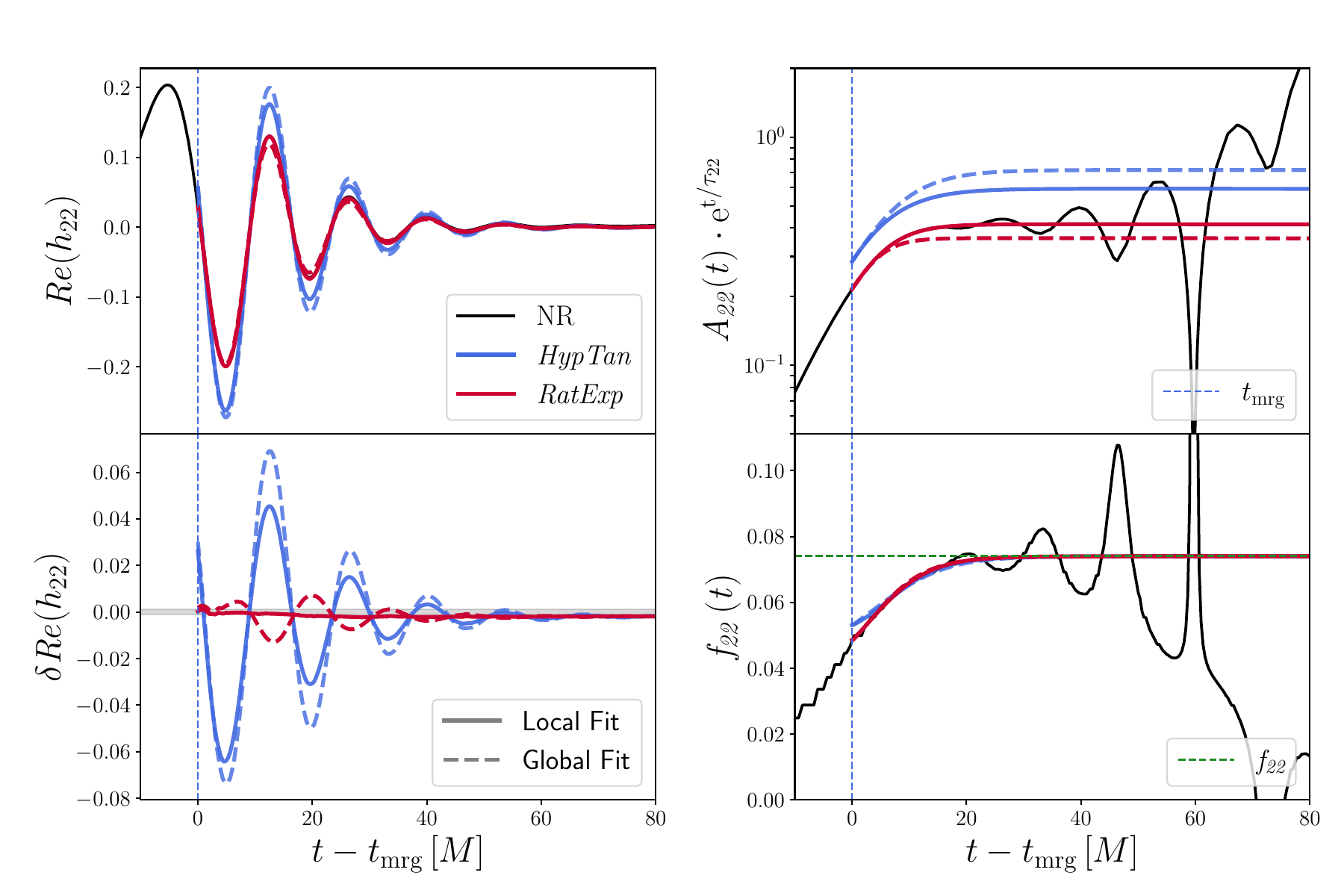}    
    \caption{Similar to Fig.~\ref{fig:single_sim_waveform}, NR strain $h_{22}$ for the illustrative case \texttt{RIT:1479}, with high eccentricity, $e_0 \simeq 0.75$ and $q \simeq 3$.
    The mismatches obtained for the local fits are $\mathcal{O}(10^{-3})$ for the \emph{HypTan} template and $\mathcal{O}(10^{-6})$ for the \emph{RatExp} template.
    The global fits for the eccentric template display a considerable improvement in performance in the presence of residual orbital eccentricity, predominantly due to the implementation of the non-circular merger data to constrain the coefficients as detailed in Sec.~\ref{sec:model}, resulting in mismatches of the order $\mathcal{O}(10^{-3})$ for the \emph{HypTan} template and $\mathcal{O}(10^{-5})$ for the \emph{RatExp} template.
    }
    \label{fig:single_sim_waveform_ecc}
\end{figure*}

We start by extracting the coefficients values for each eccentric RIT simulation using the \emph{RatExp} template Eq.~\eqref{eq:nc_amp} with coefficients $\{c_2^A, \, c_3^A, \, c_2^{\phi}, \, c_3^{\phi}, \, c_4^{\phi}\}$, via the Bayesian algorithm outlined in Sec.~\ref{sec:single_extraction}.

An illustrative case of the data quality of a simulation output with medium eccentricity ($e_0 = 0.51$) and close to equal-mass $q=1.11$, \texttt{RIT:1334}, is presented in Fig.~\ref{fig:single_sim_waveform}, where we also compare against a state-of-the-art quasi-circular global fit using the \emph{HypTan} template.
A more challenging case with higher eccentricity, in which the quasi-circular template yield a poor description, is instead shown in Fig.~\ref{fig:single_sim_waveform_ecc}.
The figure shows how the biggest improvement comes from the usage of non-circular merger-remnant data.

We repeat this procedure for all RIT simulations passing our data quality selection criteria.
The median parameters of the coefficients describing the amplitude and phase, extracted from the Bayesian fits for all simulations, are summarised in Fig.~\ref{fig:nc_fits}.
To highlight their variation within the parameter space, we plot them as a function of two non-circular parameters $\bmrg$, $\Eeffmrg$, and mass ratio $\nu$.
A non-monotonic (as expected for eccentric data), but overall smooth structure across the parameter space is observed.
A few outliers (discussed in more detail below) are also observed, suggesting undetected inaccuracies in the simulations that our data-quality filter, based on balance laws, does not capture.
The mismatch between the reconstructed waveforms and the RIT counterparts is also presented, ranging between $10^{-6}-10^{-3}$, showcasing the accuracy of the chosen template.

Finally, a note on the $c_2^A$ results.
We noticed strong bi-modalities of the $c_2^A$ nested samples symmetrically around zero.
Depending on the simulation, the posterior samples mostly fell in one of these peaks, a behavior consistently observed across multiple randomly-seeded chains or when increasing sampler settings.
This resulted in a set of $c_2^A$ posteriors from different simulations spanning symmetrically both positive and negative values.
These symmetric peaks denote a strong degeneracy in the template used between the $c_2^A$ and $c_3^A$ coefficients, suggesting that the template could be improved by removing one of the two coefficients, and possibly extending the template time-dependence to recover the more complex cases.
Since the majority of the posteriors ended up falling in the $c_2^A>0$ branch, when constructing the global fit, we decided to disregard the simulations falling within the $c_2^A<0$ branch, ensuring a monotonic fitting behaviour.
This still yields an accurate global fit (see below) since the template waveform is only very weakly affected by jumping between the two branches, namely by a sign inversion of $c_2^A$ accompanied by a corresponding shift in $c_3^A$.

A correlation was sometimes observed also for $c_2^{\phi}$ and $c_3^{\phi}$, albeit to a lesser extent.
This correlation did not instead give rise to bi-modal islands when aggregating posteriors across different simulations, which is why it did not have an impact on our global fit procedure. 
The bounds on the coefficients in Table~\ref{tab:priors} have been chosen to prevent further degeneracy issues while also avoiding overly restrictive priors, thereby ensuring both an accurate global fit, and an agnostic exploration of the template parameter space.

\subsection{Non-circular parameterization}\label{subsec:par}

Once the ringdown template coefficients are obtained from each simulation, we aim to connect them throughout the parameter space, for which we assume the parameterisation of~\cite{Carullo:2023kvj, Carullo:2024smg} in terms of the binary parameters, constructed as follows.
Starting from the ADM energy and angular momentum calculated at the beginning of the simulation $\{E^{\rm ADM}_0, \, J_0^{\rm ADM}\}$ and taking into account the losses due ot GW radiation, it is possible to compute the values of the mass-rescaled (adimensional) effective energy and angular momentum at merger $\{\Eeffmrg, \, \jmrg\}$.
These two parameters, coupled with the binary mass ratio $\nu$, are sufficient to describe the coefficient variations in the nonspinning binary parameter space for generic orbits.

Alternatively, a convenient combination of these variables in the form of an effective \textit{impact parameter} at merger, $\bmrg$, is constructed.
The latter parameter was previously considered for bound orbits, both in the test-mass~\cite{Albanesi:2023bgi} and comparable-mass limit~\cite{Carullo:2023kvj, Carullo:2024smg}.
In the latter case, accurate expressions of the merger properties and of the remnant BH parameters $\{M_f, \, a_f\}$ were obtained as a function of $\bmrg$ only.
Since in principle these quantities should be described by two parameters (e.g. $\{\Eeffmrg, \, \jmrg\}$), accurate fits in terms of the single variable $\bmrg$ was taken as indicating a ``quasi-universal'' behaviour~\cite{Carullo:2023kvj, Carullo:2024smg}.
However, ringdown late-time complex amplitudes were found to require more than a single parameter~\cite{Carullo:2024smg}.
Hence, in what follows, we consider various combinations of $\{\nu, \, \bmrg, \, \Eeffmrg, \, \jmrg\}$ to construct the global fit, aiming to find the best combination of parameters allowing for an accurate global fit.

\begin{figure*}[htp!]
    \centering
    \includegraphics[width=0.9\textwidth]{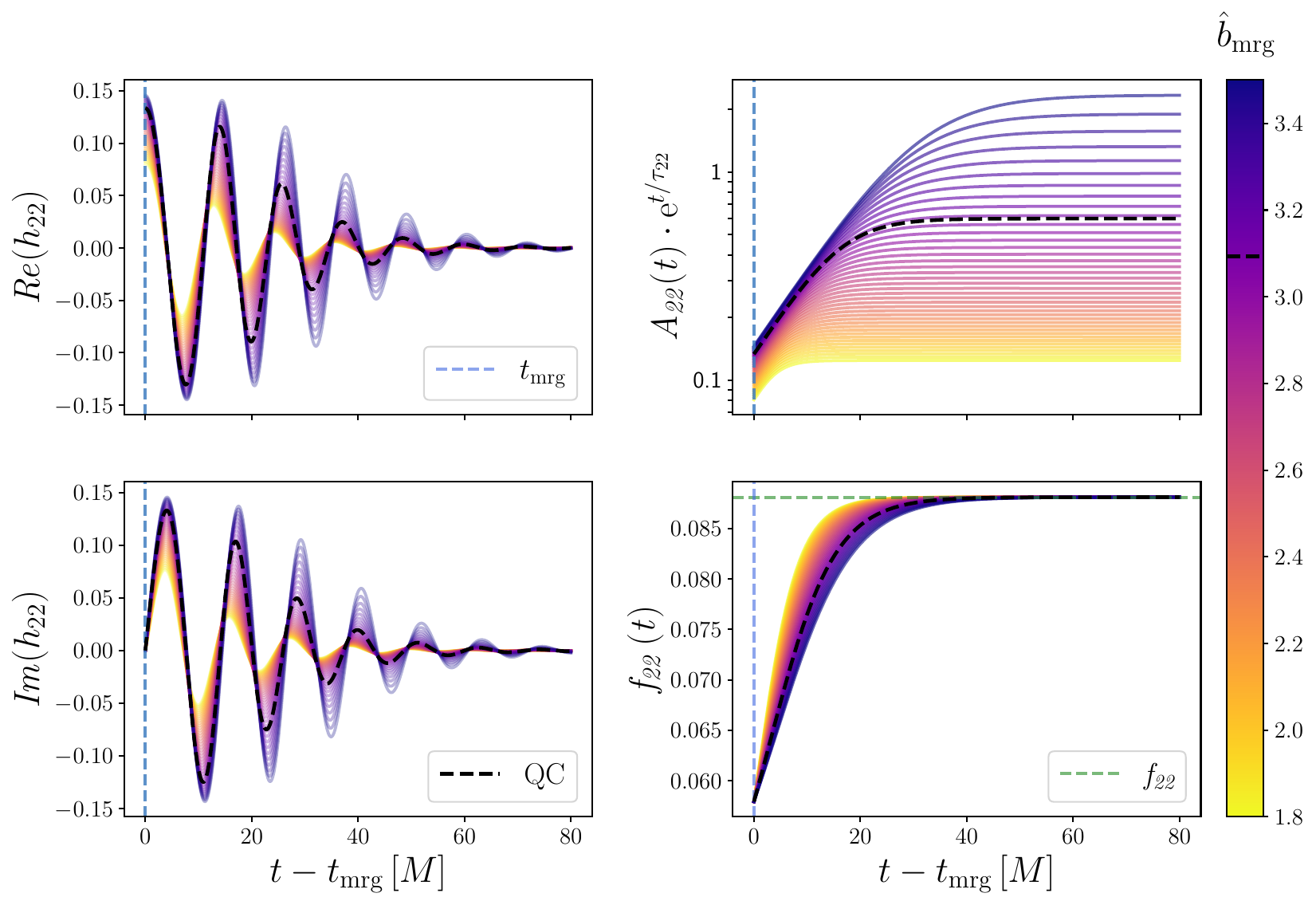}   
    \caption{
    Waveform dependency on the effective impact parameter, $\bmrg$, assuming a single-parameter ($m=1$) polynomial global cubic ($\mathsf{d}=3$) parametric fit for the amplitude and phase coefficients against NR data.
    The gray dashed line represents the quasi-circular waveform for $\bmrg\approx3.1$.
    }
    \label{fig:waveform_bmrg}
\end{figure*}

To showcase how these parameters impact the waveform template, in Fig.~\ref{fig:waveform_bmrg}, we display the waveform (both oscillatory components, and amplitude and phase) variation as a function the dynamical impact parameter $\bmrg$, when assuming a polynomial global fit (constructed below) of the amplitude and phase coefficients given by Eq.~\ref{eq:template_2D}, with $\mathsf{d}=3$.
The time-dependent amplitude (frequency) displays a smooth monotonic increase (decrease) with $\bmrg$.

\subsection{Global fits results}

\begin{figure*}[htp!]
    \centering
    \includegraphics[width=.625\textwidth]{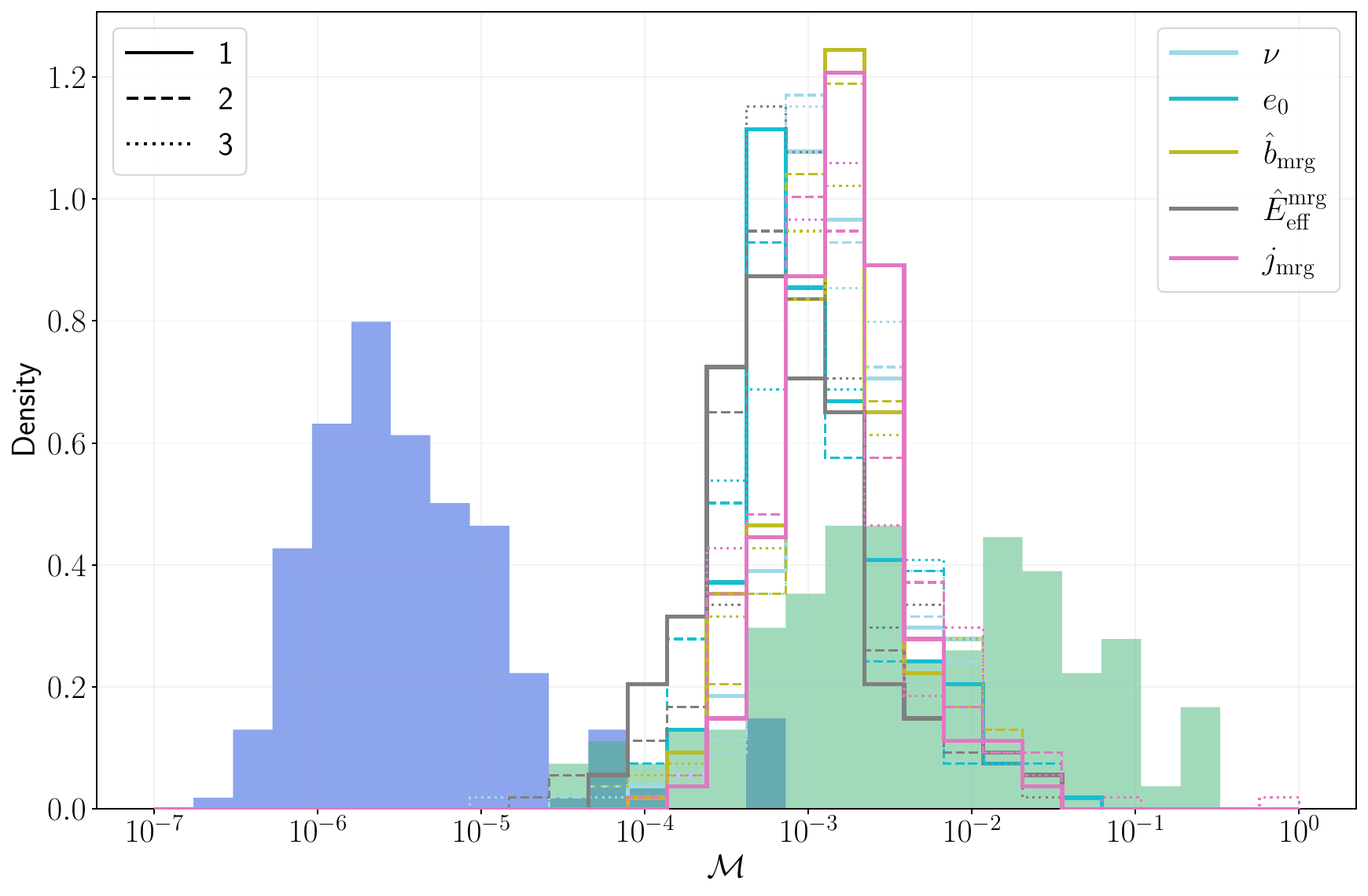}
    \includegraphics[width=.625\textwidth]{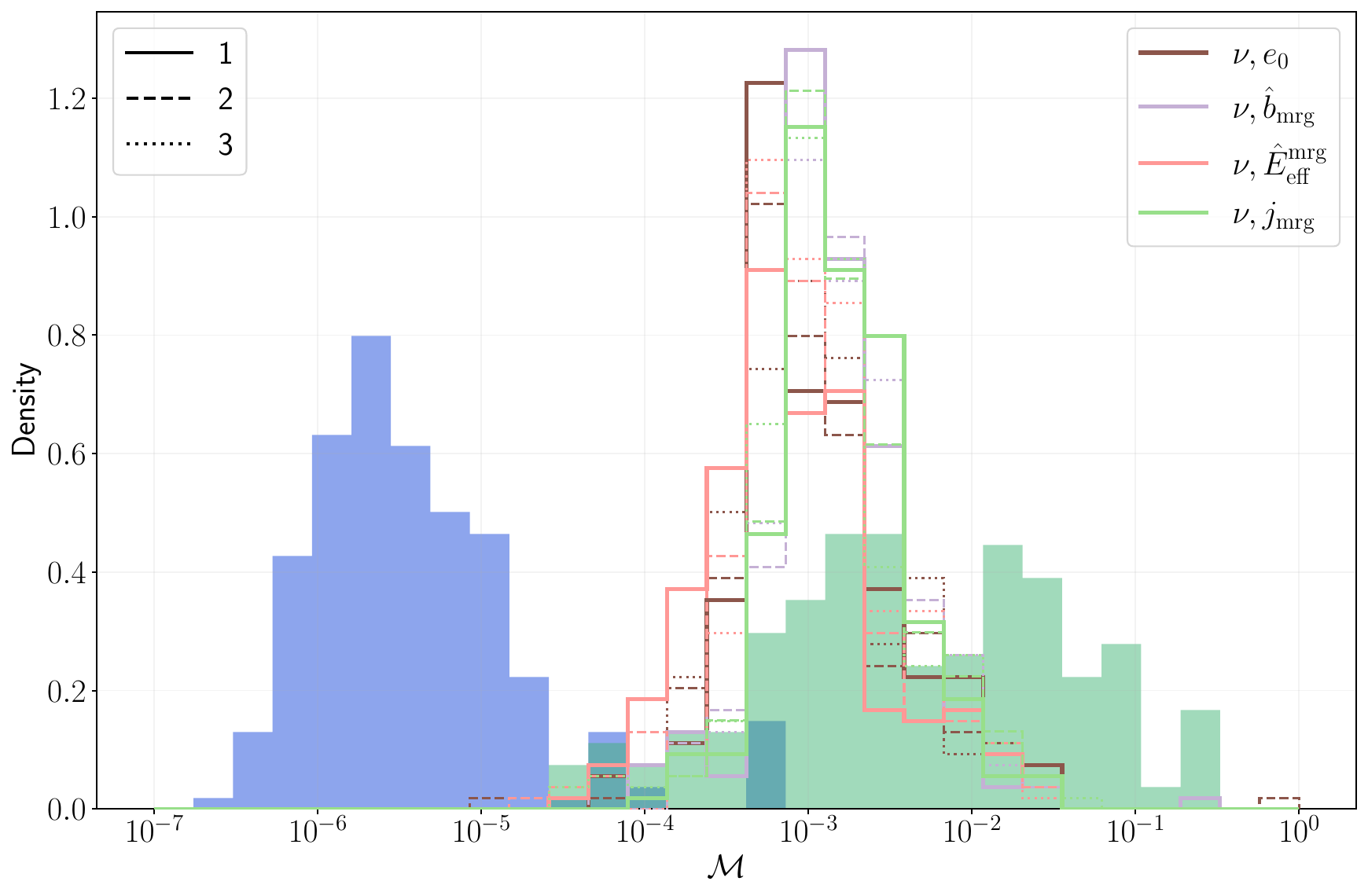}
    \includegraphics[width=.625\textwidth]{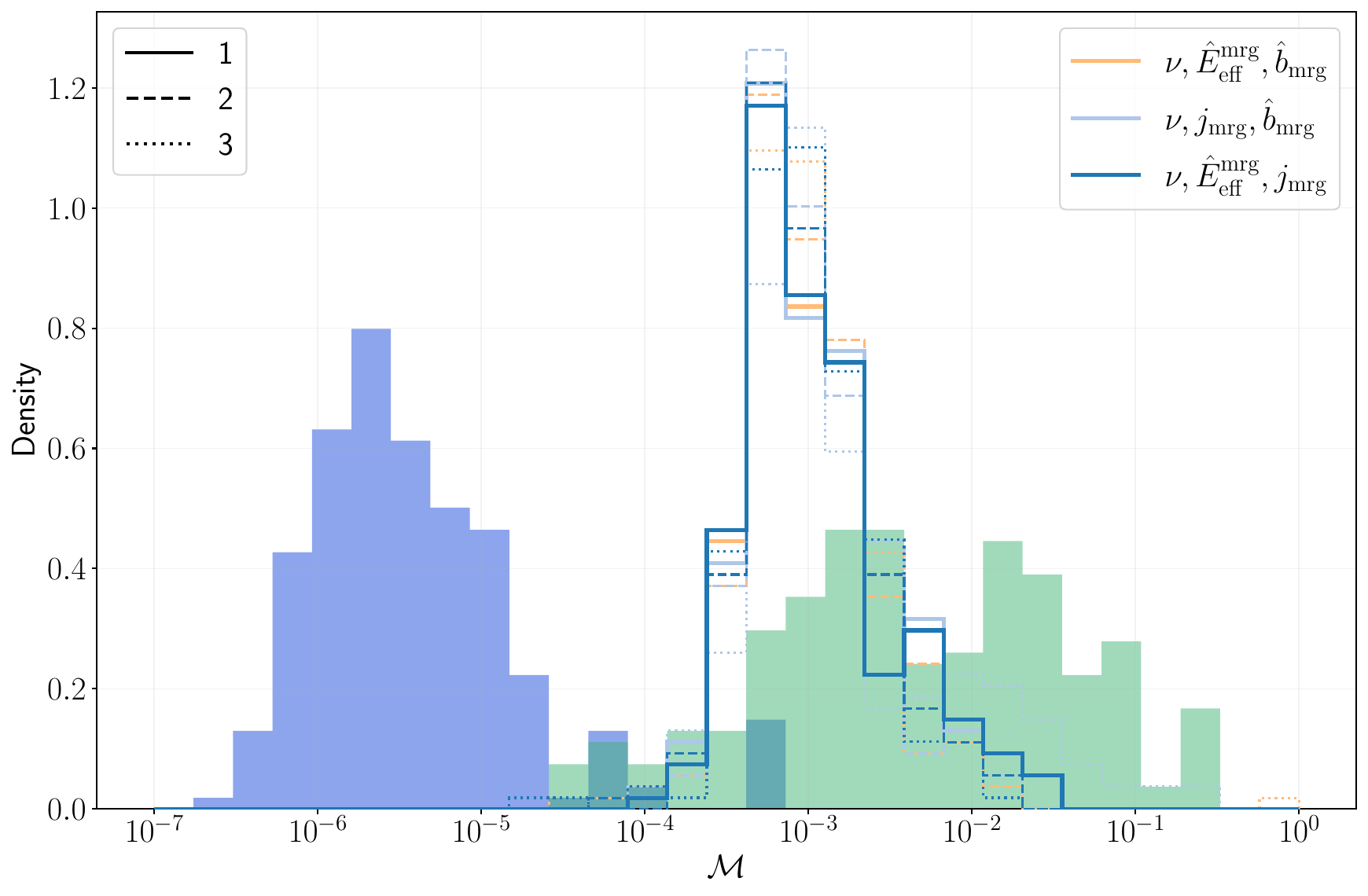}
    \caption{
    Mismatches of the \emph{RatExp} waveform template against NR simulations, when assuming global fits of the amplitude and phase coefficients as a function of the binary parameters shown in the legend.
    For each parameter combination, we present the mismatches for a linear, quadratic, and cubic maximum polynomial orders with \emph{top:} single, \emph{middle:} dual (single eccentric parameter with symmetric mass ratio), \emph{bottom:} triple parameter (dual dynamics-based parameterization with symmetric mass ratio) fits.
    The filled histogram in light blue represents the results of the corresponding local fits, while the green corresponds to the quasi-circular template.
    }
   \label{fig:mismatch_histograms_bayesian}
\end{figure*}

We generate parametric fits for the amplitude and phase coefficients $\{c_2^A, \, c_3^A, \, c_2^{\phi}, \, c_3^{\phi}, \, c_4^{\phi}\}$ extracted from single simulations, via the Bayesian algorithm outlined in Sec.~\ref{sec:single_extraction}.
We parameterize their dependence via symmetric mass ratio and a combination of non-circular parameters, specifically combinations of one, two, or three of $\{\nu, \, e_0, \, \bmrg, \, \Eeffmrg, \, \jmrg\}$, namely $m=\{1,2,3\}$ in Eq.~\ref{eq:template_2D}.
For each parameter combination, we consider linear, quadratic and cubic fits, namely $\mathsf{d}=\{1,2,3\}$ in Eq.~\ref{eq:template_2D}.
Given the result of each of these global fits, we generate the corresponding ringdown waveform templates, now only as a function of $\{\nu, \, e_0, \, \bmrg, \, \Eeffmrg, \, \jmrg\}$.
The effectiveness of the template is verified by computing mismatches between the global-fit waveform template and the NR data, fixing all parameters except for a free phase.
The results of this procedure are displayed in Fig.~\ref{fig:mismatch_histograms_bayesian}, which constitutes the main result of this work.

In all cases, the vast majority of our template mismatches lies within $[10^{-4}, 10^{-2}]$, an order of magnitude improvement with respect to existing quasi-circular fits.
As expected, the mismatches monotonically improve when increasing the number of parameters through which the fits are constructed, and generally decrease when increasing the polynomial order.
The third-order global fit for the amplitude and phase coefficients as functions of the symmetric mass ratio and two eccentric parameters, $\{ \nu, \Eeffmrg, \, \jmrg \}$ with $\mathsf{d}=3$, is found to yield the best representation of the numerical data (closely followed by the replacement of $\jmrg$ with $\bmrg$, in agreement with expectations from previous work~\cite{Carullo:2024smg}).
For this best-performing three parameter combination, no mismatches above $10^{-1}$ are present, with only a few simulations above $10^{-2}$.
Overall, mismatches increase by $\sim 3$ orders of magnitudes from single-simulations fits when imposing a smooth dependence on the binary parameter space.

\begin{figure*}[htp!]
    \centering
    \includegraphics[width=.475\textwidth]{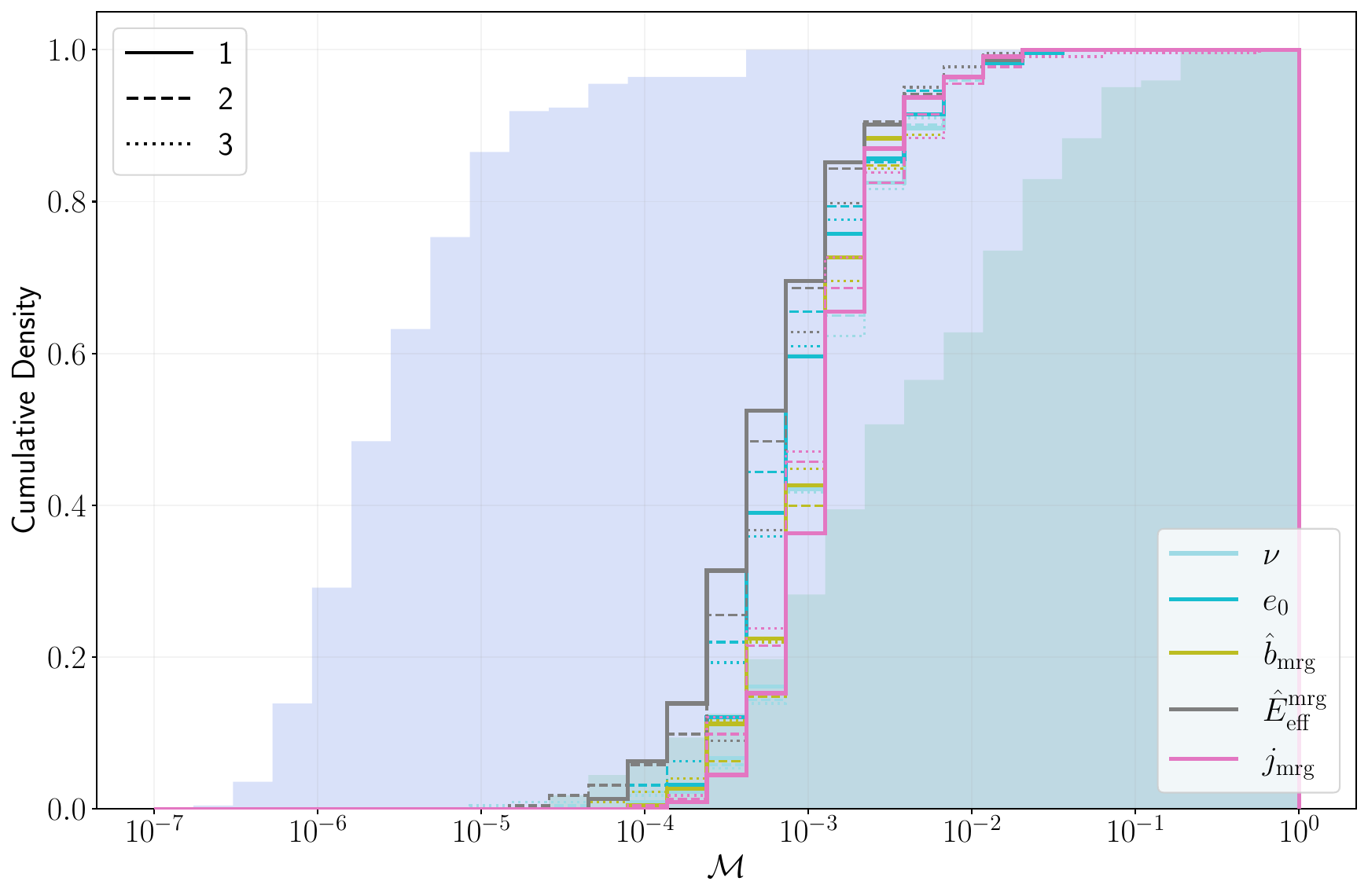}
    \includegraphics[width=.475\textwidth]{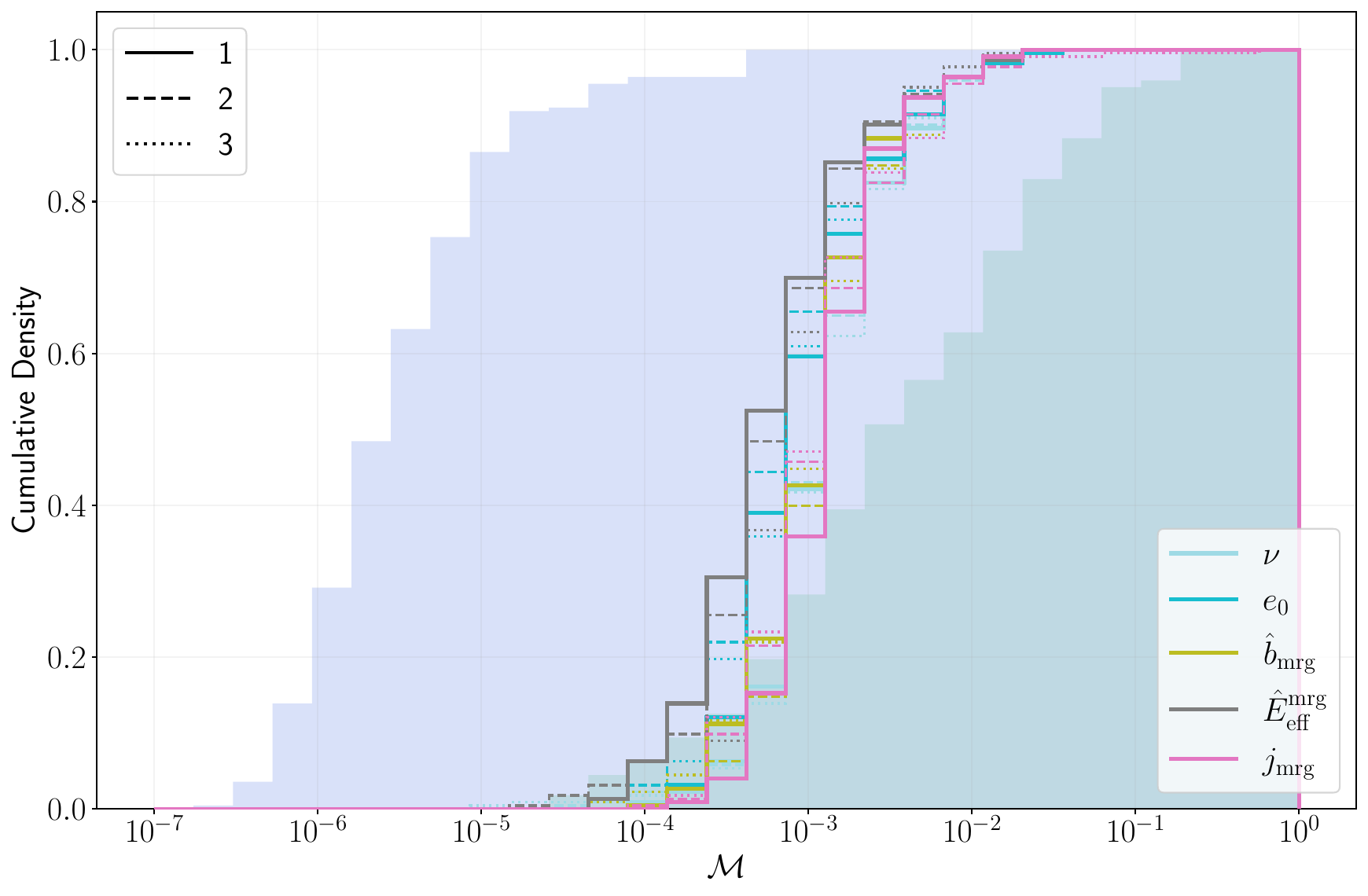}
    \includegraphics[width=.475\textwidth]{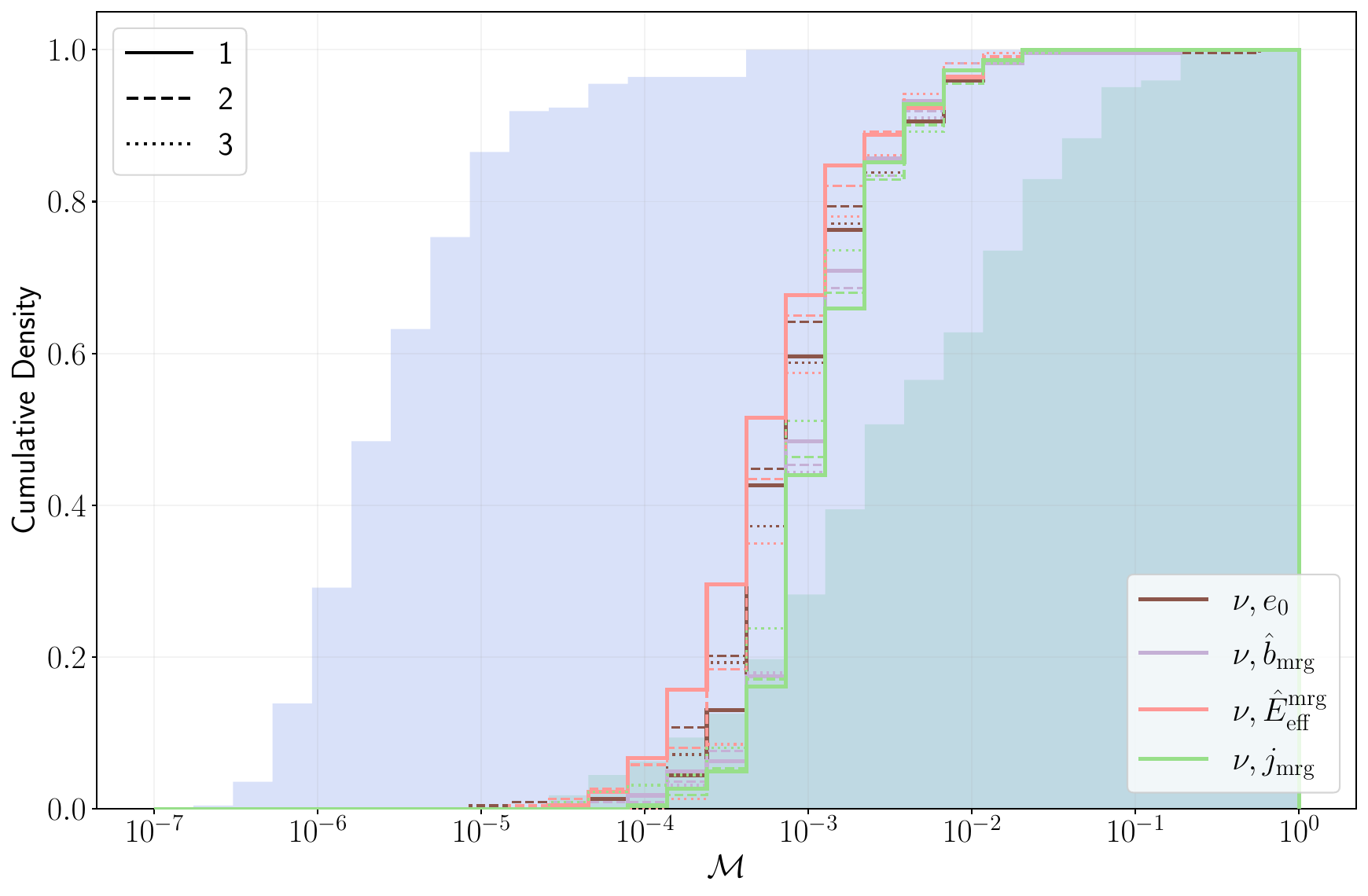}
    \includegraphics[width=.475\textwidth]{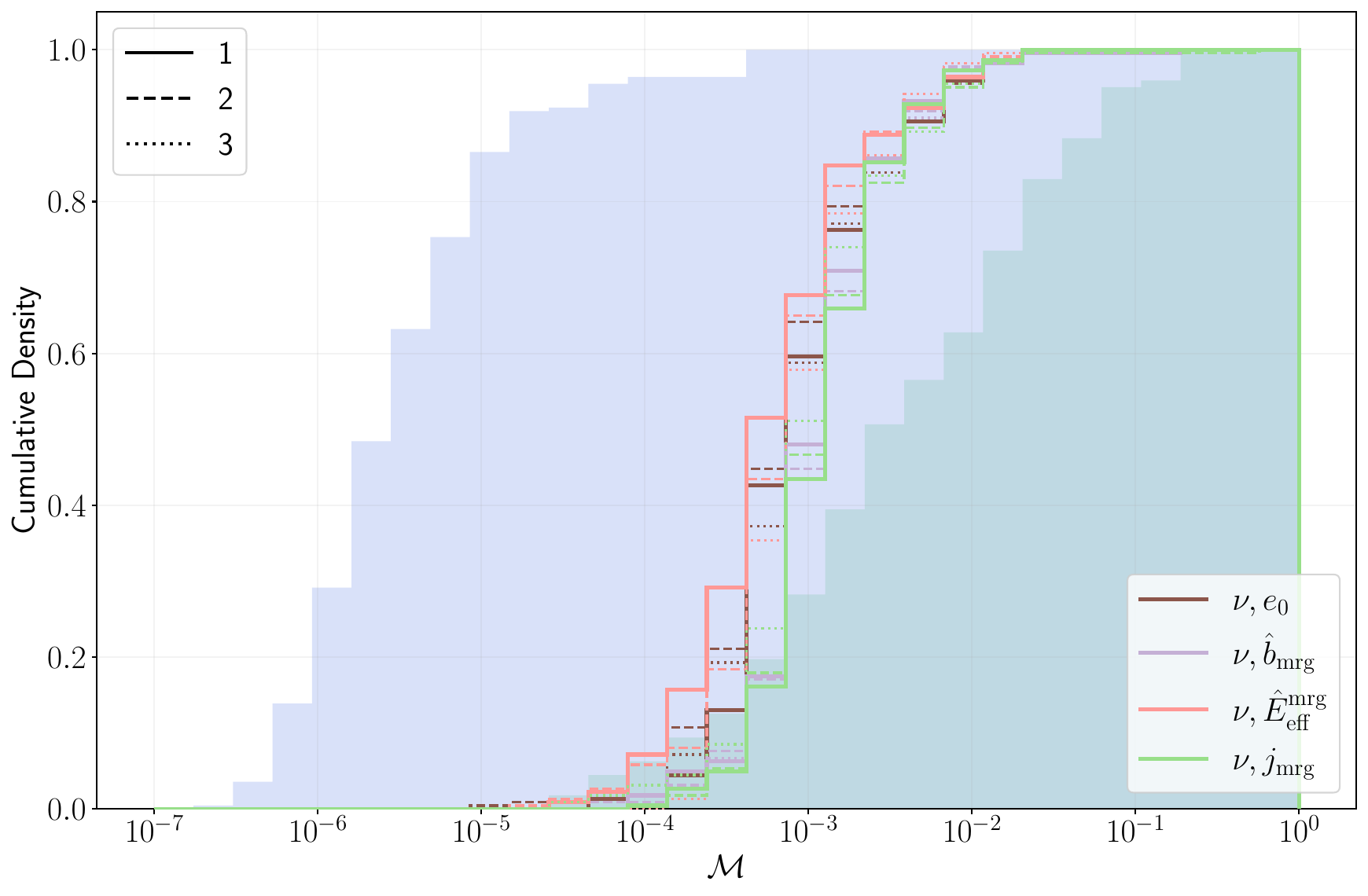}
    \includegraphics[width=.475\textwidth]{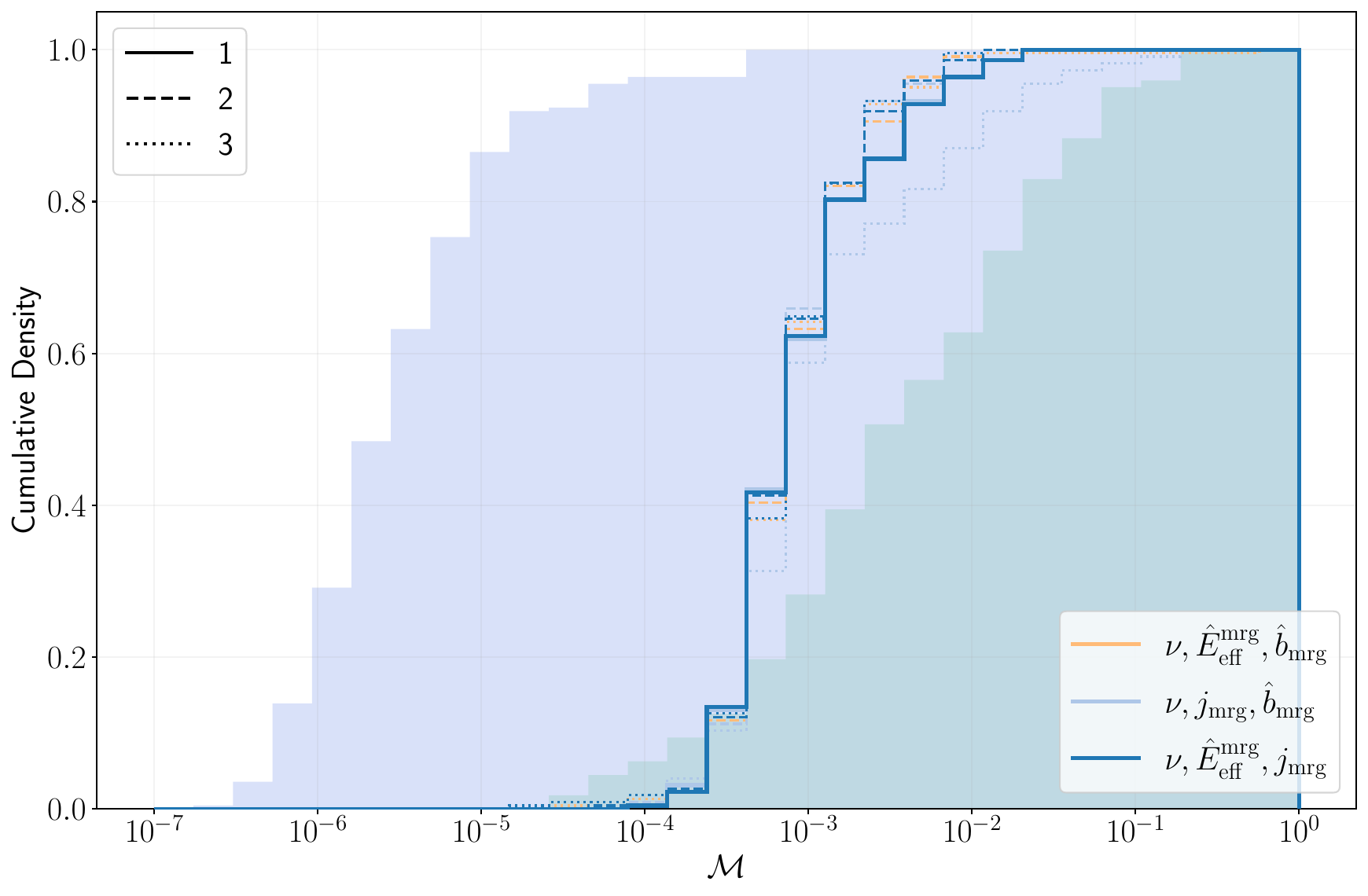}
    \includegraphics[width=.475\textwidth]{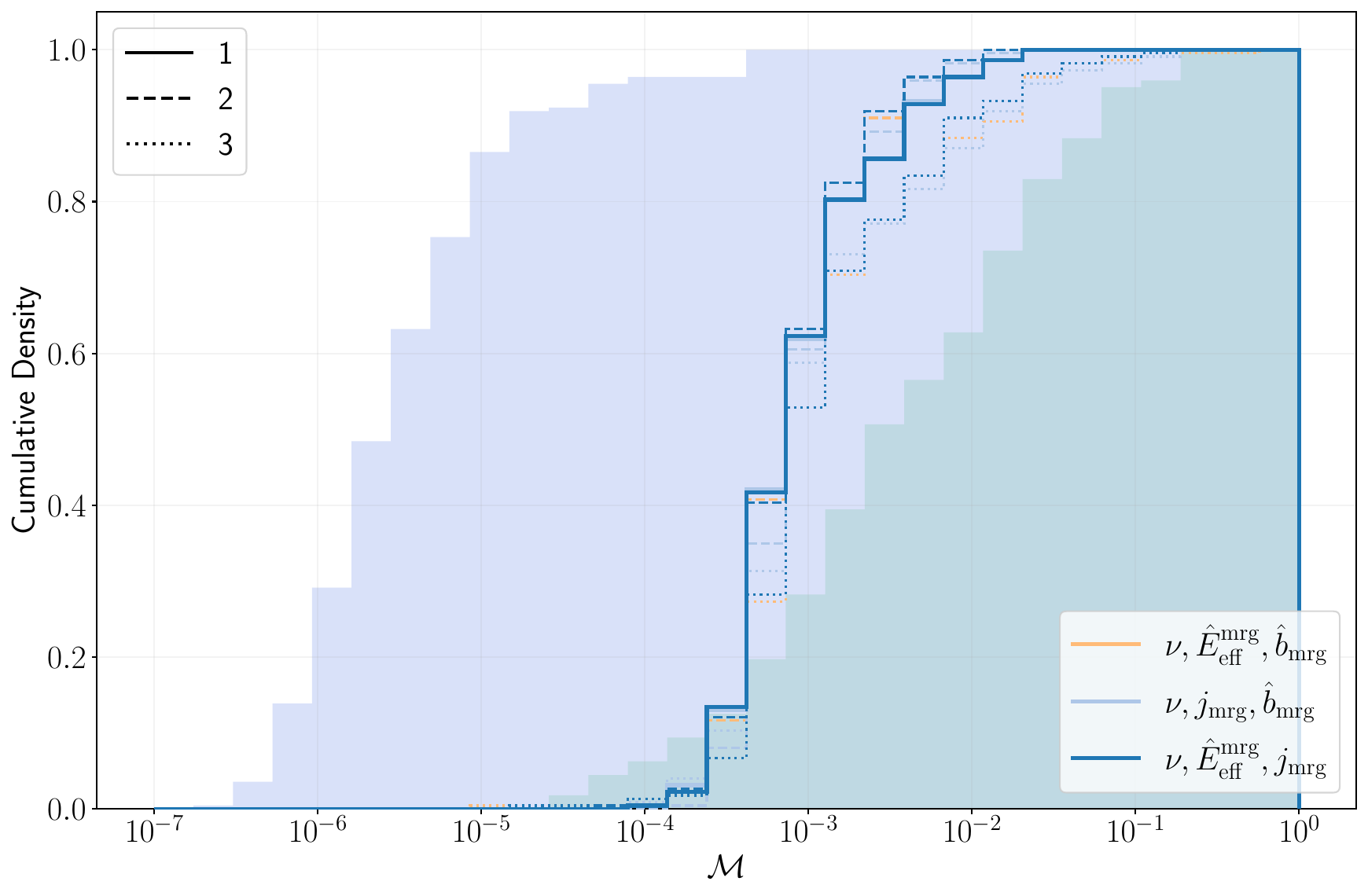}
    \caption{
    Cumulative distribution functions (CDFs) of the mismatches distributions from the global polynomial fits of amplitudes and phase coefficients using both the \emph{left:} Bayesian method (outlined in Sec.~\ref{subsec:bayesian_global_fit}) \emph{right:} Least Squares method (outlined in Appendix~\ref{app:lsq_linear}).
    Different colors correspond to different combinations of binary parameters, while different line styles to different maximum orders of the polynomial fits, similar to Fig.~\ref{fig:mismatch_histograms_bayesian}.
    The filled CDF in light blue represents the results of the corresponding local fits, while the green corresponds to the quasi-circular (\emph{HypTan}) template.}\label{fig:mismatch_histograms_cumulative}
\end{figure*}

\begin{figure*}[htp!]
    \centering
    \includegraphics[width=0.58\textwidth]{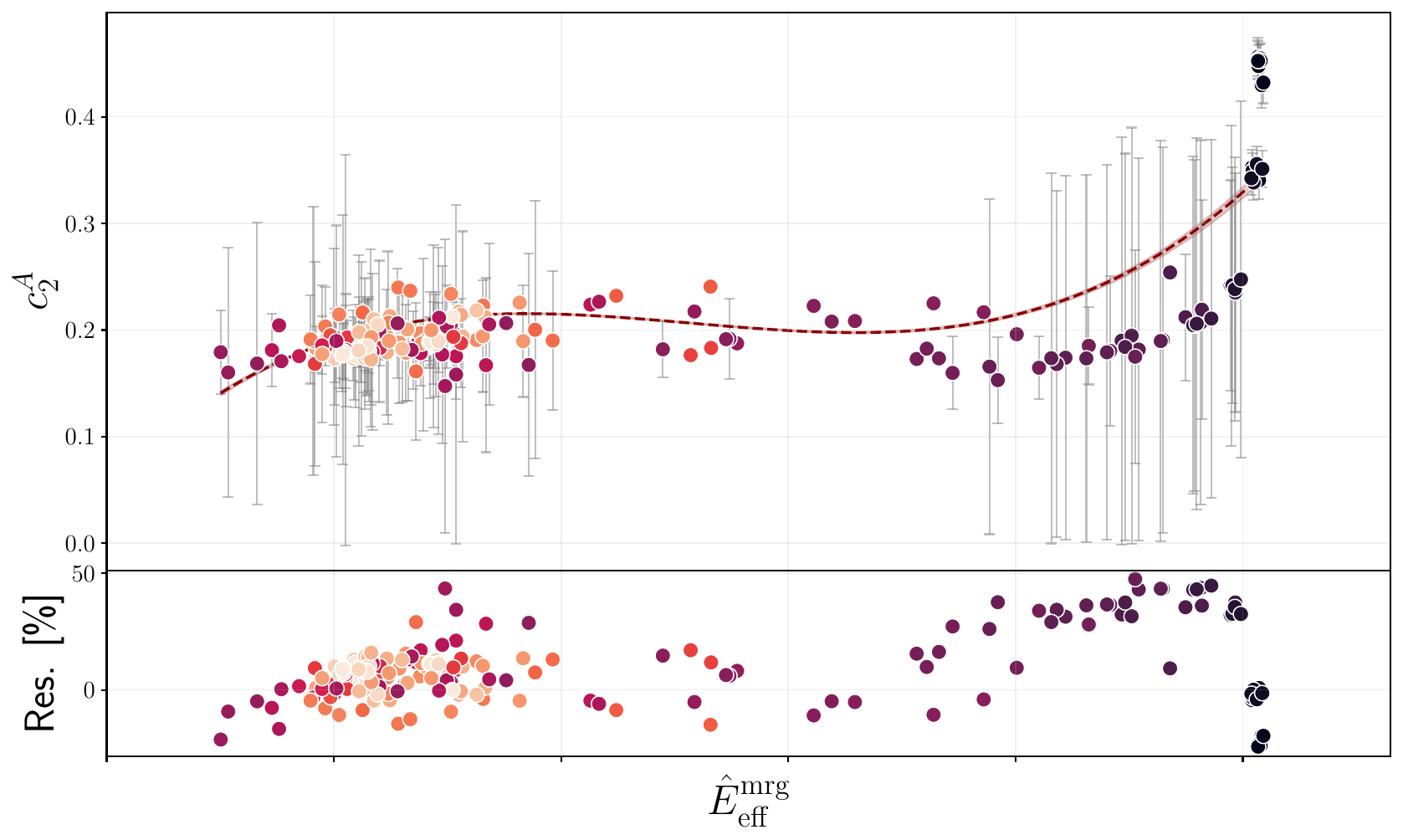}
    \includegraphics[width=0.37\textwidth]{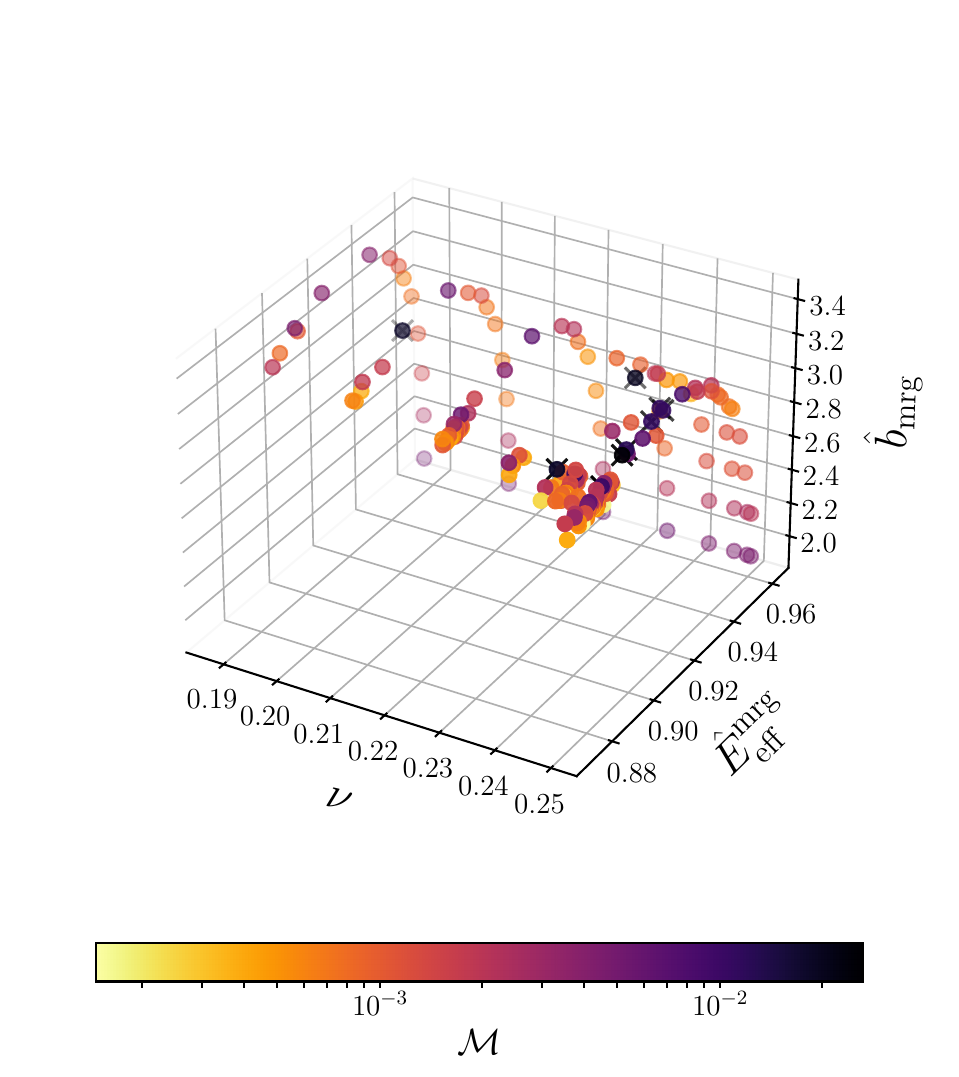}
    \caption{
    \textit{Left:} Global cubic ($\mathsf{d}=3$) fits for the amplitude coefficient, $c_2^A$, as a function of the non-circular parameters $\Eeffmrg$.
    The fit includes the standard deviation errors shown for each datapoint and the inferred Bayesian global fit median (dashed line) and the $95\%$ confidence interval (fainter bands, barely visible) are also shown.
    We observe a notable spiral pattern in the residuals, indicating potential incompleteness in the model.
    \textit{Right:} Corresponding mismatch distribution obtained by aligning the phase of the waveforms generated from the single-parameter cubic template model and the NR simulation.
    Crosses symbolize mismatches that exceed a threshold of $10^{-2}$.
    }
   \label{fig:nc_global_fits_cubic_emrg}
\end{figure*}

Fig.~\ref{fig:mismatch_histograms_cumulative} displays the cumulative distribution function of the mismatch results, which shows that, when considering only single-parameters fits, the best (worst) mismatches are obtained when fitting against $\Eeffmrg$ ($\jmrg$ or $\nu$).
The $\nu$-only fits represent the case when the non-circular dependence on the binary parameter space is ignored in the coefficients fits (albeit not in the merger data), which unsurprisingly scores among the lowest-performing parameterisations.
Linear fits serve as a computationally inexpensive and sufficiently accurate method to capture the waveform variations, but accuracy tends to improve with increasing polynomial order.
In some cases (e.g. for $\Eeffmrg$), the mismatches do not decrease monotonically as a function of the maximum polynomial order $\mathsf{d}$.
This is due to the imperfect representation of NR data via our templates, not by the coefficients' fit convergence, which has been checked with higher sampler settings.
We verified this statement by checking that the $\tilde{\chi}^2$ statistic (maximum likelihood agreement) of the coefficients fits themselves decreases monotonically for increasing $\mathsf{d}$;
the same check is also repeated for all parameters combinations discussed below. 
There exists a narrow parameter space for moderate $\Eeffmrg$, higher symmetric mass ratios, and higher $\bmrg$ that is not well-captured in the global fits, resulting in mismatches greater than $10^{-2}$.
This could be due to the incompleteness of the single-parameter model against the interplay of the mass ratio and the eccentric parameters.
In Fig.~\ref{fig:nc_global_fits_cubic_emrg}, we show the cubic fit of the dominant amplitude parameter ($c_2^A$) using $\Eeffmrg$, to understand the variation of one of the amplitude coefficients and the resultant fit employed, with the corresponding mismatches.
Despite correctly capturing the overall global trend, the datapoints' bifurcation and structure in the residual indicate the (expected) need to extend the dimensionality of the fit, including the mass ratio and full non-circular dependence.
In the same figure we also report the performance of this fit against a complete representation of the parameter space as a function of $\{\nu, \Eeffmrg, \, \bmrg\}$, showcasing that the worst mismatches ($\sim 10^{-2}$) are obtained in the equal-mass and largest impact parameter case, without a discernible $\Eeffmrg$ dependence.

\begin{figure*}[htp!]
    \centering
    \includegraphics[width=0.4\textwidth]{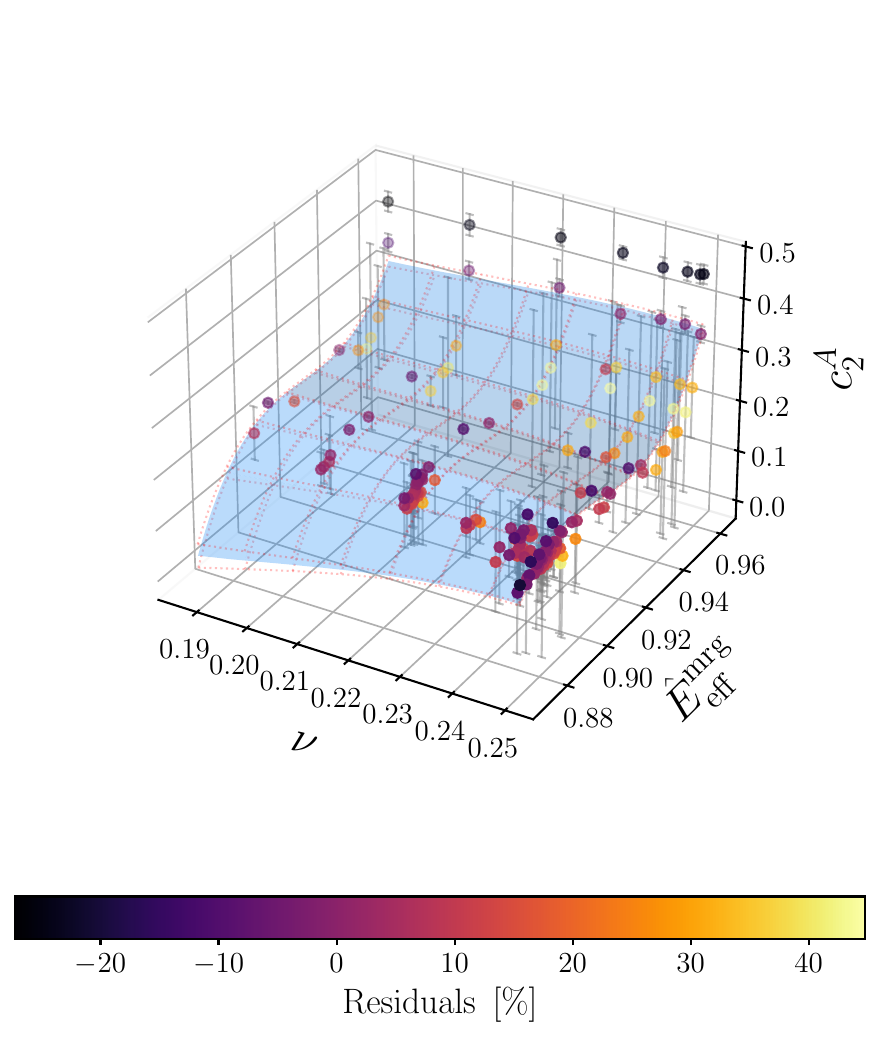}
    \includegraphics[width=0.45\textwidth]{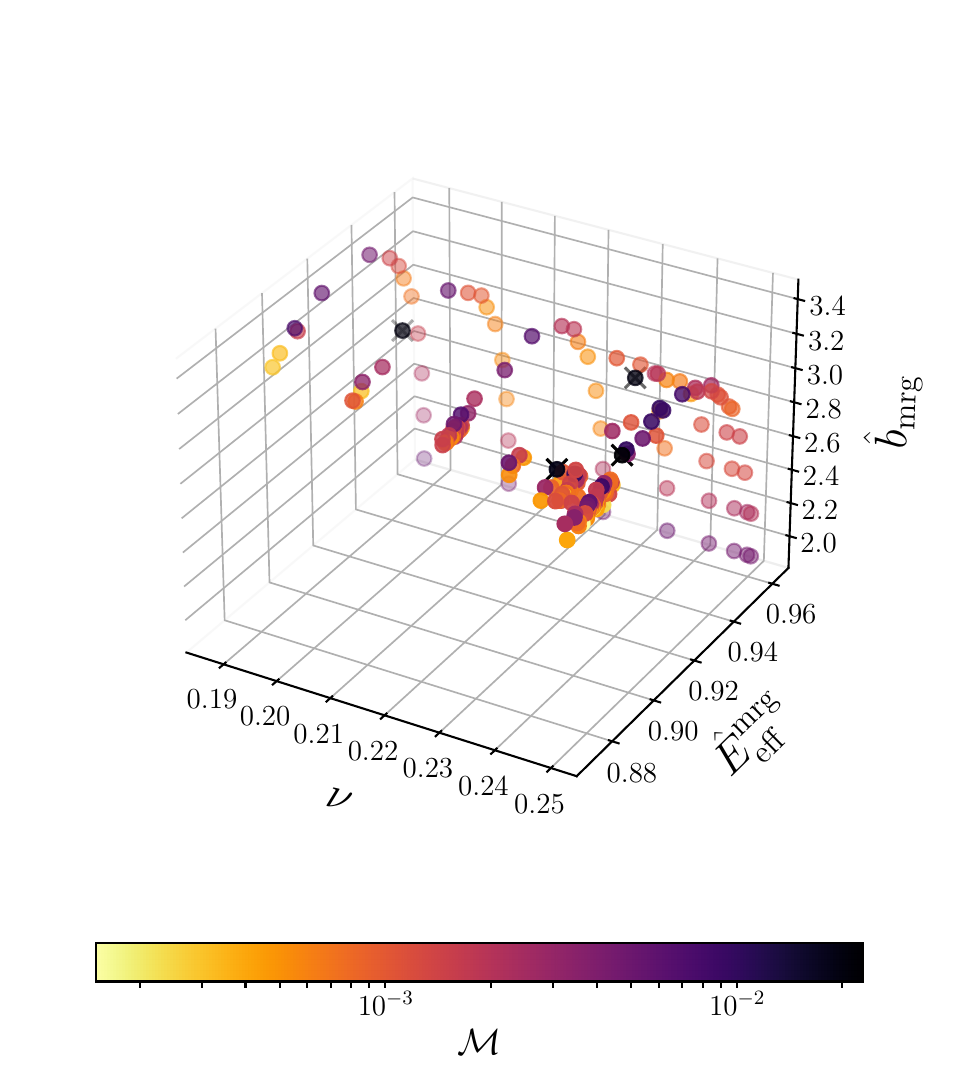}
    \caption{
    \textit{Left:} Global cubic ($\mathsf{d}=3$) fits for the amplitude coefficient, $c_2^A$, as a function of one of the non-circular parameters, $\Eeffmrg$, along with the symmetric mass ratio $\nu$.
    The fit includes the standard deviation errors shown for each datapoint, with the median (solid dashed surface) and $95\%$ confidence interval (dotted wireframe) from the Bayesian fit shown as well.
    \textit{Right:} Mismatches obtained by aligning the phase of the waveforms generated from the dual-parameter cubic template model and the NR simulation, similar to Fig.~\ref{fig:nc_global_fits_cubic_emrg}.
    }
   \label{fig:nc_global_fits_cubic_nu_emrg}
\end{figure*}

In double-parameter fits, we always include the mass-ratio dependence, augmenting it with a single non-circular parameter.
These combinations correspond to the quasi-universal parameterization explored in Ref.~\cite{Carullo:2023kvj, Carullo:2024smg}.
Mismatches markedly improve compared to the single-parameter case, and again the best (worst) results are obtained when fitting against $\Eeffmrg$ ($\jmrg$), see Fig.~\ref{fig:mismatch_histograms_cumulative}.
The two-dimensional case for $c_2^A$, now in terms of both symmetric mass ratio $\nu$ and non-circular parameter $\Eeffmrg$, is displayed in Fig.~\ref{fig:nc_global_fits_cubic_nu_emrg}.
The fits show improved mismatches, particularly capturing high eccentricity cases.
The equal-mass and large impact parameter case still shows large mismatches.
Further structure in the residuals indicates the extent of the quasi-universality violation and the necessity of adding another variable to the fit.

\begin{figure*}[htp!]
    \centering
    \includegraphics[width=0.37\textwidth]{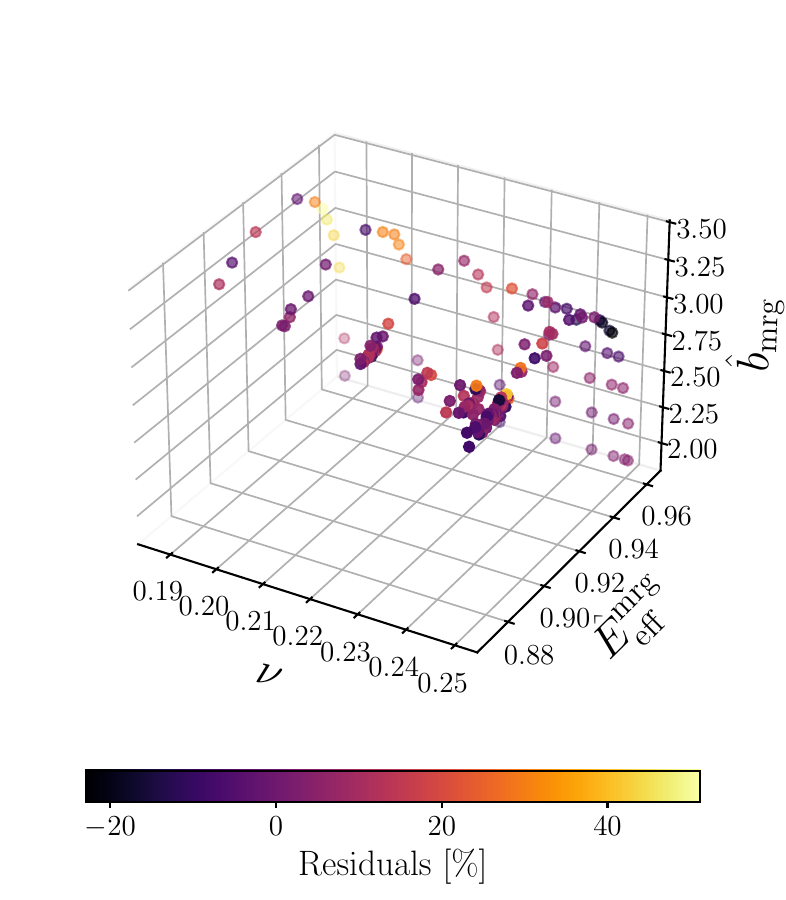}
    \includegraphics[width=0.37\textwidth]{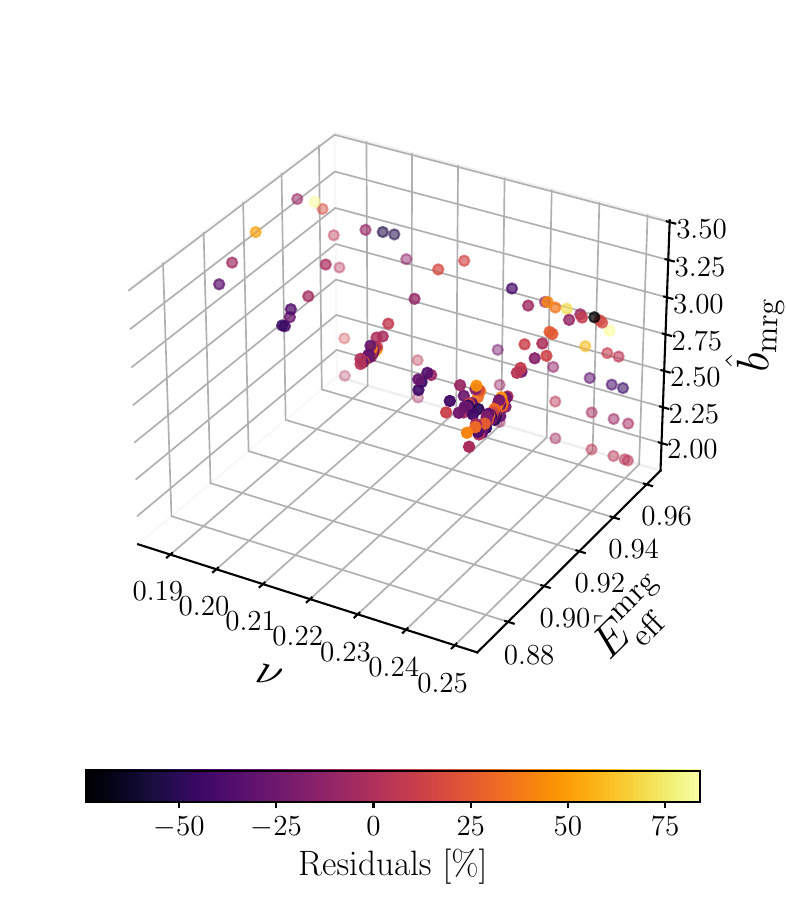}
    \includegraphics[width=0.37\textwidth]{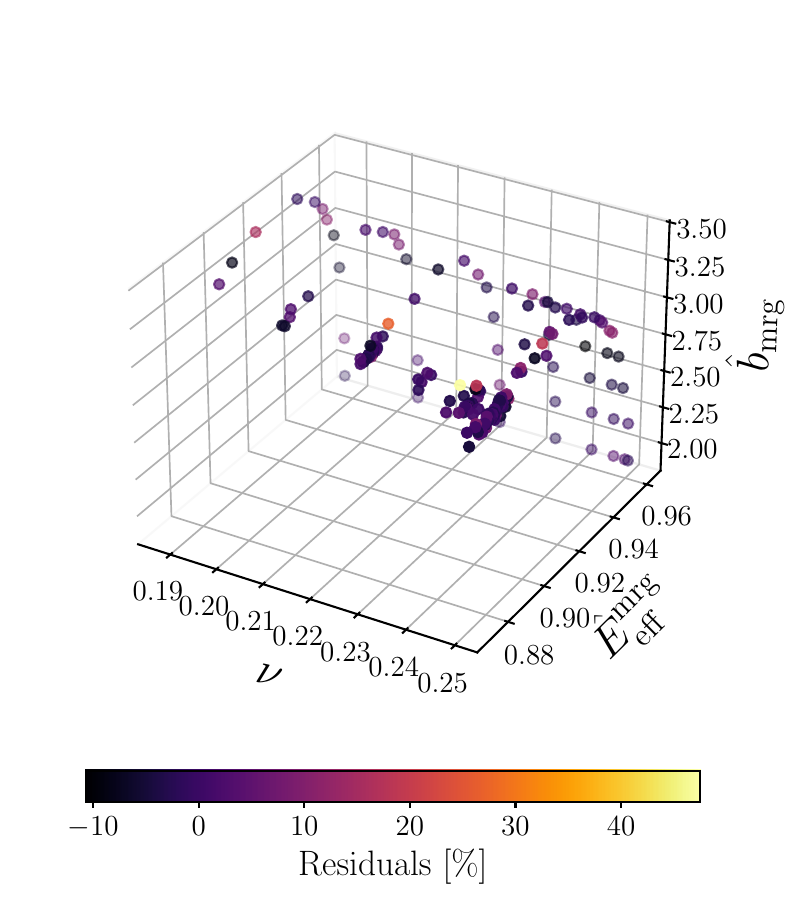}
    \includegraphics[width=0.37\textwidth]{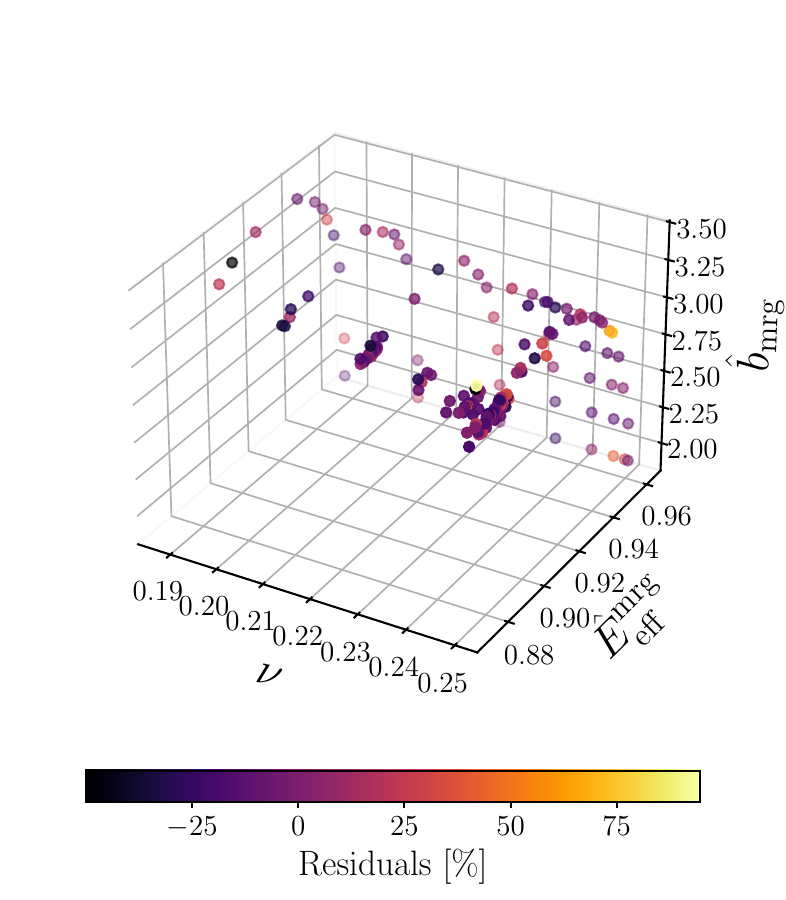}
    \includegraphics[width=0.37\textwidth]{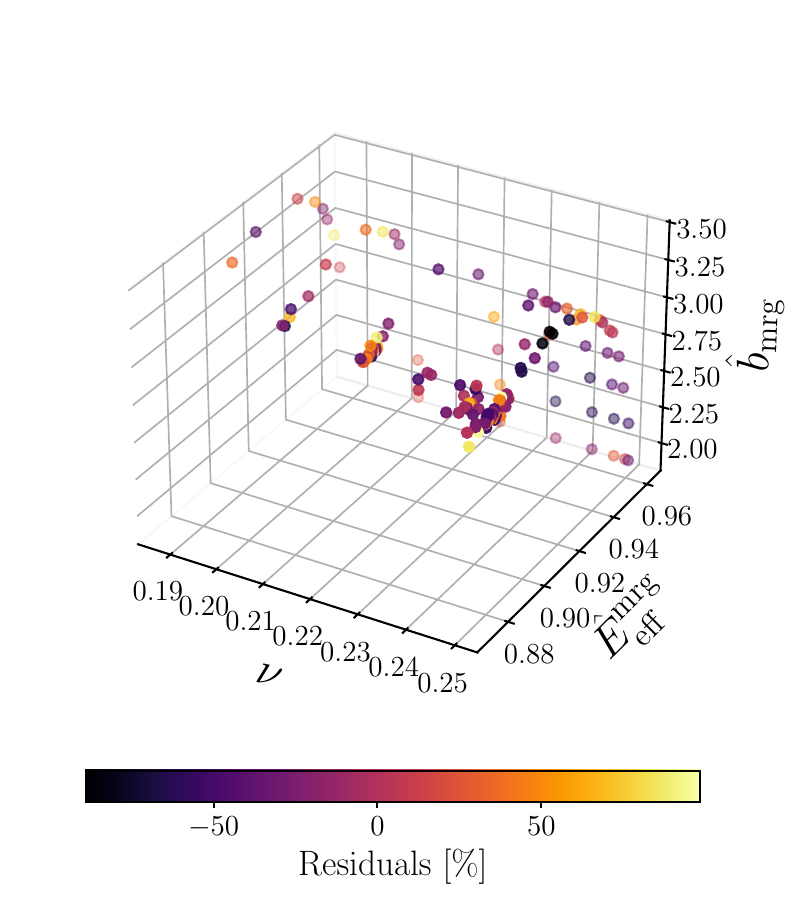}
    \includegraphics[width=0.39\textwidth]{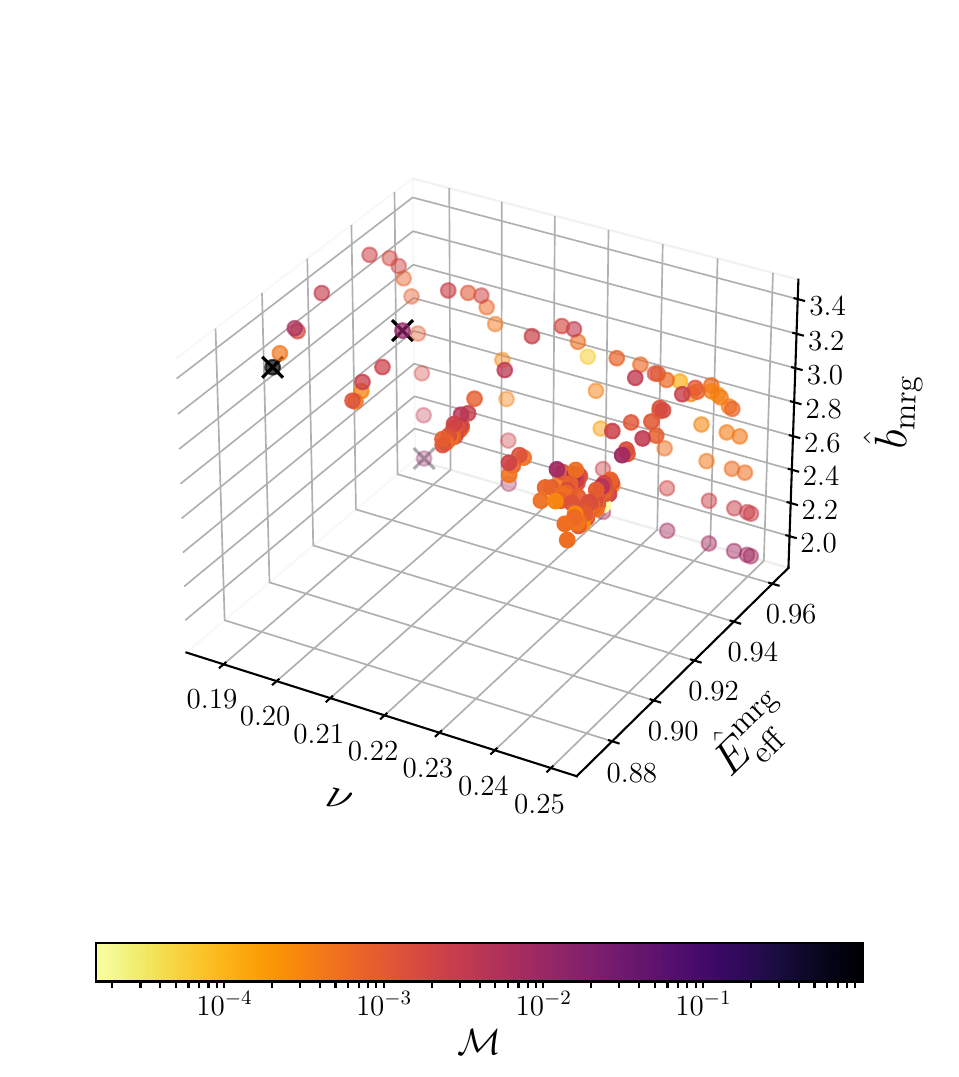}
    \caption{
    \textit{Top, Middle, Bottom Left:} Global triple-parameter cubic ($\mathsf{d}=3$) fits for the amplitude and phase coefficients $\{c_2^A, \, c_3^A, \, c_2^{\phi}, \, c_3^{\phi}, \, c_4^{\phi}\}$, as a function of two of the non-circular parameters $\bmrg$, and $\Eeffmrg$, along with the symmetric mass ratio, i.e. $\boldsymbol{Q}=\{\nu, \, \bmrg, \Eeffmrg\}$.
    \textit{Bottom:} Mismatches obtained by aligning the phase of the waveforms generated from the dual-parameter linear template model and the NR simulation, similar to Fig.~\ref{fig:nc_global_fits_cubic_emrg}.
    }
   \label{fig:nc_global_fits_cubic_nu_emrg_bmrg}
\end{figure*}

Finally, accounting for the complete dimensionality of the parameter space, we use the symmetric mass ratio $\nu$, along with two non-circular parameters combinations, in a constrained three-dimensional global fit, as demonstrated in Figs.~\ref{fig:nc_global_fits_cubic_nu_emrg_bmrg}.
As expected, mismatches further decreased compared to the two-dimensional fits, including for the challenging large impact parameter cases.
As shown in Fig.~\ref{fig:mismatch_histograms_cumulative}, the best representation is provided by the combination $\{ \nu, \Eeffmrg, \, \jmrg \}$ (closely followed by $\{ \nu, \Eeffmrg, \, \bmrg \}$) using cubic order, yielding only three cases with mismatches larger than $10^{-2}$.

\clearpage

\section{Conclusions}
\label{sec:conclusions}

We presented NR-informed parametric fits for the quadrupolar post-merger waveform emitted by non-spinning BH binaries in highly eccentric orbits, based on the non-circular dynamics parameterisation introduced in~\cite{Carullo:2023kvj}.
We found that the key ingredient to extend quasi-circular models in this regime is the inclusion of non-circular merger data obtained in Ref.~\cite{Carullo:2023kvj}.
The rational exponential ans\"atz, introduced in Ref.~\cite{Albanesi:2023bgi} for the time-dependent ringdown amplitude, also contributes to significantly improves the agreement with highly-eccentric numerical data with respect to previous hyperbolic tangent forms.
We calibrate the template against NR simulations obtained from the non-spinning, non-circular bounded RIT catalog in the range corresponding to symmetric mass ratio across $0.18\lesssim\nu\lesssim0.25$, and two dynamics-based non-circular parameters: effective energy $0.87\lesssim\Eeffmrg\lesssim0.96$, and impact parameter $1.91\lesssim\bmrg\lesssim3.41$.

A parameterisation in terms of the $\{\nu,\,\bmrg,\,\Eeffmrg\}$ allows for an accurate and smooth coverage of the binary parameter space via third-order polynomial fits.
We find that the effective energy at merger is the most informative parameter, allowing us to capture the bulk of the parameter space dependence.
Overall, waveform mismatches improve by more than an order of magnitude compared to state-of-the-art quasi-circular template belonging to the same waveform family, with the vast majority within the range $[10^{-4}, 10^{-2}]$, and only three cases with worse values.
While reproducing past results, we also found an order of magnitude improvement in the mismatch of quasi-circular non-spinning binaries due to the improved accuracy of our fitting algorithm.

Future directions include extending the fits to higher modes, the spinning case, incorporating test-mass data to ensure a smooth transition to the extreme-mass-ratio limit~\cite{Faggioli:2026alx, Albanesi:2026qtx}, and modeling the kick dependence identified in Ref.~\cite{Radia:2021hjs}. 

\acknowledgments
We are indebted to Vasco Gennari and Walter Del Pozzo for their work on the implementation of the \texttt{TEOBPM} template in \texttt{pyRing}, a key software component of our model building.
We are grateful to Simone Albanesi for conduting the internal LVK publication review, and to Keefe Mitman for elucidating the impact of BMS transformations on our formalism.
We are also grateful to Alessandro Nagar, Simone Albanesi, Sebastiano Bernuzzi, and Rossella Gamba for helpful discussions and suggestions.
N. R. further acknowledges support from Kishore Vaigyanik Protsahan Yojana (KVPY), funded by the Department of Science and Technology (DST), Government of India.
\textit{Software:} Contents of this manuscript have been derived using the publicly available \texttt{python} software packages: \texttt{cpnest, cython, matplotlib, numpy, pandas, pyRing, scipy, seaborn}~\cite{cpnest,cython,matplotlib,numpy,pandas,pandas_zenodo,pyRing,scipy,seaborn}. 
We utilize \texttt{bayRing}~\href{https://github.com/GCArullo/bayRing}{\textcolor{blue!50}{\faGithubSquare}} to obtain local fits for the coefficients that describe the implemented eccentric template.
The obtained coefficients, and the code to reproduce plots, mismatches and least-squares fits will be publicly available, at the time of journal submission, at: \texttt{nc\_ringdown}~\href{https://github.com/nishkalrao20/nc_ringdown}{\textcolor{blue!50}{\faGithubSquare}}.
The templates we developed have been incorporated within the \texttt{pyRing}~\href{https://git.ligo.org/lscsoft/pyring}{\textcolor{blue!50}{\faGithubSquare}} package.

\appendix

\section{Algorithm testing in quasi-circular case}
\label{sec:quasi_circular_testing}

\begin{figure*}[htp!]
    \centering
    \includegraphics[width=0.325\textwidth]{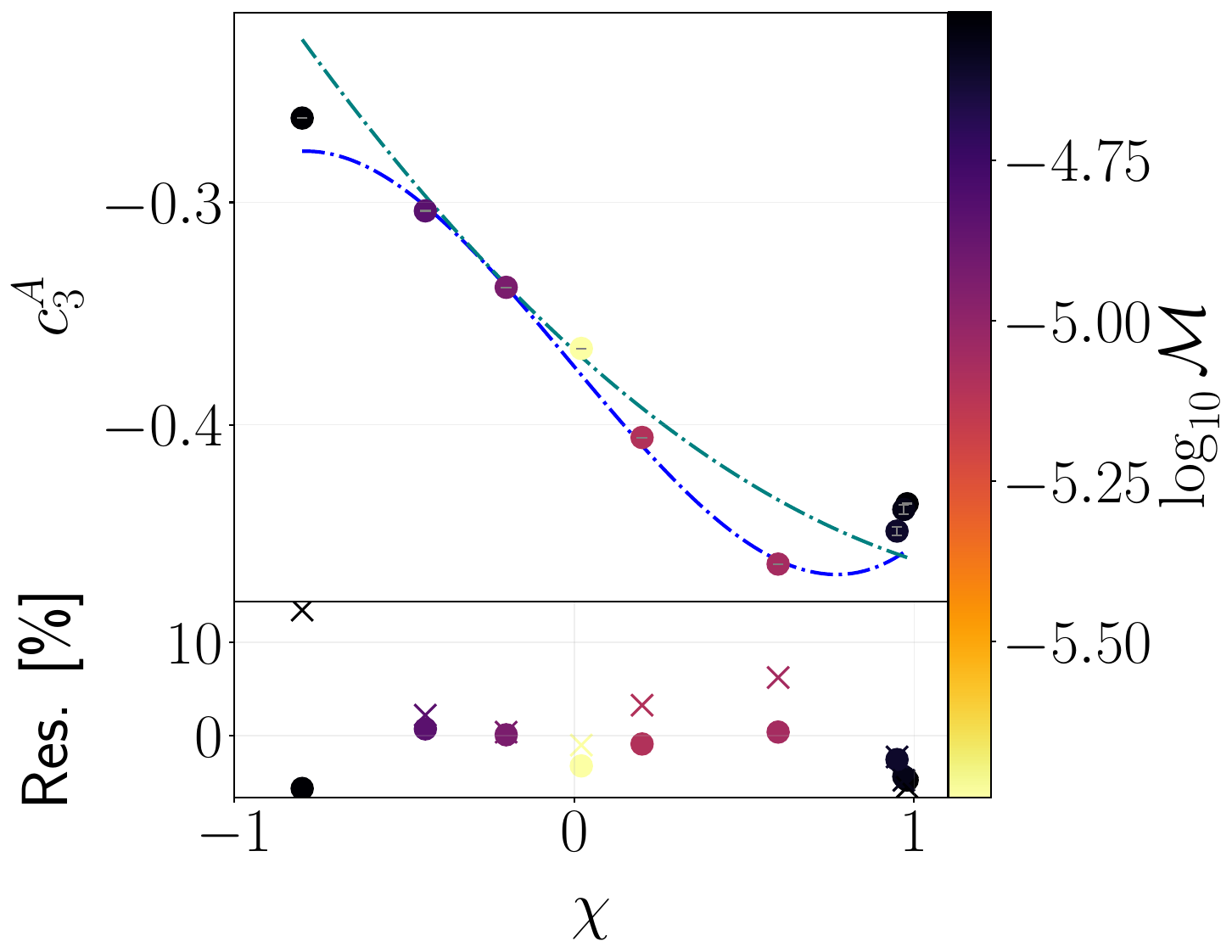}   
    \includegraphics[width=0.325\textwidth]{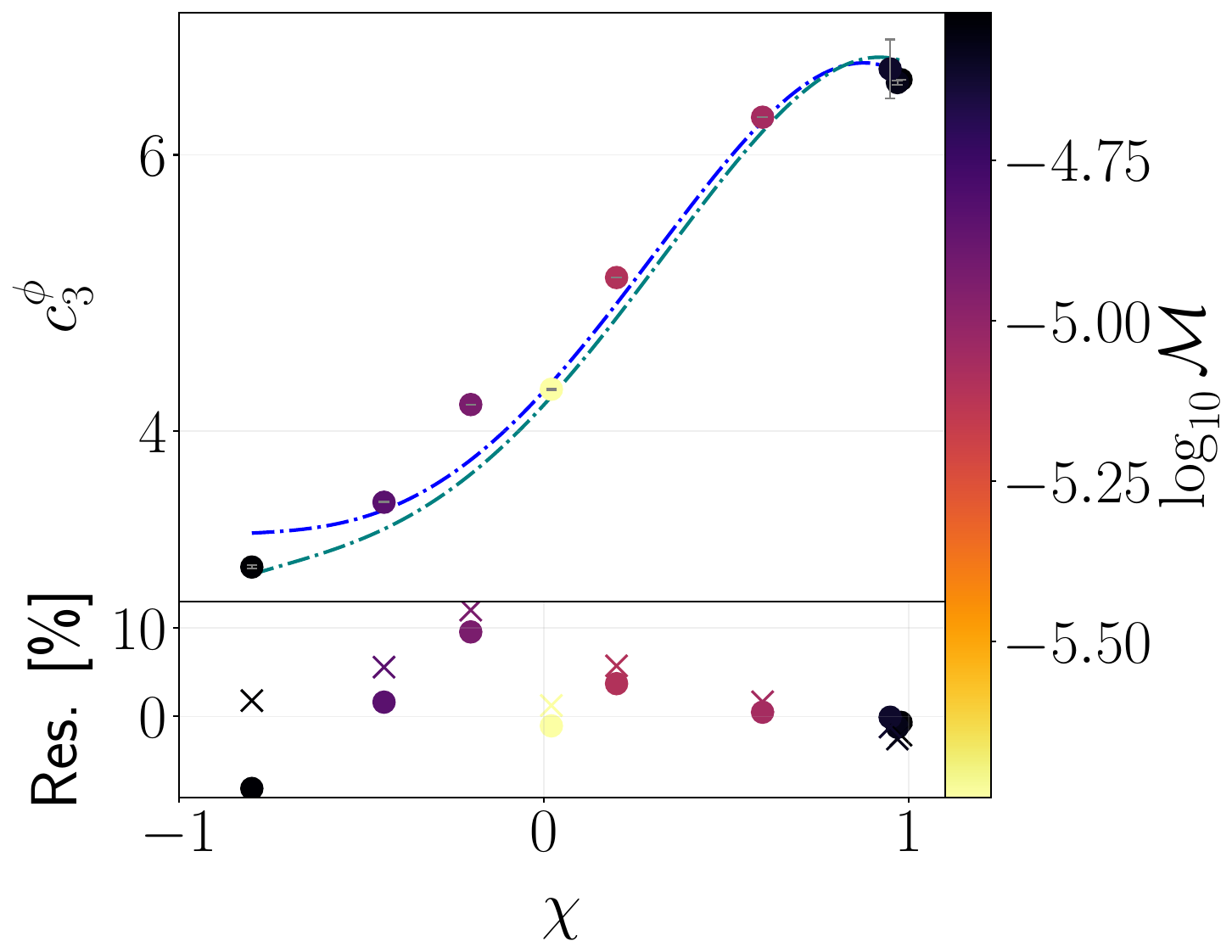}   
    \includegraphics[width=0.325\textwidth]{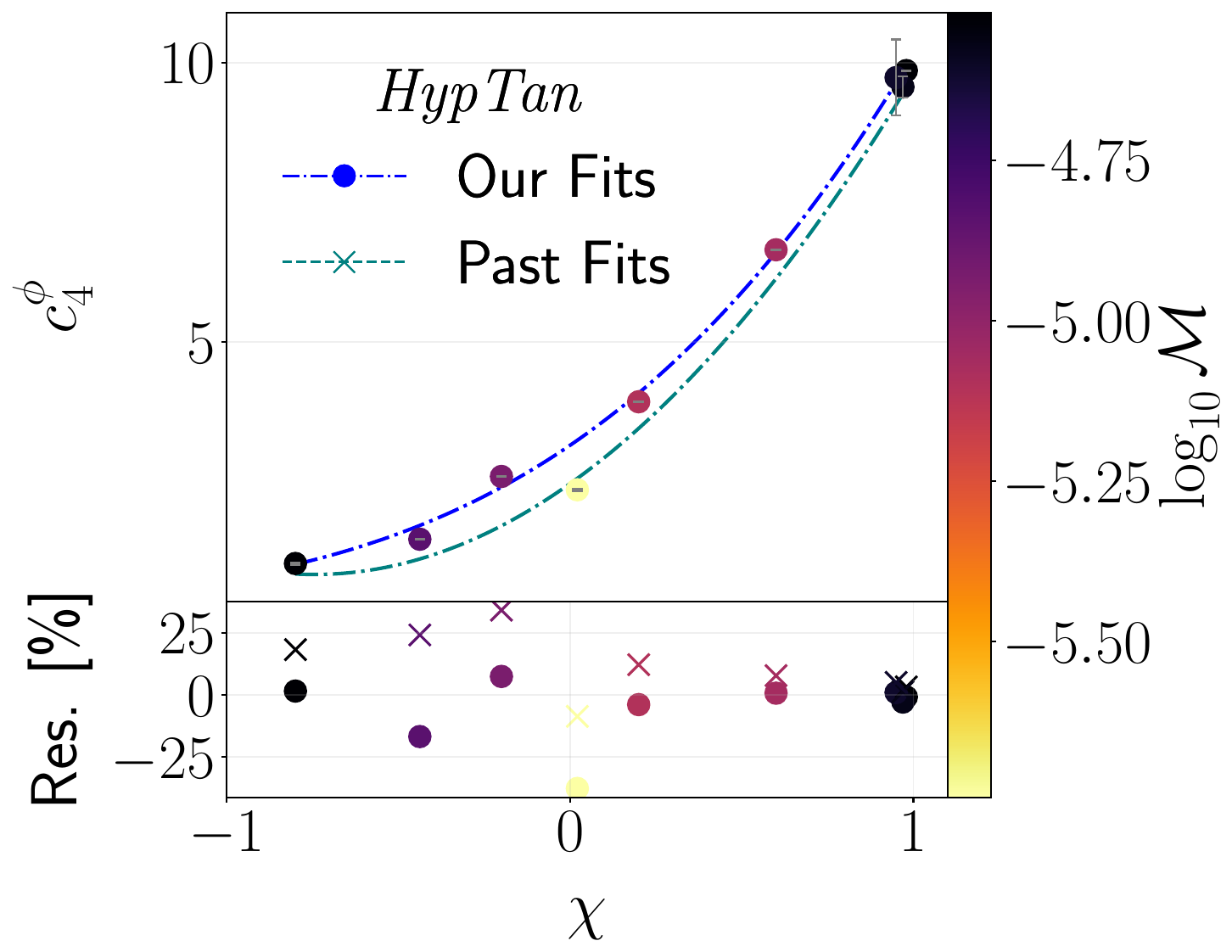}
    \caption{
    Median amplitude and phase coefficients from the Bayesian fits with the \emph{HypTan} template for the equal-mass equal-spin SXS catalog (from Ref.~\cite{Damour:2014yha}) in terms of the effective spin $\chi$.
    The colors represent the (log) mismatch of the model vs the corresponding NR simulation when assuming the median coefficients extracted.
    We present the polynomial fit from Table~II of Ref.~\cite{Damour:2014yha} in teal, with the residuals of the median coefficients against the past global fit given by the $\times$ marker.
    }
   \label{fig:qc_fits_sxs_chi}
\end{figure*}

\begin{figure*}[htp!]
    \centering
    \includegraphics[width=0.325\textwidth]{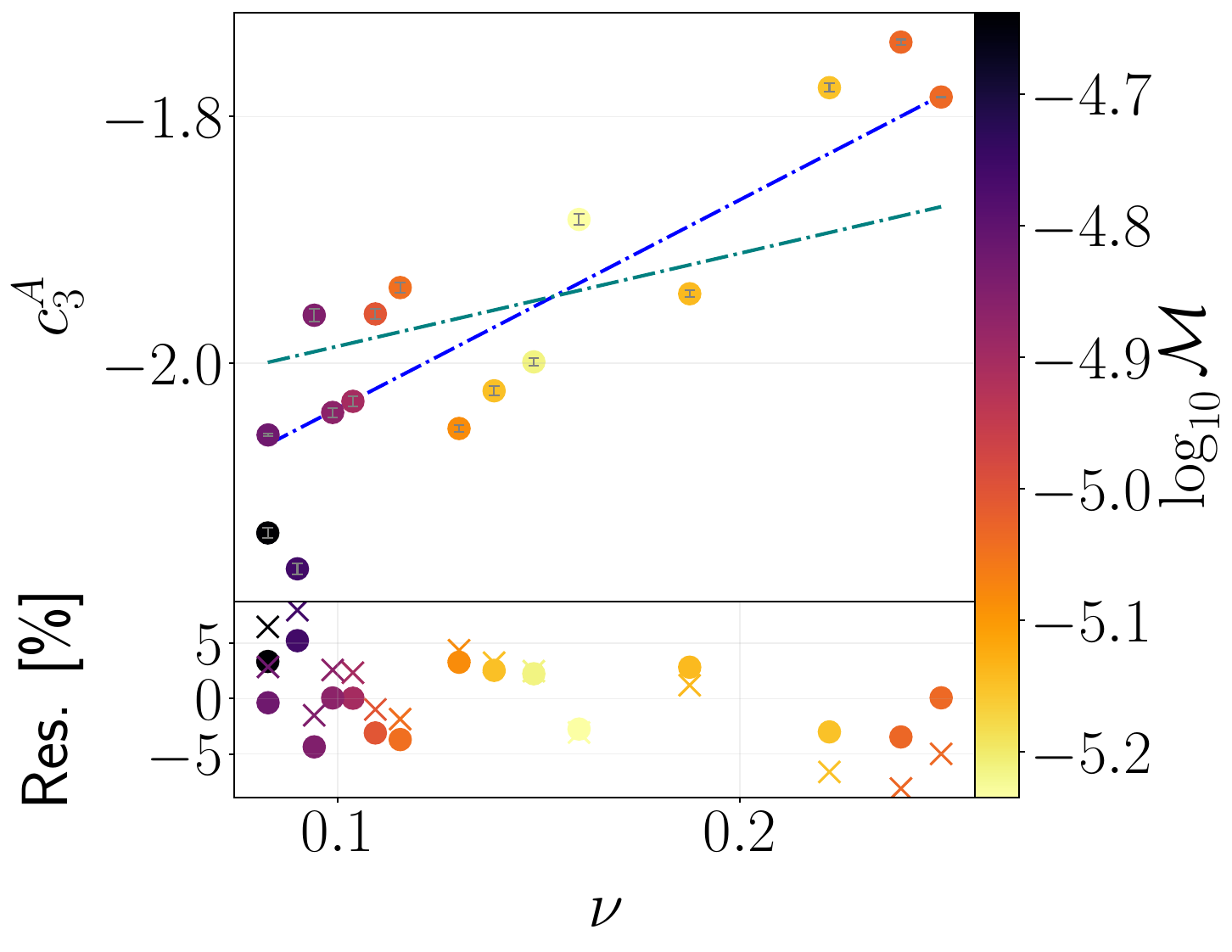}   
    \includegraphics[width=0.325\textwidth]{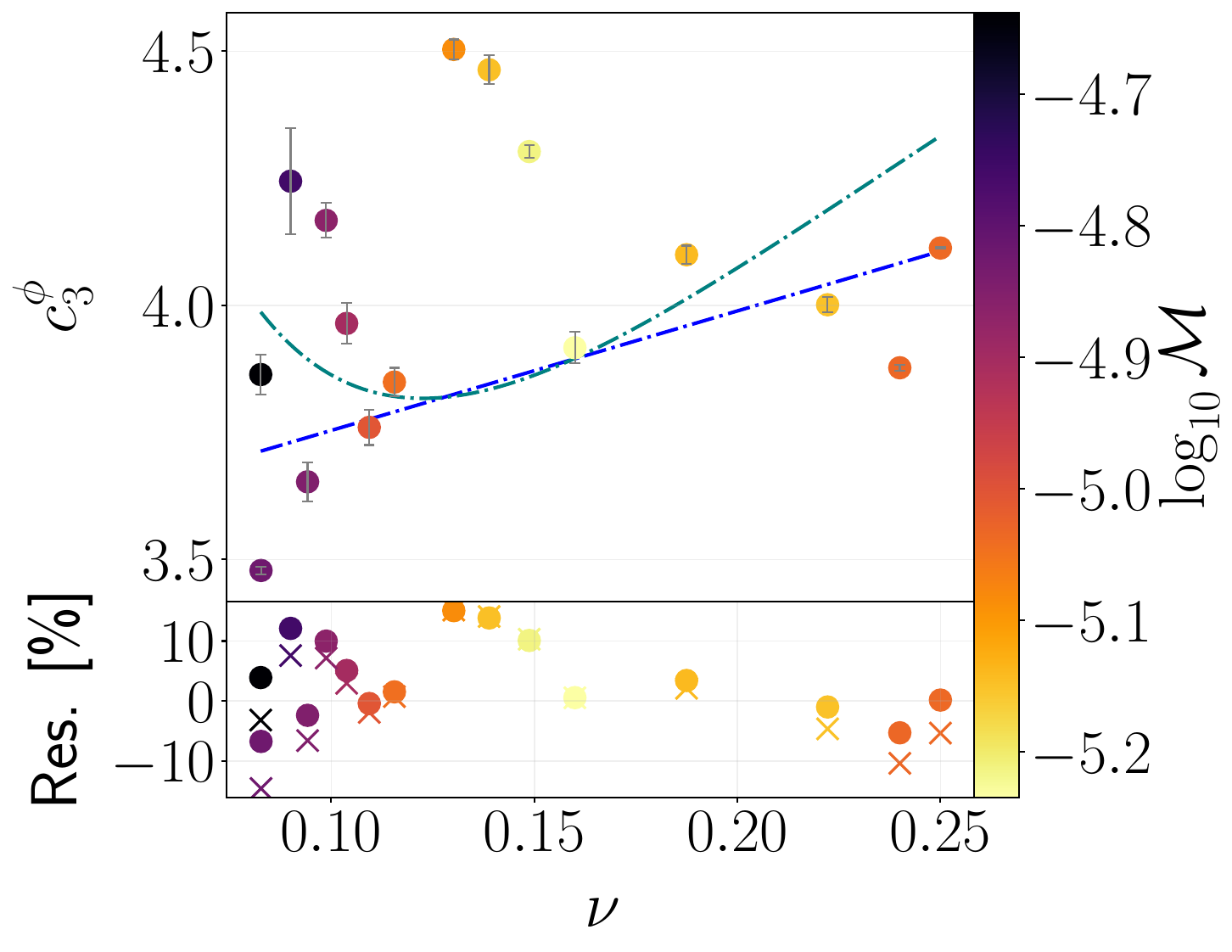}   
    \includegraphics[width=0.325\textwidth]{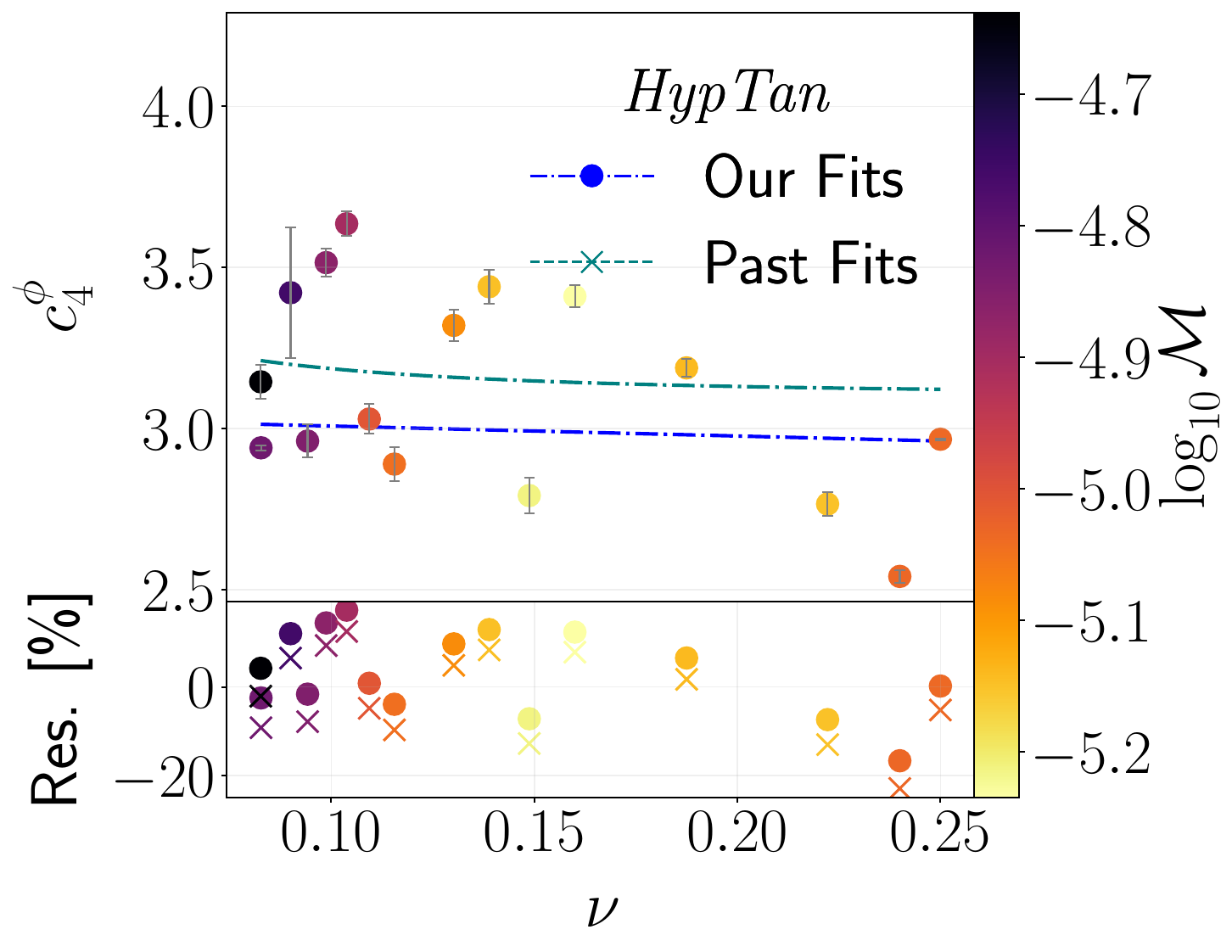}
    \caption{
    Similar to Fig.~\ref{fig:qc_fits_sxs_chi}, median amplitude and phase coefficients from the Bayesian fits for the non-spinning SXS simulations (from Ref.~\cite{Nagar:2019wds}) against symmetric mass ratio, $\nu$.
    A parsimonious linear $\nu$-dependency was chosen since we verified the global fit is only weakly sensitive to small variations in the symmetric mass ratio $\nu$. 
    The rational fit for the phase coefficients from Table~IV of Ref.~\cite{Nagar:2019wds} is represented in teal, with the residuals given by the $\times$ marker.
    }
   \label{fig:qc_fits_sxs_nu}
\end{figure*}

\begin{figure*}[htp!]
    \centering
    \includegraphics[width=0.325\textwidth]{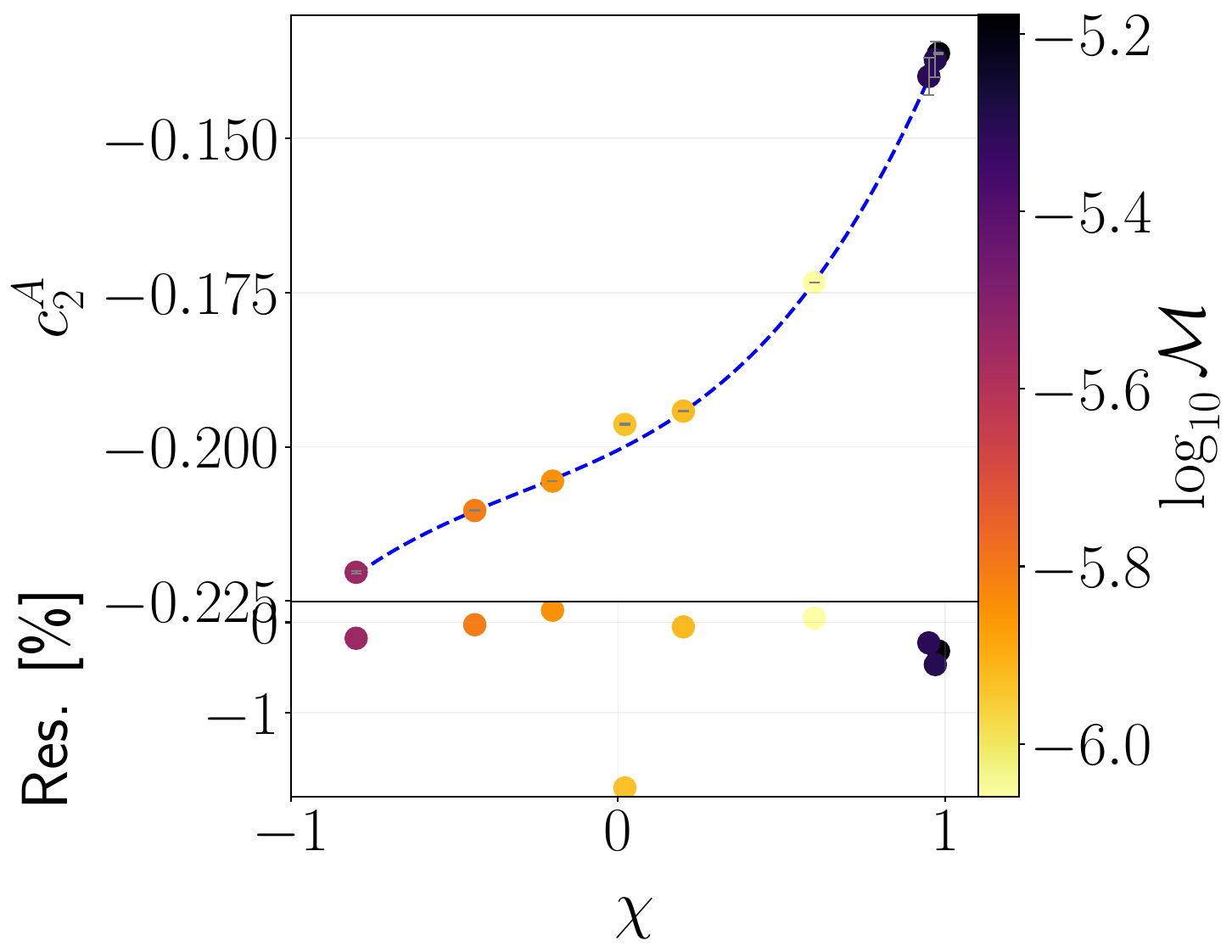}   
    \includegraphics[width=0.325\textwidth]{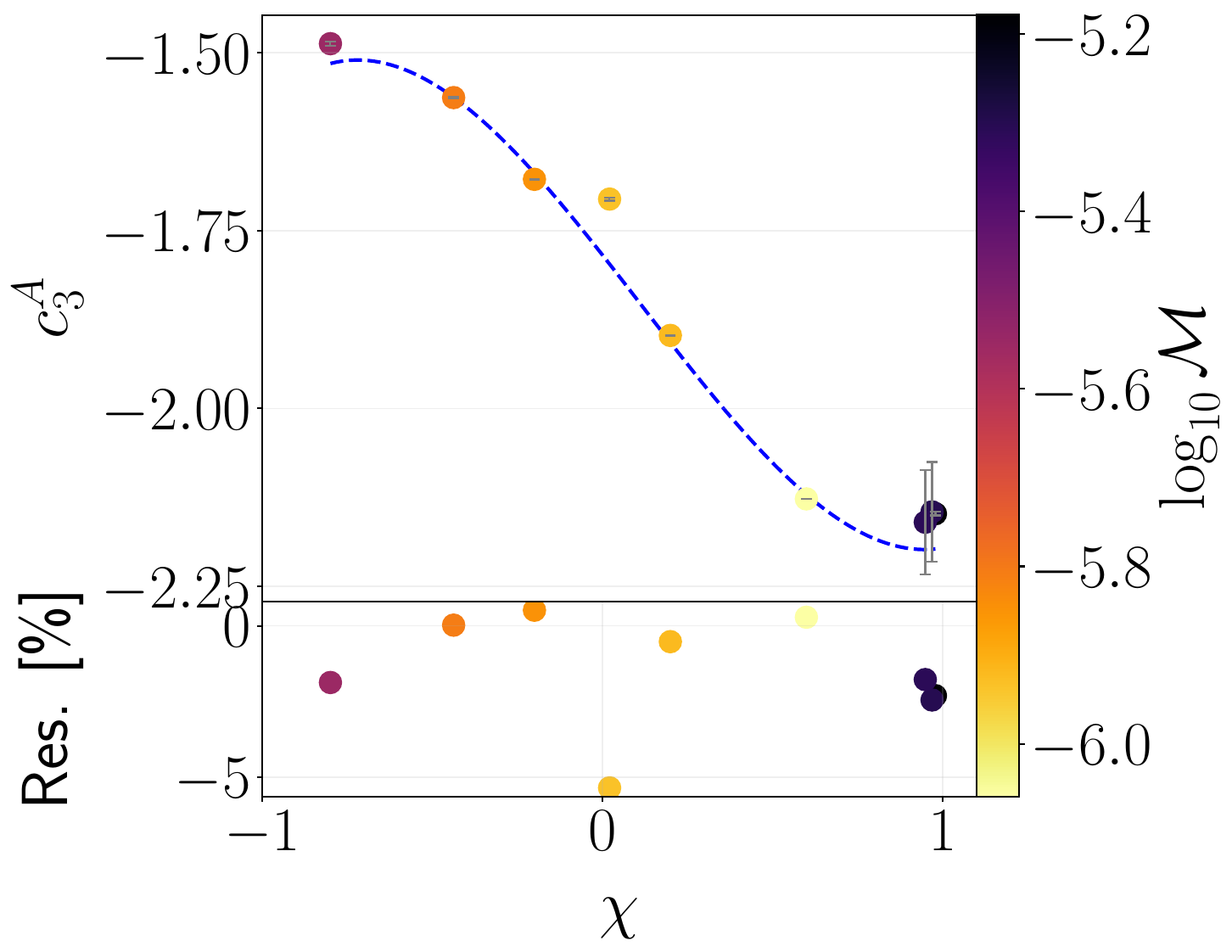}   
    \includegraphics[width=0.325\textwidth]{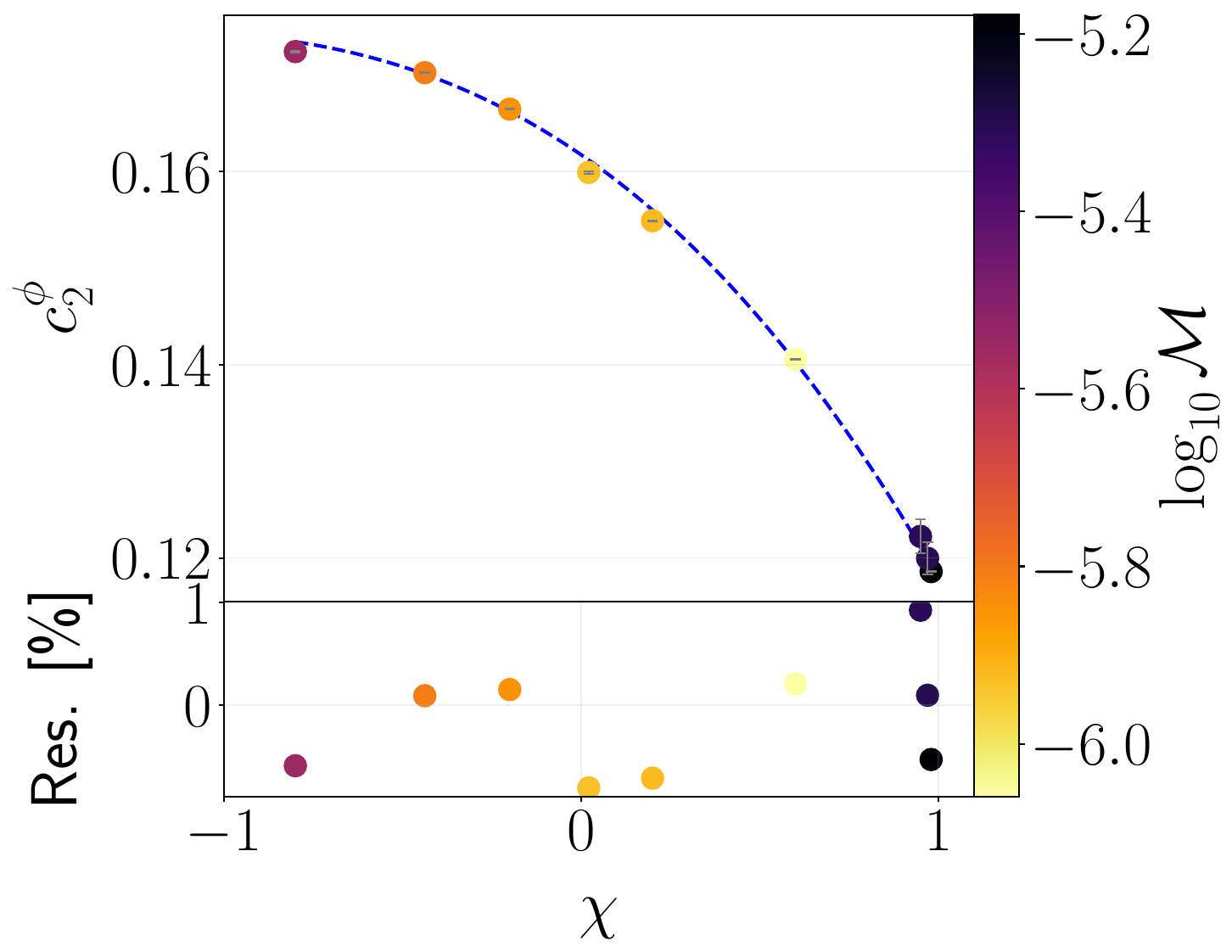}   
    \includegraphics[width=0.325\textwidth]{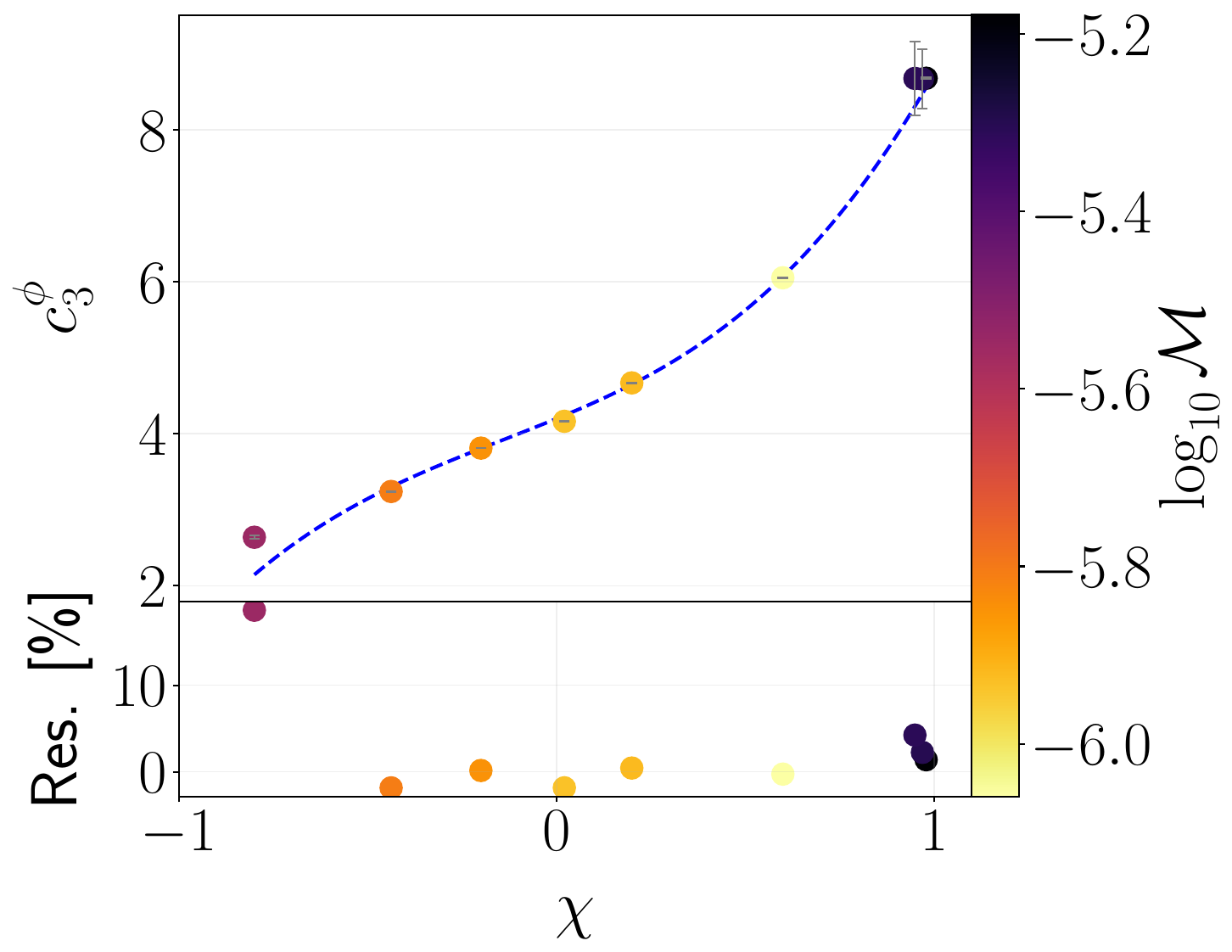}   
    \includegraphics[width=0.325\textwidth]{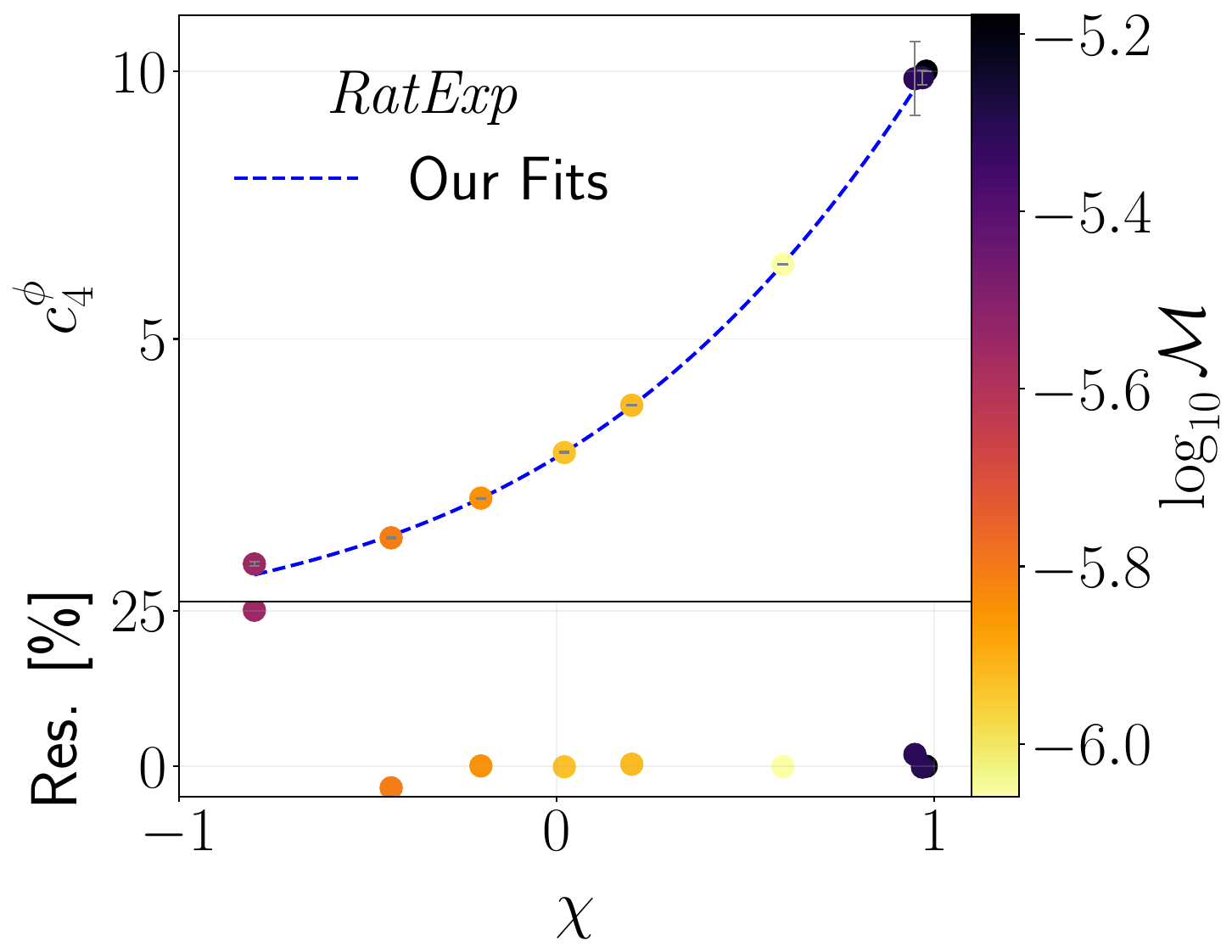}   
    \caption{
    Median amplitude and phase coefficients from the Bayesian fits with the \emph{RatExp} template for the equal-mass equal-spin SXS catalog (from Ref.~\cite{Damour:2014yha}) in terms of the effective spin $\chi$.
    The colors represent the (log) mismatch of the model vs the corresponding NR simulation when assuming the median coefficients extracted.
    }
   \label{fig:nc_fits_sxs_chi}
\end{figure*}

\begin{figure*}[htp!]
    \centering
    \includegraphics[width=0.325\textwidth]{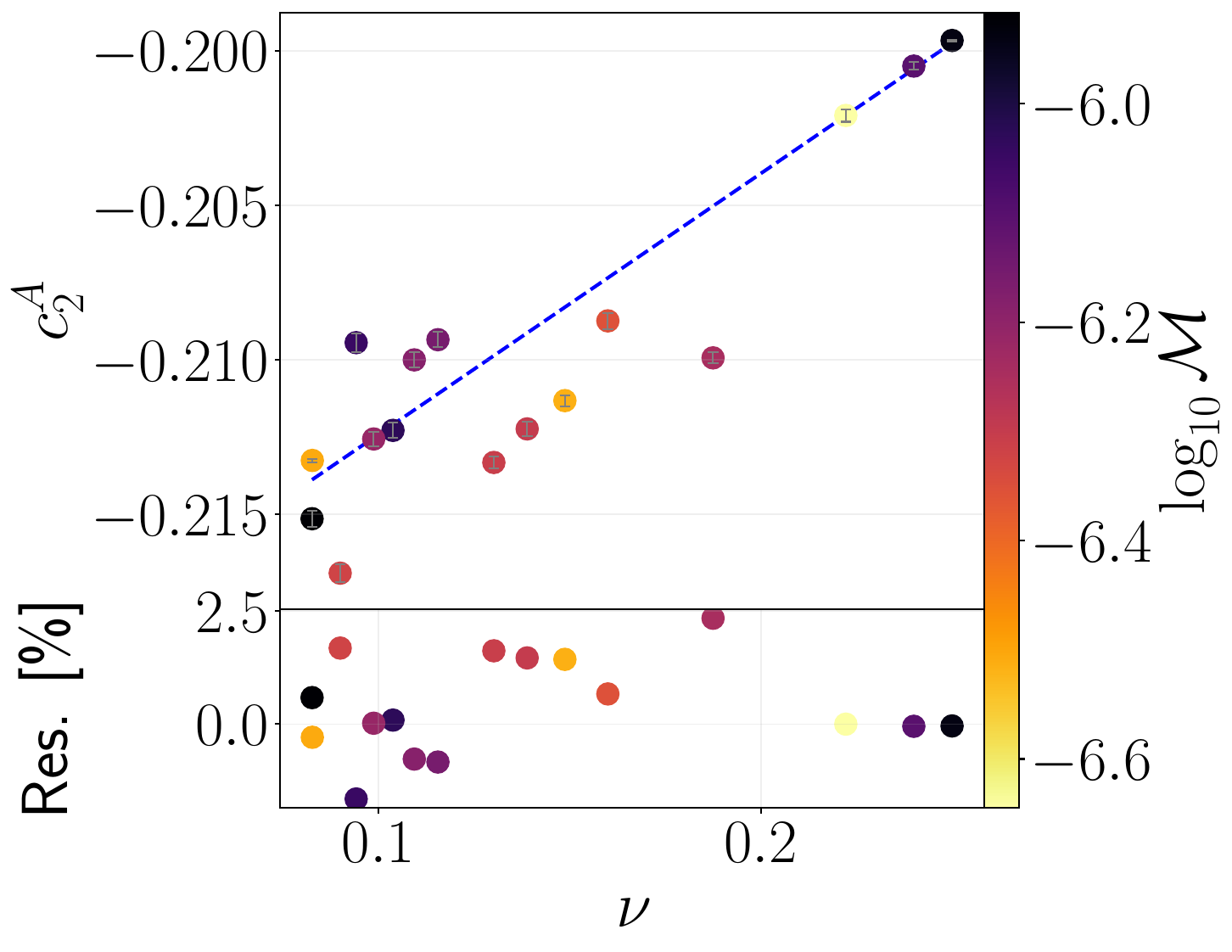}   
    \includegraphics[width=0.325\textwidth]{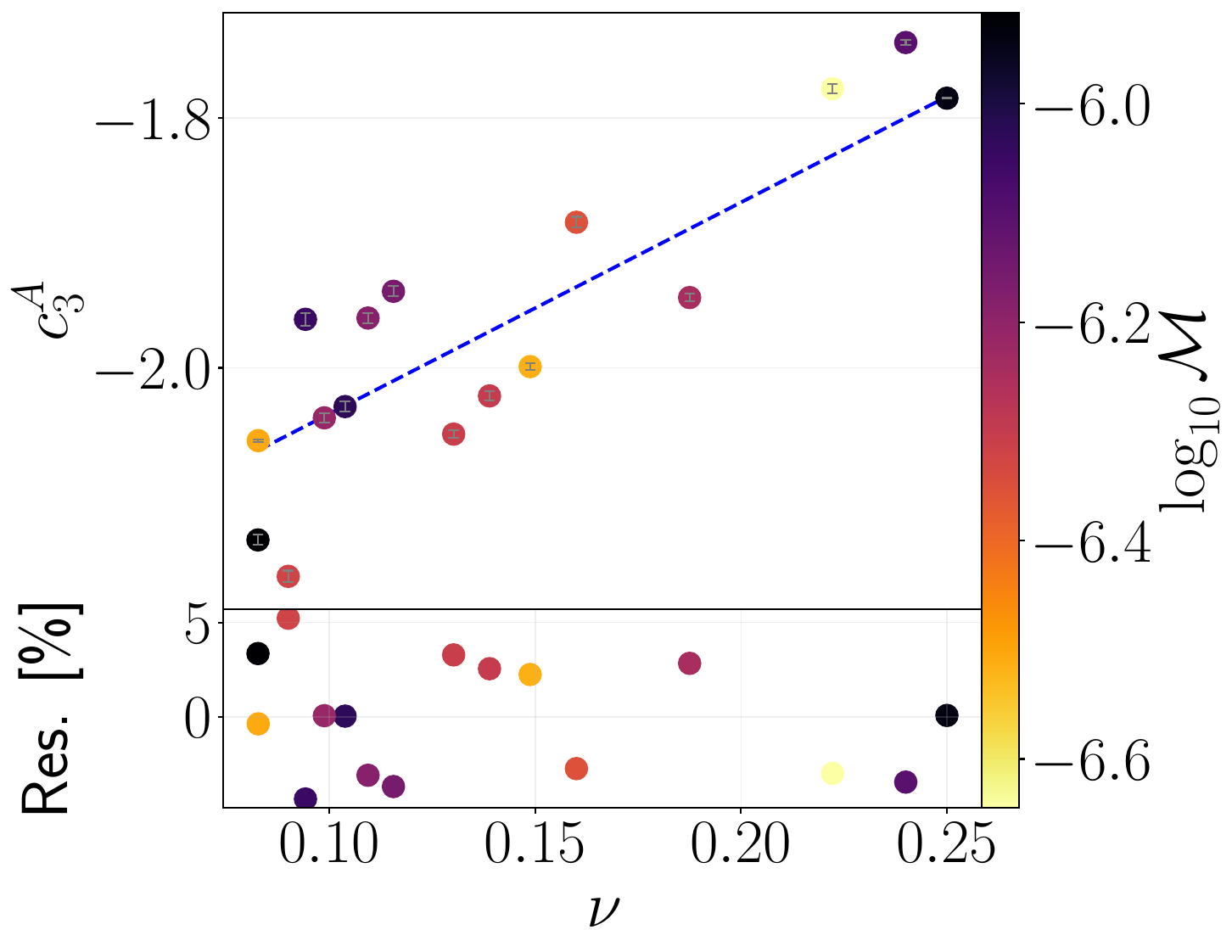}   
    \includegraphics[width=0.325\textwidth]{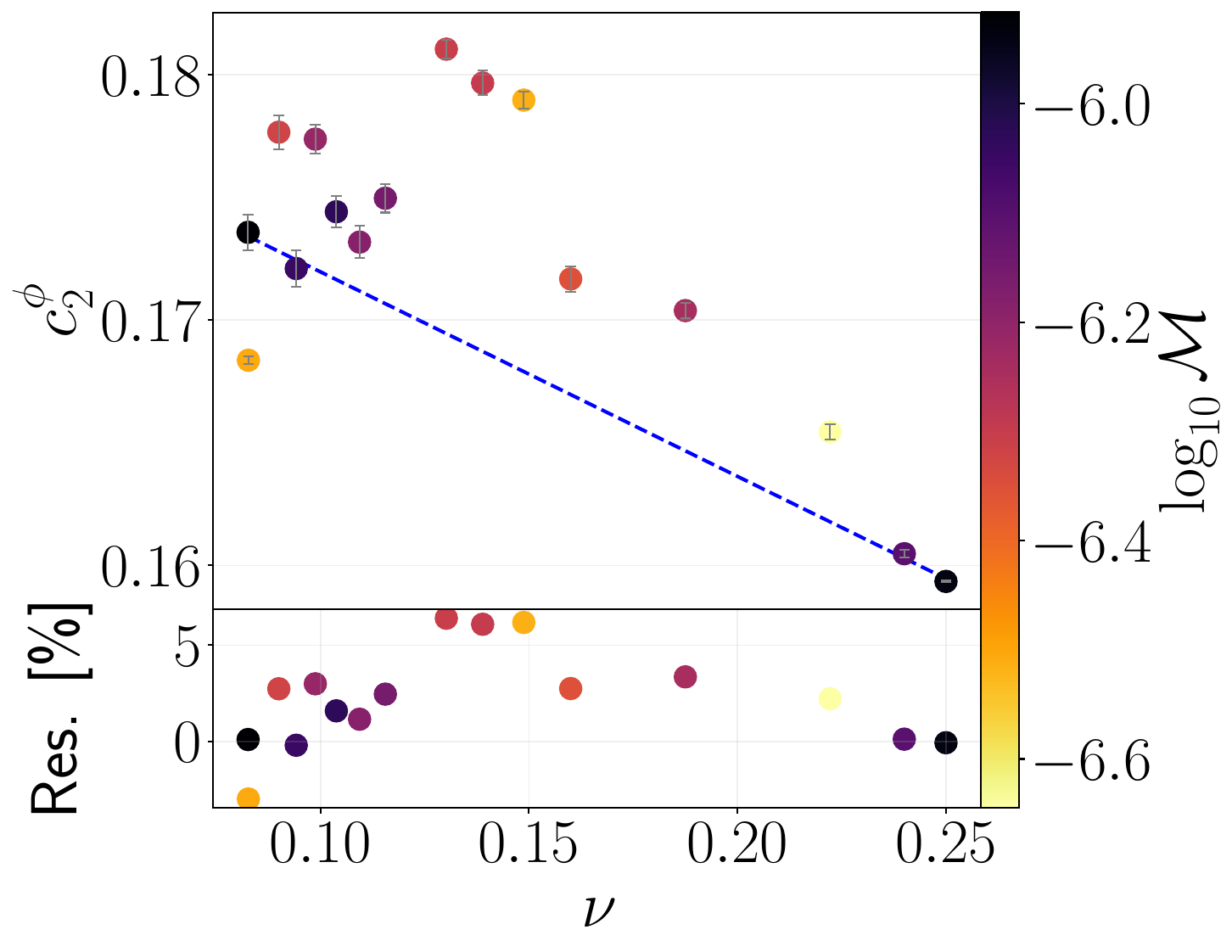}   
    \includegraphics[width=0.325\textwidth]{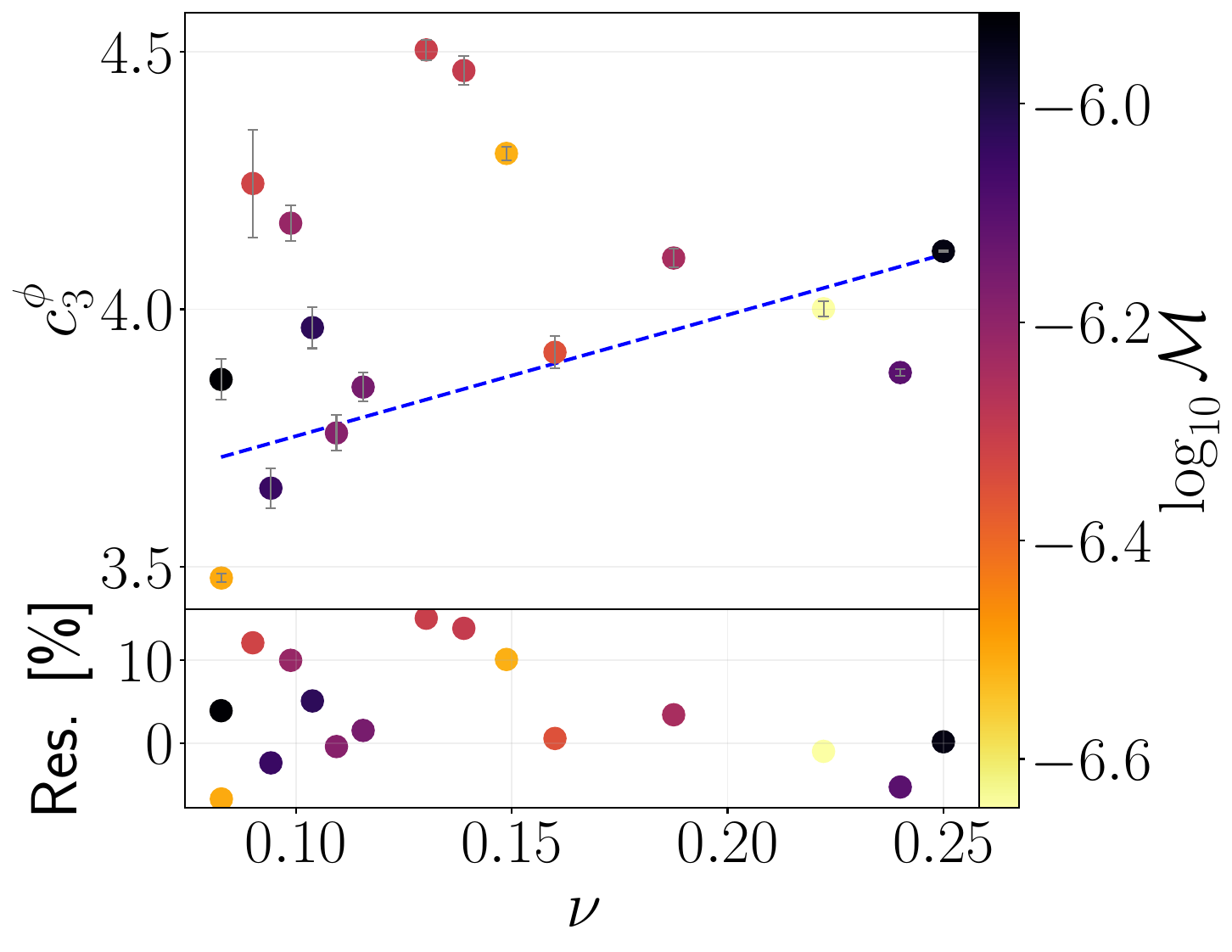}   
    \includegraphics[width=0.325\textwidth]{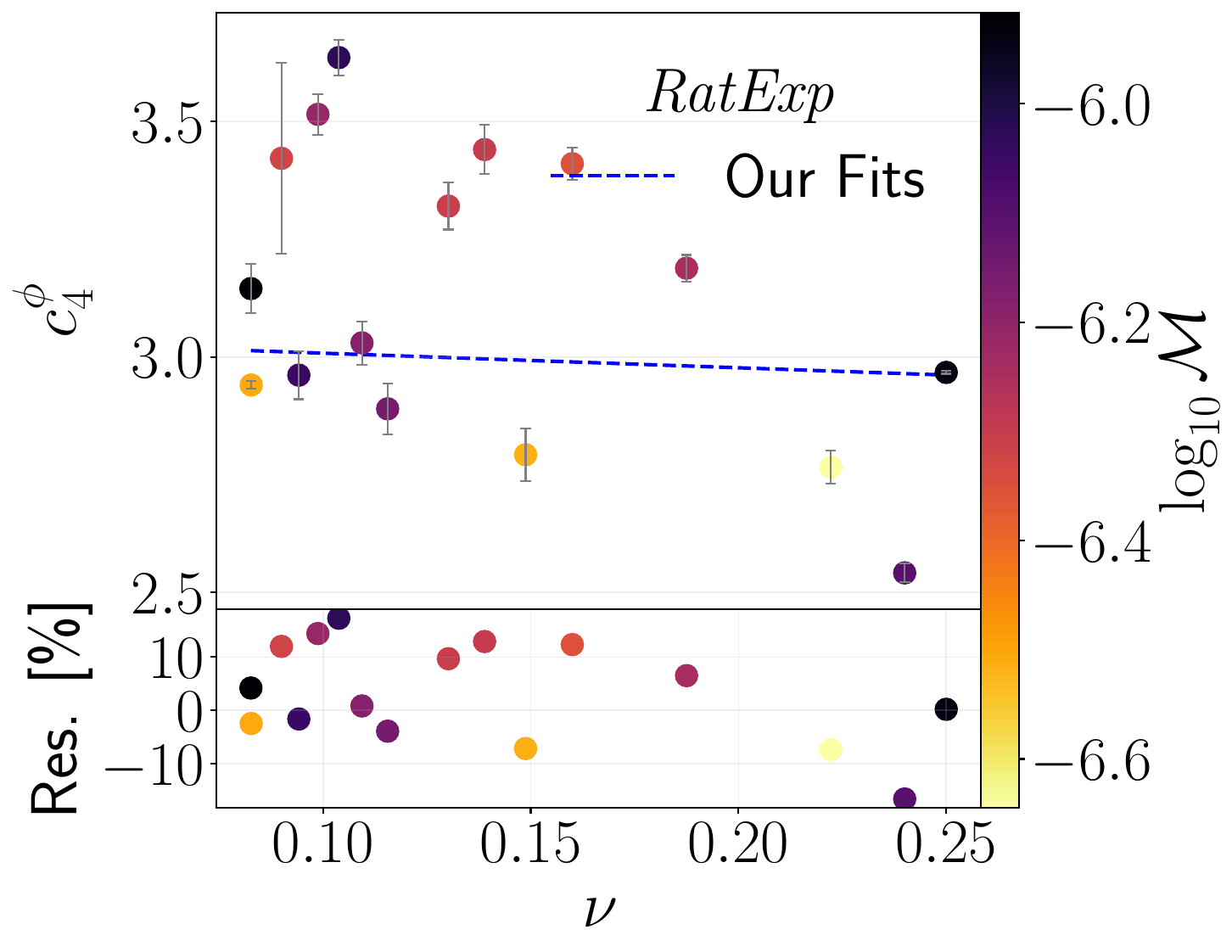}   
    \caption{
    Similar to Fig.~\ref{fig:nc_fits_sxs_chi}, median amplitude and phase coefficients from the Bayesian fits for the non-spinning SXS simulations (from Ref.~\cite{Nagar:2019wds}) against symmetric mass ratio, $\nu$.
    As before, we choose a linear $\nu$-dependency since the global fit was weakly sensitive to small variations in $\nu$. 
    }
   \label{fig:nc_fits_sxs_nu}
\end{figure*}

\begin{figure*}[htp!]
    \centering
    \includegraphics[width=0.4\textwidth]{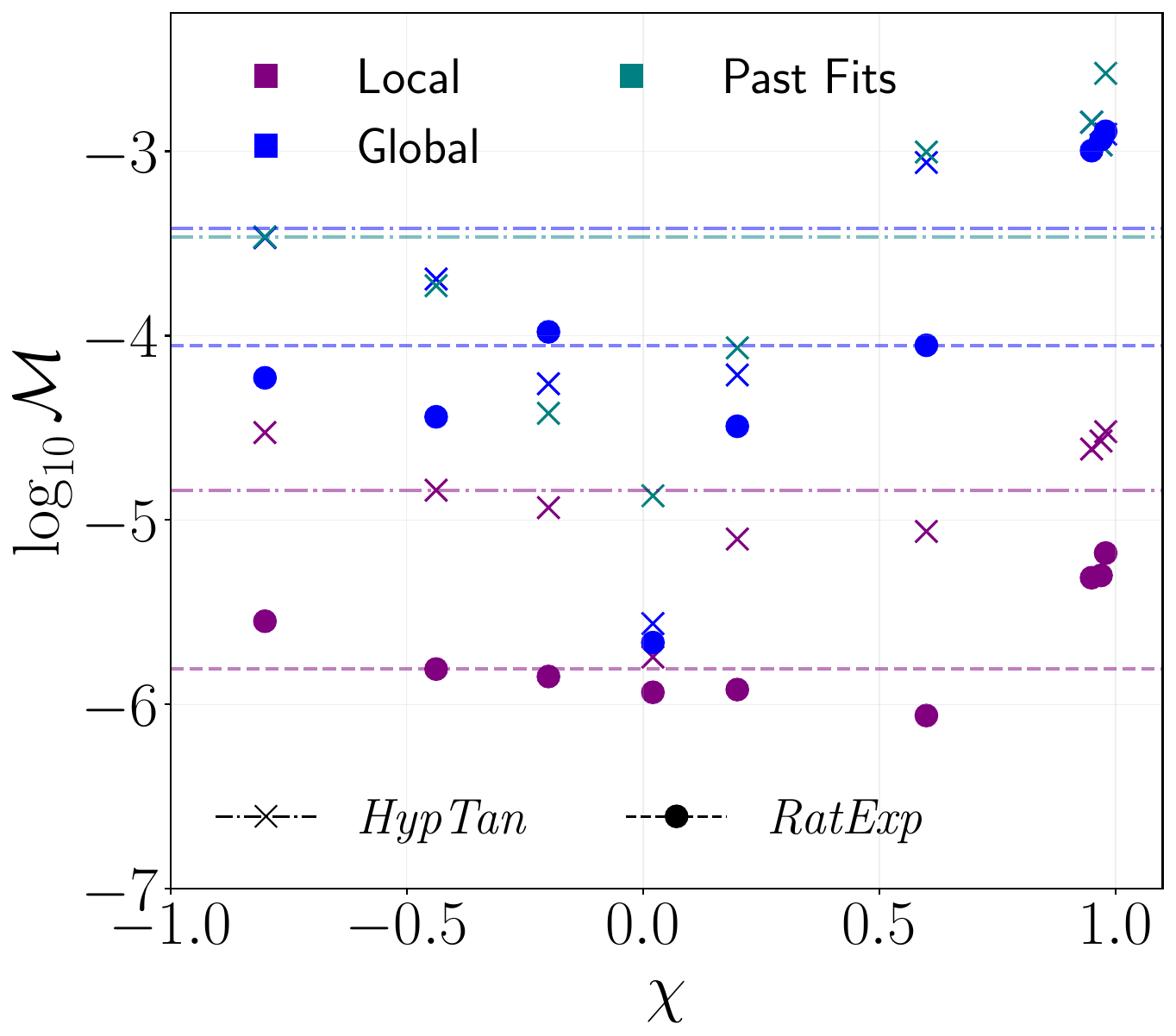}
    \includegraphics[width=0.4\textwidth]{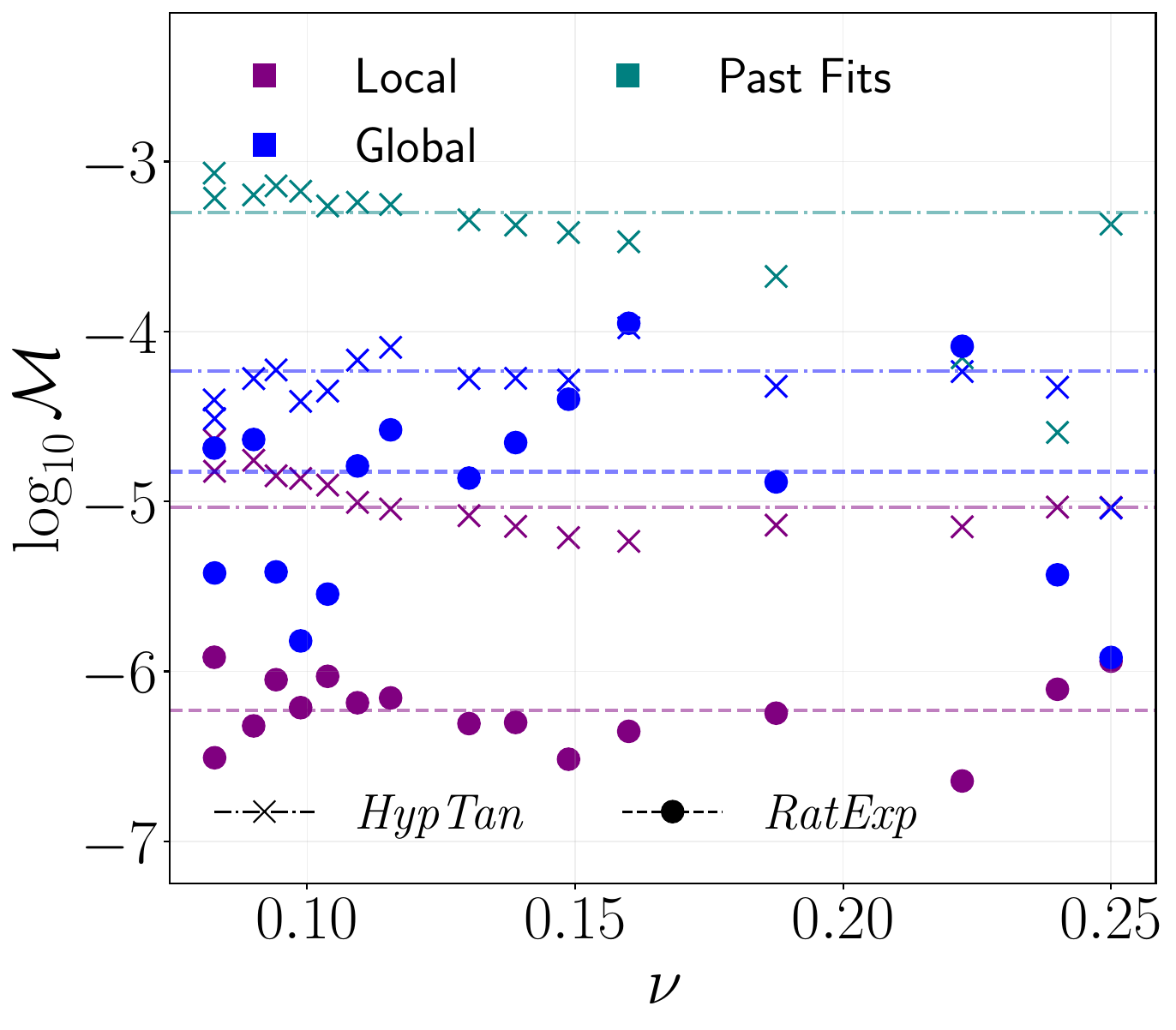}
    \caption{
    \textit{Left:} (Log) Mismatches between the NR waveform and the local fit reconstructed waveform from the median coefficients for the amplitudes and phases (purple), global fit amplitude and phase (blue), and the past fits as described in the main text (teal), plotted against $\chi$.
    The marker and the linestyle for the median mismatches represent the \emph{HypTan} ($\times$, dot-dashed) and \emph{RatExp} ($\bullet$, dashed) templates used for the fits.
    The improvement of the employed algorithm can be observed by comparing the Global and Past Fits with the \emph{HypTan} template, while considerable improvement is noted for the Global Fit with the \emph{RatExp} template, demonstrating the efficacy of the new template.
    \textit{Right:} similar result as a function of $\nu$, for the non-spinning cases.
    }
   \label{fig:mismatch_sxs_chi_nu}
\end{figure*}

We test our algorithm, parameterization and the global fit construction by reproducing previous fits in the quasi-circular case with the \emph{HypTan} template.
We begin comparing the equal-mass, equal-spin fits for the simulations in Table~I of Ref.~\cite{Damour:2014yha}.
The effective spin $\chi$~\cite{Hannam:2013oca, Ajith:2009bn, Santamaria:2010yb, Scheel:2025jct} is defined as the mass-weighted ratio of the dimensionless spin components, $\vec{\chi}_1$ and $\vec{\chi}_2$, expressed as $\chi=\left(m_1\vec{\chi}_1+m_2\vec{\chi}_2\right)/M\cdot\hat{L}$ with $\hat{L}$ being the direction of the instantaneous Newtonian orbital angular momentum.
Our fitting procedure differs both in the template and algorithm employed.
Firstly, to verify the Bayesian algorithm and inclusion of the simulation errors, we fit the same set of signals with the \emph{HypTan} template described in Eqs.~\eqref{eq:qc_amp} and~\eqref{eq:qc_phase}.
We employ polynomial fits for the extracted coefficients, but now accounting for the posterior errors in the global fits.
The median values obtained through our Bayesian fitting algorithm are presented in Fig.~\ref{fig:qc_fits_sxs_chi} and follow the same qualitative trends identified in previous work for the coefficients global fits~\cite{Damour:2014yha}.

We then repeat the non-spinning fits of Ref.~\cite{Nagar:2019wds}, based on the SXS simulations listed in their Table~II.
Results are shown in Fig.~\ref{fig:qc_fits_sxs_nu}, obtaining again a good agreement with past results.
We note that while the median coefficients for the equal-mass case are observed to be monotonic with respect to the effective spin, the non-spinning case is quite insensitive to considerable variations in the symmetric mass ratio, which explains the global fit accuracy even in the presence of large residuals of the phase coefficients.

The mismatches between the reconstructed waveforms and the SXS NR simulations are presented in Fig.~\ref{fig:mismatch_sxs_chi_nu} for both the equal-mass fits (left), and the non-spinning case (right).
The comparison showcases the improvements brought in by the fitting algorithm, here applied to the \emph{HypTan} template. 
The global fit median mismatch shows a slight improvement compared to the previous equal-mass case discussed in Ref.~\cite{Damour:2014yha}, while it indicates a significant improvement over the non-spinning fits in Ref.~\cite{Nagar:2019wds}.

Additionally, we test the \emph{RatExp} template, described in Eq.~\eqref{eq:nc_amp}, and repeat the Bayesian fits for the above simulations.
The coefficients, for the two cases against effective spin $\chi$ and symmetric mass ratio $\nu$, are displayed in Fig.~\ref{fig:nc_fits_sxs_chi} and Fig.~\ref{fig:nc_fits_sxs_nu} respectively.
The accuracy of both the algorithm and this improved template can be observed by the significantly decreased mismatches in Figs.~\ref{fig:mismatch_sxs_chi_nu}.
The \emph{RatExp} template clearly provides a significant improvement in the local and global fits as compared to the \emph{HypTan} template in both cases.

\paragraph*{Testing with Eccentricity}

\begin{figure*}[htp!]
    \centering
    \includegraphics[width=0.9\textwidth]{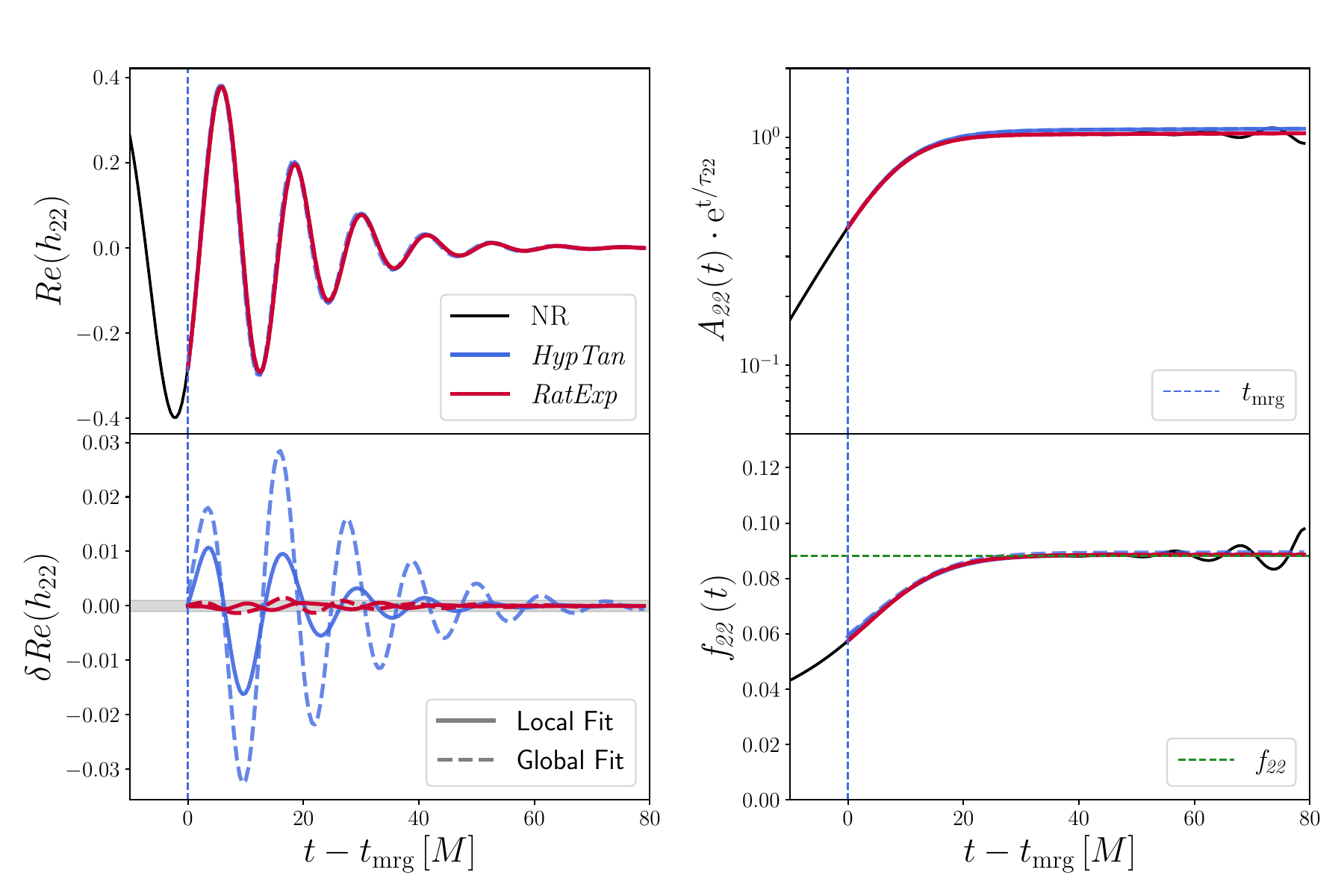}       
    \caption{
    NR strain $h_{22}$ for the illustrative case \texttt{SXS:2527}, with $e_0 \simeq 0.79$ and $q \simeq 1$.
    The red thick (brown dot-dashed) line represents the fit using the extracted coefficients for the \emph{RatExp} (\emph{HypTan}) template defined in Eq.~\eqref{eq:nc_amp} (Eqs.~\eqref{eq:qc_amp}, and~\eqref{eq:qc_phase}), fitting the numerical data only after $t_{\rm mrg}$.
    The mismatches obtained are $\mathcal{O}(10^{-4})$ for the \emph{HypTan} template and $\mathcal{O}(10^{-7})$ for the \emph{RatExp} template.
    Despite the large initial $e_0$, a longer evolution compared to RIT data implies that the residual orbital eccentricity is low by the time the system reaches the merger-ringdown.
    The post-merger thus well-aligns with a quasi-circular description, and the improvement in the mismatches with the \emph{RatExp} methodology can be attributed to the template functional form (Eq.~\eqref{eq:nc_amp}) better describing the amplitude evolution over the quasi-circular template of Eq.~\eqref{eq:qc_amp}.
    }
    \label{fig:sxs_ecc_waveform}
\end{figure*}

Although the simulations we considered for the algorithm testing had very low orbital eccentricities, we demonstrate the effectiveness of the \emph{RatExp} template in recovering a highly eccentric case from the SXS catalog, as shown in Fig.~\ref{fig:sxs_ecc_waveform}. 
As an illustrative example, the NR simulation, \texttt{SXS:2527}, features nearly equal-mass black holes with mass ratio $q\simeq 1$, is non-spinning, and has a high eccentricity ($e_0 \simeq 0.79$). 
We compared fits against both \emph{HypTan} and \emph{RatExp} templates and found that the fit was better for the \emph{RatExp} template, with a small improvement in the mismatch.
Future work will involve analyzing the entire SXS catalog.

\clearpage

\section{Constrained Least Squares Global Fit}
\label{app:lsq_linear}

\begin{figure*}[htp!]
    \centering
    \includegraphics[width=0.9\textwidth]{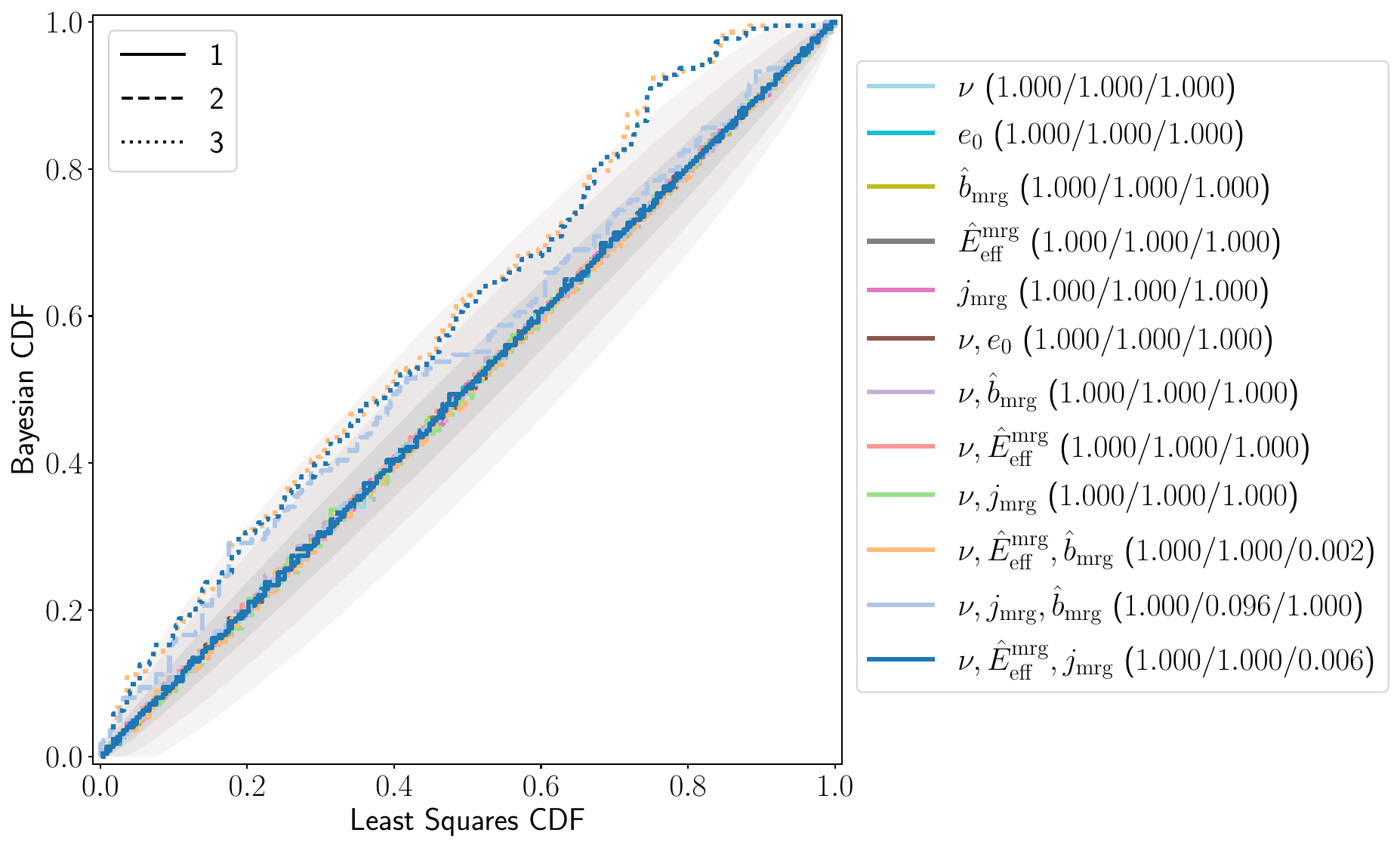}
    \caption{
    Probability-probability (P-P) plots of mismatches from Bayesian and Least Squares distributions using the CDFs.
    The deviation of the methods is predominantly seen at third polynomial order with three parameters, where the Bayesian CDF for the mismatch is peaking earlier, with lower mismatches.
    The bias indicates the limitation of the constrained global multivariate fits in higher dimensions, as compared to nested sampling.
    For the remaining cases across different parameters and orders, both methods yield consistent results.
    The $1$-, $2$- and $3$-$\sigma$ confidence intervals are indicated by the shaded regions, and $p$-values are shown for each of the parameters in successive orders (linear/quadratic/cubic).
    }
   \label{fig:pp_mismatch}
\end{figure*}

Given the high dimensionality of the parameter space, we validate the Bayesian fits by comparing them with a multivariate least-squares global-constrained fit.
The latter targets minimization of the residual functional using the \texttt{scipy.optimize.minimize} module with the Sequential Least Squares Programming (\texttt{SLSQP}) method.
An overview of the algorithm is illustrated in the Alg.~\ref{alg:poly_fit} snippet, which includes pseudocode for the major functions and modules utilized.

\begin{algorithm}[H]
    \caption{Constrained Multivariate Global fit}
    \label{alg:poly_fit}
    \begin{algorithmic}
    \Require $\mathsf{d}\in\{1, \, 2, \, 3\}$, $\boldsymbol{Q}=\{Q_i\}, \, Y,\sigma_Y$,
    \State \qquad Bounds $(p^{\min}_{\boldsymbol{\alpha}}, \, p^{\max}_{\boldsymbol{\alpha}})$, Coeff. bounds $(Y_{\min},\, Y_{\max})$
    \State
    \Function{Poly}{$\boldsymbol{Q}, \, Y_0, \, p_{\boldsymbol{\alpha}} \, \mathsf{d}$}
      \State Ensure $p^{\min}_{\boldsymbol{\alpha}}<p_{\boldsymbol{\alpha}}<p^{\max}_{\boldsymbol{\alpha}}$
      \State Define $\tilde{Y}=Y_0 + \sum_{k=1}^{\mathsf{d}}\sum_{|\boldsymbol{\alpha}|=k}p_{\boldsymbol{\alpha}}\prod_{i=1}^m Q_i^{\alpha_i}$
      \State Constrain $Y_{\min}\leq \tilde{Y}\leq Y_{\max}$
      \State \Return $\tilde{Y}$
    \EndFunction
    \State 
    \State Objective $\mathcal{L}(p_{\boldsymbol{\alpha}})=\sum_i\frac{1}{\sigma_i^2}($\Call{Poly}{$Q_i, \, Y_0, \, p_{\boldsymbol{\alpha}} \, \mathsf{d}$}$-Y_i)^2$
    \State
    \State Prepare $n_{\rm iter}=100$ Uniform random seeds in $(p^{\min}_{\boldsymbol{\alpha}}, \, p^{\max}_{\boldsymbol{\alpha}})$ 
    \State Try linear pre-fit seed via constrained $\texttt{lsq\_linear}$ module
    \State
    \State $p_{\boldsymbol{\alpha}}^{\star}\leftarrow\texttt{None}$, $\mathcal{L}^{\star}\leftarrow +\infty$
    \For{seed $p_{\boldsymbol{\alpha}}^0$}
      \State Run constrained \texttt{minimize} on $\mathcal{L}(p_{\boldsymbol{\alpha}})$ with \texttt{SLSQP} method
      \If{\emph{success}}
        \If{$\mathcal{L}(p_{\boldsymbol{\alpha}}) < \mathcal{L}^{\star}$} $p_{\boldsymbol{\alpha}}^{\star}\leftarrow p_{\boldsymbol{\alpha}}$, $\mathcal{L}^{\star}\leftarrow \mathcal{L}(p_{\boldsymbol{\alpha}})$
        \EndIf
      \Else
        \State Repeat minimization using order $\mathsf{d}'$
      \EndIf
    \EndFor
    \end{algorithmic}
\end{algorithm}

We apply a constrained optimization, bounding the amplitude and phase coefficients as prescribed by the eccentric priors in Table~\ref{tab:priors}.
The polynomial coefficient bounds are set to $p_{\boldsymbol{\alpha}}\in(-500,\, 500)$ to ensure fit stability.

Wherever possible, we attempt an (unconstrained) linear pre-fit (via \texttt{scipy.optimize.lsq\_linear}) and use that as one of the seeds for the minimization.
Algorithm convergence was enforced by repeating the fit using $100$ distinct iterations, verifying that it delivers compatible residuals, and selecting the maximum likelihood value.
If optimization fails at the requested order, we retry with decreasing polynomial order and (if successful) pad coefficients to the original size.

To prevent overfitting in our constrained polynomial fit, we apply a small amount of regularization~\cite{Ying:2019abc} to the residual objective function that we are optimizing. 
We found this addition particularly important due to the high dimensionality, especially when the coefficients are sparsely distributed.

We compare the $\tilde{\chi}^2$ statistic (proportional to the total residuals) using both the Bayesian results and the least square global fits.
We find comparable results between the two methods, with least squares underperforming at higher orders.
We further compare the two methods by constructing a probability-probability plot of the waveform template mismatches against NR data, when evaluated using the coefficients obtained using each of the two fitting methods.
The result, shown in Fig.~\ref{fig:pp_mismatch}, indicates that the Bayesian method outperforms the bounded least squares for the three-parameter fits at the third polynomial order.

\section{Model Constraints}
\label{app:constraints}

We present the fixed coefficients determined from the NR amplitude and phase at merger, determined by imposing continuity.

\paragraph*{Quasi-Circular Model:}
For the quasi-circular \emph{HypTan} model (Eqs.~\eqref{eq:qc_amp},~\eqref{eq:qc_phase}), enforcing continuity and first-derivative smoothness on the amplitude and phase yields the following constrained coefficients~\cite{Damour:2014yha}

\begin{subequations}
    \begin{align}
        c_1^A &= A_{22}^{\rm mrg} \alpha_1 \frac{\cosh^2(c_3^A)}{c_2^A} \, , \\
        c_2^A &= \frac{\alpha_2 - \alpha_1}{2} \, , \quad c_2^{\phi} = \alpha_2 - \alpha_1 \, , \\
        c_4^A &= A_{22}^{\rm mrg} - c_1^A \tanh(c_3^A) \, , \\
        c_1^{\phi} &= \frac{1 + c_3^{\phi} + c_4^{\phi}}{c_2^{\phi}(c_3^{\phi} + 2c_4^{\phi})} \Delta\omega \, .
    \end{align}
\end{subequations}

where $\Delta\omega=\omega_1-M\omega_{22}^{\rm mrg}$, determined by the QNM-rescaled ringdown phase, $\phi_{\bar{h}}(\tau)=\omega_1\tau-\phi_{22}(\tau)+\phi_{22}^{\rm mrg}$.
The remaining coefficients $\{c_3^A, \, c_3^{\phi}, \, c_4^{\phi}\}$ act as free parameters, which are subsequently determined by the global parameter-space fits.

\paragraph*{Non-Circular Model:}
To capture the more complex transient features characteristic of highly eccentric orbits, the \emph{RatExp} template (Eq.~\eqref{eq:nc_amp}) extends the amplitude matching to second-derivative continuity, while maintaining first-derivative continuity for the phase~\cite{Albanesi:2023bgi}. 
Solving this matching system yields the following constraints

\begin{subequations}
    \begin{align}
        c_1^A &= \frac{c_5^A \alpha_1}{c_2^A} (A_{22}^{\rm mrg})^{c_5^A} \e^{-c_3^A} (1 + \e^{c_3^A})^2 \, , \\
        c_4^A &= (A_{22}^{\rm mrg})^{c_5^A} - \frac{c_1^A}{1 + \e^{c_3^A}} \, , \\
        c_5^A &= -\frac{\ddot{A}_{22}^{\rm mrg}}{A_{22}^{\rm mrg} \alpha_1^2} + \frac{c_2^A}{\alpha_1} \frac{\e^{c_3^A} - 1}{\e^{c_3^A} + 1} \, , \\
        c_1^{\phi} &= \frac{1 + c_3^{\phi} + c_4^{\phi}}{c_2^{\phi}(c_3^{\phi} + 2c_4^{\phi})} \Delta\omega \, .
    \end{align}
\end{subequations}

Here, the parameters $\{c_2^A, \, c_3^A, \, c_2^{\phi}, \, c_3^{\phi}, \, c_4^{\phi}\}$ remain unconstrained by the boundary conditions.

\clearpage

\bibliography{bibliography}

@article{Pompili:2023tna,
    author = "Pompili, Lorenzo and others",
    title = "{Laying the foundation of the effective-one-body waveform models SEOBNRv5: Improved accuracy and efficiency for spinning nonprecessing binary black holes}",
    eprint = "2303.18039",
    archivePrefix = "arXiv",
    primaryClass = "gr-qc",
    doi = "10.1103/PhysRevD.108.124035",
    journal = "Phys. Rev. D",
    volume = "108",
    number = "12",
    pages = "124035",
    year = "2023"
}

@article{Jan:2025zcm,
    author = "Jan, Aasim and Nicolella, Sophia and Shoemaker, Deirdre and O'Shaughnessy, Richard",
    title = "{Measuring Eccentricity and Addressing Waveform Systematics in GW231123}",
    eprint = "2512.20060",
    archivePrefix = "arXiv",
    primaryClass = "gr-qc",
    month = "12",
    year = "2025",
    journal="",
}

@article{Chatziioannou:2024hju,
    author = {Chatziioannou, K. and Dent, T. and Fishbach, M. and Ohme, F. and P{\"u}rrer, M. and Raymond, V. and Veitch, J.},
    title = "{Compact binary coalescences: gravitational-wave astronomy with ground-based detectors}",
    eprint = "2409.02037",
    archivePrefix = "arXiv",
    primaryClass = "gr-qc",
    month = "9",
    year = "2024",
    journal="",
}

@article{LIGOScientific:2025rid,
    author = "Abac, A. G. and others",
    collaboration = "LIGO Scientific, Virgo, KAGRA",
    title = "{GW250114: Testing Hawking{\textquoteright}s Area Law and the Kerr Nature of Black Holes}",
    eprint = "2509.08054",
    archivePrefix = "arXiv",
    primaryClass = "gr-qc",
    reportNumber = "LIGO-P2500421",
    doi = "10.1103/kw5g-d732",
    journal = "Phys. Rev. Lett.",
    volume = "135",
    number = "11",
    pages = "111403",
    year = "2025"
}

@article{LIGOScientific:2025rsn,
    author = "Abac, A. G. and others",
    collaboration = "LIGO Scientific, VIRGO, KAGRA",
    title = "{GW231123: a Binary Black Hole Merger with Total Mass 190-265 $M_{\odot}$}",
    eprint = "2507.08219",
    archivePrefix = "arXiv",
    primaryClass = "astro-ph.HE",
    reportNumber = "DCC: P2500026-v6",
    month = "7",
    year = "2025",
    journal="",
}

@article{LIGOScientific:2020iuh,
    author = "Abbott, R. and others",
    collaboration = "LIGO Scientific, Virgo",
    title = "{GW190521: A Binary Black Hole Merger with a Total Mass of $150  M_{\odot}$}",
    eprint = "2009.01075",
    archivePrefix = "arXiv",
    primaryClass = "gr-qc",
    doi = "10.1103/PhysRevLett.125.101102",
    journal = "Phys. Rev. Lett.",
    volume = "125",
    number = "10",
    pages = "101102",
    year = "2020"
}

@article{LIGOScientific:2020ufj,
    author = "Abbott, R. and others",
    collaboration = "LIGO Scientific, Virgo",
    title = "{Properties and Astrophysical Implications of the 150 M$_\odot$ Binary Black Hole Merger GW190521}",
    eprint = "2009.01190",
    archivePrefix = "arXiv",
    primaryClass = "astro-ph.HE",
    reportNumber = "LIGO-P2000021",
    doi = "10.3847/2041-8213/aba493",
    journal = "Astrophys. J. Lett.",
    volume = "900",
    number = "1",
    pages = "L13",
    year = "2020"
}

@article{LIGOScientific:2025slb,
    author = "Abac, A. G. and others",
    collaboration = "LIGO Scientific, VIRGO, KAGRA",
    title = "{GWTC-4.0: Updating the Gravitational-Wave Transient Catalog with Observations from the First Part of the Fourth LIGO-Virgo-KAGRA Observing Run}",
    eprint = "2508.18082",
    archivePrefix = "arXiv",
    primaryClass = "gr-qc",
    reportNumber = "LIGO-P2400386",
    month = "8",
    year = "2025",
    journal="",
}

@article{Gayathri:2020coq,
    author = "Gayathri, V. and Healy, J. and Lange, J. and O'Brien, B. and Szczepanczyk, M. and Bartos, Imre and Campanelli, M. and Klimenko, S. and Lousto, C. O. and O'Shaughnessy, R.",
    title = "{Eccentricity estimate for black hole mergers with numerical relativity simulations}",
    eprint = "2009.05461",
    archivePrefix = "arXiv",
    primaryClass = "astro-ph.HE",
    doi = "10.1038/s41550-021-01568-w",
    journal = "Nature Astron.",
    volume = "6",
    number = "3",
    pages = "344--349",
    year = "2022"
}

@article{Berti:2025hly,
    author = "Berti, Emanuele and others",
    title = "{Black hole spectroscopy: from theory to experiment}",
    eprint = "2505.23895",
    archivePrefix = "arXiv",
    primaryClass = "gr-qc",
    month = "5",
    year = "2025",
    journal="",
}

@article{LIGOScientific:2016lio,
    author = "Abbott, B. P. and others",
    collaboration = "LIGO Scientific, Virgo",
    title = "{Tests of general relativity with GW150914}",
    eprint = "1602.03841",
    archivePrefix = "arXiv",
    primaryClass = "gr-qc",
    reportNumber = "LIGO-P1500213",
    doi = "10.1103/PhysRevLett.116.221101",
    journal = "Phys. Rev. Lett.",
    volume = "116",
    number = "22",
    pages = "221101",
    year = "2016",
    note = "[Erratum: Phys.Rev.Lett. 121, 129902 (2018)]"
}

@misc{LIGOT1800044,
  author       = {{LIGO Scientific Collaboration}},
  title        = {Instrument Science White Paper 2018},
  year         = {2018},
  note         = {\url{https://dcc.ligo.org/LIGO-T1800044/public}},
  howpublished = {LIGO Document T1800044}
}

@article{LIGOScientific:2014pky,
    author = "Aasi, J. and others",
    collaboration = "LIGO Scientific",
    title = "{Advanced LIGO}",
    eprint = "1411.4547",
    archivePrefix = "arXiv",
    primaryClass = "gr-qc",
    doi = "10.1088/0264-9381/32/7/074001",
    journal = "Class. Quant. Grav.",
    volume = "32",
    pages = "074001",
    year = "2015"
}

@article{VIRGO:2014yos,
    author = "Acernese, F. and others",
    collaboration = "VIRGO",
    title = "{Advanced Virgo: a second-generation interferometric gravitational wave detector}",
    eprint = "1408.3978",
    archivePrefix = "arXiv",
    primaryClass = "gr-qc",
    doi = "10.1088/0264-9381/32/2/024001",
    journal = "Class. Quant. Grav.",
    volume = "32",
    number = "2",
    pages = "024001",
    year = "2015"
}

@article{KAGRA:2018plz,
    author = "Akutsu, T. and others",
    collaboration = "KAGRA",
    title = "{KAGRA: 2.5 Generation Interferometric Gravitational Wave Detector}",
    eprint = "1811.08079",
    archivePrefix = "arXiv",
    primaryClass = "gr-qc",
    reportNumber = "JGW-P1809243",
    doi = "10.1038/s41550-018-0658-y",
    journal = "Nature Astron.",
    volume = "3",
    number = "1",
    pages = "35--40",
    year = "2019"
}

@article{Vishveshwara:1970cc,
    author = "Vishveshwara, C. V.",
    title = "{Stability of the schwarzschild metric}",
    doi = "10.1103/PhysRevD.1.2870",
    journal = "Phys. Rev. D",
    volume = "1",
    pages = "2870--2879",
    year = "1970"
}

@article{Press:1971wr,
    author = "Press, William H.",
    title = "{Long Wave Trains of Gravitational Waves from a Vibrating Black Hole}",
    reportNumber = "OAP-262",
    doi = "10.1086/180849",
    journal = "Astrophys. J. Lett.",
    volume = "170",
    pages = "L105--L108",
    year = "1971"
}

@article{Teukolsky:1973ha,
    author = "Teukolsky, Saul A.",
    title = "{Perturbations of a rotating black hole. 1. Fundamental equations for gravitational electromagnetic and neutrino field perturbations}",
    doi = "10.1086/152444",
    journal = "Astrophys. J.",
    volume = "185",
    pages = "635--647",
    year = "1973"
}

@article{Press:1973zz,
    author = "Press, William H. and Teukolsky, Saul A.",
    title = "{Perturbations of a Rotating Black Hole. II. Dynamical Stability of the Kerr Metric}",
    doi = "10.1086/152445",
    journal = "Astrophys. J.",
    volume = "185",
    pages = "649--674",
    year = "1973"
}

@article{Chandrasekhar:1975zza,
    author = "Chandrasekhar, S. and Detweiler, Steven L.",
    title = "{The quasi-normal modes of the Schwarzschild black hole}",
    doi = "10.1098/rspa.1975.0112",
    journal = "Proc. Roy. Soc. Lond. A",
    volume = "344",
    pages = "441--452",
    year = "1975"
}

@article{Chandrasekhar:1975zz,
    author = "Chandrasekhar, S. and Detweiler, Steven L.",
    title = "{Equations governing axisymmetric perturbations of the Kerr black-hole}",
    doi = "10.1098/rspa.1975.0130",
    journal = "Proc. Roy. Soc. Lond. A",
    volume = "345",
    pages = "145--167",
    year = "1975"
}

@article{Chandrasekhar:1976zz,
    author = "Chandrasekhar, S. and Detweiler, Steven L.",
    title = "{Equations governing gravitational perturbations of the Kerr black-hole}",
    doi = "10.1098/rspa.1976.0101",
    journal = "Proc. Roy. Soc. Lond. A",
    volume = "350",
    pages = "165--174",
    year = "1976"
}

@article{Berti:2009kk,
    author = "Berti, Emanuele and Cardoso, Vitor and Starinets, Andrei O.",
    title = "{Quasinormal modes of black holes and black branes}",
    eprint = "0905.2975",
    archivePrefix = "arXiv",
    primaryClass = "gr-qc",
    doi = "10.1088/0264-9381/26/16/163001",
    journal = "Class. Quant. Grav.",
    volume = "26",
    pages = "163001",
    year = "2009"
}

@inbook{Mapelli:2021taw,
    author = "Mapelli, Michela",
    title = "{Formation Channels of Single and Binary Stellar-Mass Black Holes}",
    eprint = "2106.00699",
    archivePrefix = "arXiv",
    primaryClass = "astro-ph.HE",
    doi = "10.1007/978-981-15-4702-7_16-1",
    year = "2021"
}

@article{Samsing:2013kua,
    author = "Samsing, Johan and MacLeod, Morgan and Ramirez-Ruiz, Enrico",
    title = "{The Formation of Eccentric Compact Binary Inspirals and the Role of Gravitational Wave Emission in Binary-Single Stellar Encounters}",
    eprint = "1308.2964",
    archivePrefix = "arXiv",
    primaryClass = "astro-ph.HE",
    doi = "10.1088/0004-637X/784/1/71",
    journal = "Astrophys. J.",
    volume = "784",
    pages = "71",
    year = "2014"
}

@article{Zevin:2021rtf,
    author = "Zevin, Michael and Romero-Shaw, Isobel M. and Kremer, Kyle and Thrane, Eric and Lasky, Paul D.",
    title = "{Implications of Eccentric Observations on Binary Black Hole Formation Channels}",
    eprint = "2106.09042",
    archivePrefix = "arXiv",
    primaryClass = "astro-ph.HE",
    doi = "10.3847/2041-8213/ac32dc",
    journal = "Astrophys. J. Lett.",
    volume = "921",
    number = "2",
    pages = "L43",
    year = "2021"
}

@article{Mandel:2018hfr,
    author = "Mandel, Ilya and Farmer, Alison",
    title = "{Merging stellar-mass binary black holes}",
    eprint = "1806.05820",
    archivePrefix = "arXiv",
    primaryClass = "astro-ph.HE",
    doi = "10.1016/j.physrep.2022.01.003",
    journal = "Phys. Rept.",
    volume = "955",
    pages = "1--24",
    year = "2022"
}

@article{Peters:1963ux,
    author = "Peters, P. C. and Mathews, J.",
    title = "{Gravitational radiation from point masses in a Keplerian orbit}",
    doi = "10.1103/PhysRev.131.435",
    journal = "Phys. Rev.",
    volume = "131",
    pages = "435--439",
    year = "1963"
}

@article{Buonanno:1998gg,
    author = "Buonanno, A. and Damour, T.",
    title = "{Effective one-body approach to general relativistic two-body dynamics}",
    eprint = "gr-qc/9811091",
    archivePrefix = "arXiv",
    reportNumber = "IHES-P-98-74",
    doi = "10.1103/PhysRevD.59.084006",
    journal = "Phys. Rev. D",
    volume = "59",
    pages = "084006",
    year = "1999"
}

@article{Damour:2007xr,
    author = "Damour, Thibault and Nagar, Alessandro",
    title = "{Faithful effective-one-body waveforms of small-mass-ratio coalescing black-hole binaries}",
    eprint = "0705.2519",
    archivePrefix = "arXiv",
    primaryClass = "gr-qc",
    doi = "10.1103/PhysRevD.76.064028",
    journal = "Phys. Rev. D",
    volume = "76",
    pages = "064028",
    year = "2007"
}

@article{Damour:2001tu,
    author = "Damour, Thibault",
    title = "{Coalescence of two spinning black holes: an effective one-body approach}",
    eprint = "gr-qc/0103018",
    archivePrefix = "arXiv",
    reportNumber = "IHES-P-01-11",
    doi = "10.1103/PhysRevD.64.124013",
    journal = "Phys. Rev. D",
    volume = "64",
    pages = "124013",
    year = "2001"
}

@article{Buonanno:2000ef,
    author = "Buonanno, Alessandra and Damour, Thibault",
    title = "{Transition from inspiral to plunge in binary black hole coalescences}",
    eprint = "gr-qc/0001013",
    archivePrefix = "arXiv",
    reportNumber = "IHES-P-99-90, GRP-99-521",
    doi = "10.1103/PhysRevD.62.064015",
    journal = "Phys. Rev. D",
    volume = "62",
    pages = "064015",
    year = "2000"
}

@article{Peters:1964zz,
    author = "Peters, P. C.",
    title = "{Gravitational Radiation and the Motion of Two Point Masses}",
    doi = "10.1103/PhysRev.136.B1224",
    journal = "Phys. Rev.",
    volume = "136",
    pages = "B1224--B1232",
    year = "1964"
}

@article{Carullo:2023kvj,
    author = "Carullo, Gregorio and Albanesi, Simone and Nagar, Alessandro and Gamba, Rossella and Bernuzzi, Sebastiano and Andrade, Tomas and Trenado, Juan",
    title = "{Unveiling the Merger Structure of Black Hole Binaries in Generic Planar Orbits}",
    eprint = "2309.07228",
    archivePrefix = "arXiv",
    primaryClass = "gr-qc",
    reportNumber = "VIR-0804A-23",
    doi = "10.1103/PhysRevLett.132.101401",
    journal = "Phys. Rev. Lett.",
    volume = "132",
    number = "10",
    pages = "101401",
    year = "2024"
}

@article{Narayan:2023vhm,
    author = "Narayan, Purnima and Johnson-McDaniel, Nathan K. and Gupta, Anuradha",
    title = "{Effect of ignoring eccentricity in testing general relativity with gravitational waves}",
    eprint = "2306.04068",
    archivePrefix = "arXiv",
    primaryClass = "gr-qc",
    doi = "10.1103/PhysRevD.108.064003",
    journal = "Phys. Rev. D",
    volume = "108",
    number = "6",
    pages = "064003",
    year = "2023"
}

@article{Regge:1957td,
    author = "Regge, Tullio and Wheeler, John A.",
    title = "{Stability of a Schwarzschild singularity}",
    doi = "10.1103/PhysRev.108.1063",
    journal = "Phys. Rev.",
    volume = "108",
    pages = "1063--1069",
    year = "1957"
}

@article{Zerilli:1970wzz,
    author = "Zerilli, F. J.",
    title = "{Gravitational field of a particle falling in a schwarzschild geometry analyzed in tensor harmonics}",
    doi = "10.1103/PhysRevD.2.2141",
    journal = "Phys. Rev. D",
    volume = "2",
    pages = "2141--2160",
    year = "1970"
}

@article{Zerilli:1970se,
    author = "Zerilli, Frank J.",
    title = "{Effective potential for even parity Regge-Wheeler gravitational perturbation equations}",
    doi = "10.1103/PhysRevLett.24.737",
    journal = "Phys. Rev. Lett.",
    volume = "24",
    pages = "737--738",
    year = "1970"
}

@article{Teukolsky:1972my,
    author = "Teukolsky, S. A.",
    title = "{Rotating black holes - separable wave equations for gravitational and electromagnetic perturbations}",
    reportNumber = "OAP-291",
    doi = "10.1103/PhysRevLett.29.1114",
    journal = "Phys. Rev. Lett.",
    volume = "29",
    pages = "1114--1118",
    year = "1972"
}

@article{Teukolsky:1974yv,
    author = "Teukolsky, S. A. and Press, W. H.",
    title = "{Perturbations of a rotating black hole. III - Interaction of the hole with gravitational and electromagnetic radiation}",
    doi = "10.1086/153180",
    journal = "Astrophys. J.",
    volume = "193",
    pages = "443--461",
    year = "1974"
}

@article{Nollert:1999ji,
    author = "Nollert, Hans-Peter",
    title = "{TOPICAL REVIEW: Quasinormal modes: the characteristic `sound' of black holes and neutron stars}",
    doi = "10.1088/0264-9381/16/12/201",
    journal = "Class. Quant. Grav.",
    volume = "16",
    pages = "R159--R216",
    year = "1999"
}

@article{Ferrari:2007dd,
    author = "Ferrari, Valeria and Gualtieri, Leonardo",
    title = "{Quasi-Normal Modes and Gravitational Wave Astronomy}",
    eprint = "0709.0657",
    archivePrefix = "arXiv",
    primaryClass = "gr-qc",
    doi = "10.1007/s10714-007-0585-1",
    journal = "Gen. Rel. Grav.",
    volume = "40",
    pages = "945--970",
    year = "2008"
}

@article{Echeverria:1989hg,
    author = "Echeverria, F.",
    title = "{Gravitational Wave Measurements of the Mass and Angular Momentum of a Black Hole}",
    doi = "10.1103/PhysRevD.40.3194",
    journal = "Phys. Rev. D",
    volume = "40",
    pages = "3194--3203",
    year = "1989"
}

@article{Gamba:2021ydi,
    author = "Gamba, Rossella and Ak{\c{c}}ay, Sarp and Bernuzzi, Sebastiano and Williams, Jake",
    title = "{Effective-one-body waveforms for precessing coalescing compact binaries with post-Newtonian twist}",
    eprint = "2111.03675",
    archivePrefix = "arXiv",
    primaryClass = "gr-qc",
    doi = "10.1103/PhysRevD.106.024020",
    journal = "Phys. Rev. D",
    volume = "106",
    number = "2",
    pages = "024020",
    year = "2022"
}

@article{Damour:2000we,
    author = "Damour, Thibault and Jaranowski, Piotr and Schaefer, Gerhard",
    title = "{On the determination of the last stable orbit for circular general relativistic binaries at the third postNewtonian approximation}",
    eprint = "gr-qc/0005034",
    archivePrefix = "arXiv",
    doi = "10.1103/PhysRevD.62.084011",
    journal = "Phys. Rev. D",
    volume = "62",
    pages = "084011",
    year = "2000"
}

@article{Damour:2015isa,
    author = {Damour, Thibault and Jaranowski, Piotr and Sch{\"a}fer, Gerhard},
    title = "{Fourth post-Newtonian effective one-body dynamics}",
    eprint = "1502.07245",
    archivePrefix = "arXiv",
    primaryClass = "gr-qc",
    doi = "10.1103/PhysRevD.91.084024",
    journal = "Phys. Rev. D",
    volume = "91",
    number = "8",
    pages = "084024",
    year = "2015"
}

@article{Dreyer:2003bv,
    author = "Dreyer, Olaf and Kelly, Bernard J. and Krishnan, Badri and Finn, Lee Samuel and Garrison, David and Lopez-Aleman, Ramon",
    title = "{Black hole spectroscopy: Testing general relativity through gravitational wave observations}",
    eprint = "gr-qc/0309007",
    archivePrefix = "arXiv",
    doi = "10.1088/0264-9381/21/4/003",
    journal = "Class. Quant. Grav.",
    volume = "21",
    pages = "787--804",
    year = "2004"
}

@article{Berti:2005ys,
    author = "Berti, Emanuele and Cardoso, Vitor and Will, Clifford M.",
    title = "{On gravitational-wave spectroscopy of massive black holes with the space interferometer LISA}",
    eprint = "gr-qc/0512160",
    archivePrefix = "arXiv",
    doi = "10.1103/PhysRevD.73.064030",
    journal = "Phys. Rev. D",
    volume = "73",
    pages = "064030",
    year = "2006"
}

@article{Gossan:2011ha,
    author = "Gossan, S. and Veitch, J. and Sathyaprakash, B. S.",
    title = "{Bayesian model selection for testing the no-hair theorem with black hole ringdowns}",
    eprint = "1111.5819",
    archivePrefix = "arXiv",
    primaryClass = "gr-qc",
    doi = "10.1103/PhysRevD.85.124056",
    journal = "Phys. Rev. D",
    volume = "85",
    pages = "124056",
    year = "2012"
}

@article{Berti:2015itd,
    author = "Berti, Emanuele and others",
    title = "{Testing General Relativity with Present and Future Astrophysical Observations}",
    eprint = "1501.07274",
    archivePrefix = "arXiv",
    primaryClass = "gr-qc",
    doi = "10.1088/0264-9381/32/24/243001",
    journal = "Class. Quant. Grav.",
    volume = "32",
    pages = "243001",
    year = "2015"
}

@article{Berti:2016lat,
    author = "Berti, Emanuele and Sesana, Alberto and Barausse, Enrico and Cardoso, Vitor and Belczynski, Krzysztof",
    title = "{Spectroscopy of Kerr black holes with Earth- and space-based interferometers}",
    eprint = "1605.09286",
    archivePrefix = "arXiv",
    primaryClass = "gr-qc",
    doi = "10.1103/PhysRevLett.117.101102",
    journal = "Phys. Rev. Lett.",
    volume = "117",
    number = "10",
    pages = "101102",
    year = "2016"
}

@article{Baibhav:2017jhs,
    author = "Baibhav, Vishal and Berti, Emanuele and Cardoso, Vitor and Khanna, Gaurav",
    title = "{Black Hole Spectroscopy: Systematic Errors and Ringdown Energy Estimates}",
    eprint = "1710.02156",
    archivePrefix = "arXiv",
    primaryClass = "gr-qc",
    doi = "10.1103/PhysRevD.97.044048",
    journal = "Phys. Rev. D",
    volume = "97",
    number = "4",
    pages = "044048",
    year = "2018"
}

@article{Baibhav:2018rfk,
    author = "Baibhav, Vishal and Berti, Emanuele",
    title = "{Multimode black hole spectroscopy}",
    eprint = "1809.03500",
    archivePrefix = "arXiv",
    primaryClass = "gr-qc",
    doi = "10.1103/PhysRevD.99.024005",
    journal = "Phys. Rev. D",
    volume = "99",
    number = "2",
    pages = "024005",
    year = "2019"
}

@phdthesis{Carullo:2022qak,
    author = "Carullo, Gregorio",
    title = "{Black Hole Spectroscopy: from a mathematical problem to an observational reality}",
    school = "U. Pisa (main)",
    year = "2022"
}

@article{Antonini:2013tea,
    author = "Antonini, Fabio and Murray, Norman and Mikkola, Seppo",
    title = "{Black hole triple dynamics: breakdown of the orbit average approximation and implications for gravitational wave detections}",
    eprint = "1308.3674",
    archivePrefix = "arXiv",
    primaryClass = "astro-ph.HE",
    doi = "10.1088/0004-637X/781/1/45",
    journal = "Astrophys. J.",
    volume = "781",
    pages = "45",
    year = "2014"
}

@article{Samsing:2017xmd,
    author = "Samsing, Johan",
    title = "{Eccentric Black Hole Mergers Forming in Globular Clusters}",
    eprint = "1711.07452",
    archivePrefix = "arXiv",
    primaryClass = "astro-ph.HE",
    doi = "10.1103/PhysRevD.97.103014",
    journal = "Phys. Rev. D",
    volume = "97",
    number = "10",
    pages = "103014",
    year = "2018"
}

@article{Rodriguez:2018pss,
    author = "Rodriguez, Carl L. and Amaro-Seoane, Pau and Chatterjee, Sourav and Kremer, Kyle and Rasio, Frederic A. and Samsing, Johan and Ye, Claire S. and Zevin, Michael",
    title = "{Post-Newtonian Dynamics in Dense Star Clusters: Formation, Masses, and Merger Rates of Highly-Eccentric Black Hole Binaries}",
    eprint = "1811.04926",
    archivePrefix = "arXiv",
    primaryClass = "astro-ph.HE",
    doi = "10.1103/PhysRevD.98.123005",
    journal = "Phys. Rev. D",
    volume = "98",
    number = "12",
    pages = "123005",
    year = "2018"
}

@article{Zevin:2018kzq,
    author = "Zevin, Michael and Samsing, Johan and Rodriguez, Carl and Haster, Carl-Johan and Ramirez-Ruiz, Enrico",
    title = "{Eccentric Black Hole Mergers in Dense Star Clusters: The Role of Binary{\textendash}Binary Encounters}",
    eprint = "1810.00901",
    archivePrefix = "arXiv",
    primaryClass = "astro-ph.HE",
    reportNumber = "LIGO-P1800275",
    doi = "10.3847/1538-4357/aaf6ec",
    journal = "Astrophys. J.",
    volume = "871",
    number = "1",
    pages = "91",
    year = "2019"
}

@article{Chattopadhyay:2023pil,
    author = "Chattopadhyay, Debatri and Stegmann, Jakob and Antonini, Fabio and Barber, Jordan and Romero-Shaw, Isobel M.",
    title = "{Double black hole mergers in nuclear star clusters: eccentricities, spins, masses, and the growth of massive seeds}",
    eprint = "2308.10884",
    archivePrefix = "arXiv",
    primaryClass = "astro-ph.HE",
    doi = "10.1093/mnras/stad3048",
    journal = "Mon. Not. Roy. Astron. Soc.",
    volume = "526",
    number = "4",
    pages = "4908--4928",
    year = "2023"
}

@article{Liu:2025yok,
    author = "Liu, Bin and Lai, Dong",
    title = "{Hierarchical Black Hole Mergers in Nuclear Star Clusters: A Combined Dynamical-Secular Channel for GW231123-like Events}",
    eprint = "2511.13820",
    archivePrefix = "arXiv",
    journal = "",
    primaryClass = "astro-ph.HE",
    month = "11",
    year = "2025"
}

@article{Islam:2026yxx,
    author = "Islam, Tousif and Wadekar, Digvijay and Kritos, Konstantinos",
    title = "{Kick matters: The impact of a new recoil model on the retention of hierarchical black-hole remnants in globular clusters}",
    eprint = "2603.10170",
    archivePrefix = "arXiv",
    journal = "",
    primaryClass = "astro-ph.HE",
    month = "3",
    year = "2026"
}

@article{Lai:2026yvm,
    author = "Lai, Qi and Lan, Qing-Yu and Wang, Zhan-He and Piao, Yun-Song",
    title = "{Testing the wormhole echo hypothesis for GW231123}",
    eprint = "2602.01615",
    archivePrefix = "arXiv",
    journal = "",
    primaryClass = "gr-qc",
    month = "2",
    year = "2026"
}

@article{Tenorio:2026dcc,
    author = "Tenorio, Rodrigo and Gerosa, Davide",
    title = "{On the exceptionality of exceptional gravitational-wave events}",
    eprint = "2601.02467",
    archivePrefix = "arXiv",
    journal = "",
    primaryClass = "astro-ph.HE",
    month = "1",
    year = "2026"
}

@article{Croon:2025gol,
    author = "Croon, Djuna and Gerosa, Davide and Sakstein, Jeremy",
    title = "{Can GW231123 have a stellar origin?}",
    eprint = "2508.10088",
    archivePrefix = "arXiv",
    primaryClass = "astro-ph.HE",
    reportNumber = "IPPP/25/52",
    doi = "10.1093/mnras/stag073",
    journal = "Mon. Not. Roy. Astron. Soc.",
    volume = "546",
    number = "3",
    pages = "stag073",
    year = "2026"
}

@article{Cuceu:2025fzi,
    author = "Cuceu, Iuliu and Bizouard, Marie Anne and Christensen, Nelson and Sakellariadou, Mairi",
    title = "{GW231123: Binary black hole merger or cosmic string?}",
    eprint = "2507.20778",
    archivePrefix = "arXiv",
    primaryClass = "gr-qc",
    reportNumber = "KCL-PH-TH/2025-34",
    doi = "10.1103/zd8m-tzxd",
    journal = "Phys. Rev. D",
    volume = "113",
    number = "2",
    pages = "L021302",
    year = "2026"
}

@article{Shan:2025dcd,
    author = "Shan, Xikai and Yang, Huan and Mao, Shude",
    title = "{GW231123: A Case for Binary Microlensing in a Strong Lensing Field}",
    eprint = "2512.19118",
    archivePrefix = "arXiv",
    journal = "",
    primaryClass = "astro-ph.GA",
    reportNumber = "LIGO document P2500747",
    month = "12",
    year = "2025"
}

@article{Chakraborty:2025pxt,
    author = "Chakraborty, Aniruddha and Mukherjee, Suvodip",
    title = "{The First Model-Independent Upper Bound on Micro-lensing Signature of the Highest Mass Binary Black Hole Event GW231123}",
    eprint = "2512.19077",
    archivePrefix = "arXiv",
    journal = "",
    primaryClass = "gr-qc",
    month = "12",
    year = "2025"
}

@article{DeLuca:2025fln,
    author = "De Luca, Valerio and Franciolini, Gabriele and Riotto, Antonio",
    title = "{GW231123: a Possible Primordial Black Hole Origin}",
    eprint = "2508.09965",
    archivePrefix = "arXiv",
    journal = "",
    primaryClass = "astro-ph.CO",
    reportNumber = "CERN-TH-2025-163",
    month = "8",
    year = "2025"
}

@article{Damour:2011fu,
    author = "Damour, Thibault and Nagar, Alessandro and Pollney, Denis and Reisswig, Christian",
    title = "{Energy versus Angular Momentum in Black Hole Binaries}",
    eprint = "1110.2938",
    archivePrefix = "arXiv",
    primaryClass = "gr-qc",
    doi = "10.1103/PhysRevLett.108.131101",
    journal = "Phys. Rev. Lett.",
    volume = "108",
    pages = "131101",
    year = "2012"
}

@article{Hopper:2022rwo,
    author = "Hopper, Seth and Nagar, Alessandro and Rettegno, Piero",
    title = "{Strong-field scattering of two spinning black holes: Numerics versus analytics}",
    eprint = "2204.10299",
    archivePrefix = "arXiv",
    primaryClass = "gr-qc",
    doi = "10.1103/PhysRevD.107.124034",
    journal = "Phys. Rev. D",
    volume = "107",
    number = "12",
    pages = "124034",
    year = "2023"
}

@article{Stegmann:2025cja,
    author = "Stegmann, Jakob and Olejak, Aleksandra and de Mink, Selma E.",
    title = "{Resolving Black Hole Family Issues among the Massive Ancestors of Very High-spin Gravitational-wave Events like GW231123}",
    eprint = "2507.15967",
    archivePrefix = "arXiv",
    primaryClass = "astro-ph.HE",
    doi = "10.3847/2041-8213/ae0e5f",
    journal = "Astrophys. J. Lett.",
    volume = "992",
    number = "2",
    pages = "L26",
    year = "2025"
}

@article{Hu:2025lhv,
    author = "Hu, Qian and Narola, Harsh and Heynen, Jef and Wright, Mick and Veitch, John and Janquart, Justin and Van Den Broeck, Chris",
    title = "{GW231123: Overlapping Gravitational Wave Signals?}",
    eprint = "2512.17550",
    archivePrefix = "arXiv",
    journal = "",
    primaryClass = "gr-qc",
    reportNumber = "LIGO-P2500697",
    month = "12",
    year = "2025"
}

@article{Gu:2023eaa,
    author = "Gu, Hua-Peng and Wang, Hai-Tian and Shao, Lijing",
    title = "{Constraints on charged black holes from merger-ringdown signals in GWTC-3 and prospects for the Einstein Telescope}",
    eprint = "2310.10447",
    archivePrefix = "arXiv",
    primaryClass = "gr-qc",
    doi = "10.1103/PhysRevD.109.024058",
    journal = "Phys. Rev. D",
    volume = "109",
    number = "2",
    pages = "024058",
    year = "2024"
}

@article{Boyle:2009vi,
    author = "Boyle, Michael and Mroue, Abdul H.",
    title = "{Extrapolating gravitational-wave data from numerical simulations}",
    eprint = "0905.3177",
    archivePrefix = "arXiv",
    primaryClass = "gr-qc",
    doi = "10.1103/PhysRevD.80.124045",
    journal = "Phys. Rev. D",
    volume = "80",
    pages = "124045",
    year = "2009"
}

@article{Phukon:2024amh,
    author = "Phukon, Khun Sang and Schmidt, Patricia and Pratten, Geraint",
    title = "{Geometric template bank for the detection of spinning low-mass compact binaries with moderate orbital eccentricity}",
    eprint = "2412.06433",
    archivePrefix = "arXiv",
    primaryClass = "gr-qc",
    reportNumber = "LIGO-P2400566",
    doi = "10.1103/PhysRevD.111.043040",
    journal = "Phys. Rev. D",
    volume = "111",
    number = "4",
    pages = "043040",
    year = "2025"
}

@article{Shaikh:2025tae,
    author = "Shaikh, Md Arif and Varma, Vijay and Ramos-Buades, Antoni and Pfeiffer, Harald P. and Boyle, Michael and Kidder, Lawrence E. and Scheel, Mark A.",
    title = "{Defining eccentricity for spin-precessing binaries}",
    eprint = "2507.08345",
    archivePrefix = "arXiv",
    primaryClass = "gr-qc",
    doi = "10.1088/1361-6382/ae085d",
    journal = "Class. Quant. Grav.",
    volume = "42",
    number = "19",
    pages = "195012",
    year = "2025"
}

@article{Tiwari:2025fua,
    author = "Tiwari, Avinash and Bhat, Sajad A. and Shaikh, Md Arif and Kapadia, Shasvath J.",
    title = "{Testing the Nature of GW200105 by Probing the Frequency Evolution of Eccentricity}",
    eprint = "2509.26152",
    archivePrefix = "arXiv",
    primaryClass = "astro-ph.HE",
    doi = "10.3847/1538-4357/ae1d74",
    journal = "Astrophys. J.",
    volume = "995",
    number = "1",
    pages = "48",
    year = "2025"
}

@article{Shaikh:2024wyn,
    author = "Shaikh, Md Arif and Bhat, Sajad A. and Kapadia, Shasvath J.",
    title = "{A study of the inspiral-merger-ringdown consistency test with gravitational-wave signals from compact binaries in eccentric orbits}",
    eprint = "2402.15110",
    archivePrefix = "arXiv",
    primaryClass = "gr-qc",
    doi = "10.1103/PhysRevD.110.024030",
    journal = "Phys. Rev. D",
    volume = "110",
    number = "2",
    pages = "024030",
    year = "2024"
}

@article{Carullo:2021oxn,
    author = "Carullo, Gregorio and Laghi, Danny and Johnson-McDaniel, Nathan K. and Del Pozzo, Walter and Dias, Oscar J. C. and Godazgar, Mahdi and Santos, Jorge E.",
    title = "{Constraints on Kerr-Newman black holes from merger-ringdown gravitational-wave observations}",
    eprint = "2109.13961",
    archivePrefix = "arXiv",
    primaryClass = "gr-qc",
    doi = "10.1103/PhysRevD.105.062009",
    journal = "Phys. Rev. D",
    volume = "105",
    number = "6",
    pages = "062009",
    year = "2022"
}

@article{LIGOScientific:2026qni,
    author = "Abac, A. G. and others",
    collaboration = "LIGO Scientific, VIRGO, KAGRA",
    title = "{GWTC-4.0: Tests of General Relativity. I. Overview and General Tests}",
    eprint = "2603.19019",
    archivePrefix = "arXiv",
    primaryClass = "gr-qc",
    reportNumber = "LIGO-P2500065",
    month = "3",
    year = "2026",
        journal="",
}

@article{LIGOScientific:2025jau,
    author = "Abac, A. G. and others",
    collaboration = "LIGO Scientific, VIRGO, KAGRA",
    title = "{GWTC-4.0: Constraints on the Cosmic Expansion Rate and Modified Gravitational-wave Propagation}",
    eprint = "2509.04348",
    archivePrefix = "arXiv",
    primaryClass = "astro-ph.CO",
    reportNumber = "LIGO-P2400152",
    month = "9",
    year = "2025",
        journal="",
}

@article{LIGOScientific:2025pvj,
    author = "Abac, A. G. and others",
    collaboration = "LIGO Scientific, VIRGO, KAGRA",
    title = "{GWTC-4.0: Population Properties of Merging Compact Binaries}",
    eprint = "2508.18083",
    archivePrefix = "arXiv",
    primaryClass = "astro-ph.HE",
    reportNumber = "LIGO-P2400004",
    month = "8",
    year = "2025",
    journal="",
}

@article{LIGOScientific:2026fcf,
    author = "Abac, A. G. and others",
    collaboration = "LIGO Scientific, VIRGO, KAGRA",
    title = "{GWTC-4.0: Tests of General Relativity. II. Parameterized Tests}",
    eprint = "2603.19020",
    archivePrefix = "arXiv",
    primaryClass = "gr-qc",
    reportNumber = "LIGO-P2500066",
    month = "3",
    year = "2026",
        journal="",
}

@article{LIGOScientific:2026wpt,
    author = "Abac, A. G. and others",
    collaboration = "LIGO Scientific, VIRGO, KAGRA",
    title = "{GWTC-4.0: Tests of General Relativity. III. Tests of the Remnants}",
    eprint = "2603.19021",
    archivePrefix = "arXiv",
    primaryClass = "gr-qc",
    reportNumber = "LIGO-P2500067",
    month = "3",
    year = "2026",
    journal="",
}

@article{Goyal:2025eqo,
    author = "Goyal, Srashti and Villarrubia-Rojo, Hector and Zumalacarregui, Miguel",
    title = "{Across the Universe: GW231123 as a magnified and diffracted black hole merger}",
    eprint = "2512.17631",
    archivePrefix = "arXiv",
    journal = "",
    primaryClass = "astro-ph.GA",
    month = "12",
    year = "2025"
}

@article{Chan:2025kyu,
    author = "Chan, Juno C. L. and Ezquiaga, Jose Mar{\'\i}a and Lo, Rico K. L. and Bowman, Joey and Maga{\~n}a Zertuche, Lorena and Vujeva, Luka",
    title = "{Discovering gravitational waveform distortions from lensing: a deep dive into GW231123}",
    eprint = "2512.16916",
    archivePrefix = "arXiv",
    journal = "",
    primaryClass = "gr-qc",
    month = "12",
    year = "2025"
}

@article{LIGOScientific:2025cwb,
    author = "Abac, A. G. and others",
    collaboration = "LIGO Scientific, VIRGO, KAGRA",
    title = "{GWTC-4.0: Searches for Gravitational-Wave Lensing Signatures}",
    eprint = "2512.16347",
    archivePrefix = "arXiv",
    journal = "",
    primaryClass = "gr-qc",
    reportNumber = "LIGO-P2500419",
    month = "12",
    year = "2025"
}

@article{Delfavero:2025lup,
    author = "Delfavero, V. and Ray, S. and Cook, H. E. and Nathaniel, K. and McKernan, B. and Ford, K. E. S. and Postiglione, J. and McPike, E. and O'Shaughnessy, R.",
    title = "{Prospects for the formation of GW231123 from the AGN channel}",
    eprint = "2508.13412",
    archivePrefix = "arXiv",
    journal = "",
    primaryClass = "gr-qc",
    month = "8",
    year = "2025"
}

@article{Passenger:2025acb,
    author = "Passenger, Lachlan and Banagiri, Sharan and Thrane, Eric and Lasky, Paul D. and Borchers, Angela and Fishbach, Maya and Ye, Claire S.",
    title = "{Is GW231123 a Hierarchical Merger?}",
    eprint = "2510.14363",
    archivePrefix = "arXiv",
    primaryClass = "astro-ph.HE",
    doi = "10.3847/1538-4357/ae4358",
    journal = "Astrophys. J.",
    volume = "999",
    number = "2",
    pages = "236",
    year = "2026"
}

@article{Ray:2025rtt,
    author = "Ray, Anarya and Banagiri, Sharan and Thrane, Eric and Lasky, Paul D.",
    title = "{GW231123: extreme spins or microglitches?}",
    eprint = "2510.07228",
    archivePrefix = "arXiv",
    journal = "",
    primaryClass = "gr-qc",
    reportNumber = "LIGO-P2500613",
    month = "10",
    year = "2025"
}

@article{Chatterjee:2025avc,
    author = "Chatterjee, Chayan and McGowan, Kaylah and Deshmukh, Suyash and Tyler-Howard, Nicholas and Jani, Karan",
    title = "{Machine Learning Confirms GW231123 is a {\textquotedblleft}Lite{\textquotedblright} Intermediate Mass Black Hole Merger}",
    eprint = "2509.09161",
    archivePrefix = "arXiv",
    primaryClass = "astro-ph.HE",
    doi = "10.3847/2041-8213/ae1a5f",
    journal = "Astrophys. J. Lett.",
    volume = "995",
    number = "1",
    pages = "L6",
    year = "2025"
}

@article{Li:2025pyo,
    author = "Li, Guo-Peng and Fan, Xi-Long",
    title = "{The Hierarchical Merger Scenario for GW231123}",
    eprint = "2509.08298",
    archivePrefix = "arXiv",
    journal = "",
    primaryClass = "astro-ph.HE",
    month = "9",
    year = "2025"
}

@article{Baumgarte:2025syh,
    author = "Baumgarte, Thomas W. and Shapiro, Stuart L.",
    title = "{Can Premature Collapse Form Black Holes in the Upper and Lower Mass Gaps?}",
    eprint = "2509.04574",
    archivePrefix = "arXiv",
    primaryClass = "astro-ph.HE",
    doi = "10.1103/26yd-1mhd",
    journal = "Phys. Rev. Lett.",
    volume = "135",
    number = "19",
    pages = "191401",
    year = "2025"
}

@article{Popa:2025dpz,
    author = "Popa, Silvia A. and de Mink, Selma E.",
    title = "{Very Massive, Rapidly Spinning Binary Black Hole Progenitors through Chemically Homogeneous Evolution{\textemdash}The Case of GW231123}",
    eprint = "2509.00154",
    archivePrefix = "arXiv",
    primaryClass = "astro-ph.HE",
    doi = "10.3847/2041-8213/ae20f1",
    journal = "Astrophys. J. Lett.",
    volume = "995",
    number = "2",
    pages = "L76",
    year = "2025"
}

@article{DallAmico:2023neb,
    author = "Dall'Amico, Marco and Mapelli, Michela and Torniamenti, Stefano and Sedda, Manuel Arca",
    title = "{Eccentric black hole mergers via three-body interactions in young, globular, and nuclear star clusters}",
    eprint = "2303.07421",
    archivePrefix = "arXiv",
    primaryClass = "astro-ph.HE",
    doi = "10.1051/0004-6361/202348745",
    journal = "Astron. Astrophys.",
    volume = "683",
    pages = "A186",
    year = "2024"
}

@article{Carullo:2025oms,
    author = "Carullo, Gregorio",
    title = "{Black hole spectroscopy: status report}",
    doi = "10.1007/s10714-025-03408-y",
    journal = "Gen. Rel. Grav.",
    volume = "57",
    number = "5",
    pages = "76",
    year = "2025"
}

@article{Blanchet:2013haa,
    author = "Blanchet, Luc",
    title = "{Post-Newtonian Theory for Gravitational Waves}",
    eprint = "1310.1528",
    archivePrefix = "arXiv",
    primaryClass = "gr-qc",
    doi = "10.12942/lrr-2014-2",
    journal = "Living Rev. Rel.",
    volume = "17",
    pages = "2",
    year = "2014"
}

@article{Bhagwat:2017tkm,
    author = "Bhagwat, Swetha and Okounkova, Maria and Ballmer, Stefan W. and Brown, Duncan A. and Giesler, Matthew and Scheel, Mark A. and Teukolsky, Saul A.",
    title = "{On choosing the start time of binary black hole ringdowns}",
    eprint = "1711.00926",
    archivePrefix = "arXiv",
    primaryClass = "gr-qc",
    doi = "10.1103/PhysRevD.97.104065",
    journal = "Phys. Rev. D",
    volume = "97",
    number = "10",
    pages = "104065",
    year = "2018"
}

@article{Akcay:2018yyh,
    author = "Akcay, Sarp and Bernuzzi, Sebastiano and Messina, Francesco and Nagar, Alessandro and Ortiz, N{\'e}stor and Rettegno, Piero",
    title = "{Effective-one-body multipolar waveform for tidally interacting binary neutron stars up to merger}",
    eprint = "1812.02744",
    archivePrefix = "arXiv",
    primaryClass = "gr-qc",
    doi = "10.1103/PhysRevD.99.044051",
    journal = "Phys. Rev. D",
    volume = "99",
    number = "4",
    pages = "044051",
    year = "2019"
}

@article{Akcay:2020qrj,
    author = "Akcay, Sarp and Gamba, Rossella and Bernuzzi, Sebastiano",
    title = "{Hybrid post-Newtonian effective-one-body scheme for spin-precessing compact-binary waveforms up to merger}",
    eprint = "2005.05338",
    archivePrefix = "arXiv",
    primaryClass = "gr-qc",
    doi = "10.1103/PhysRevD.103.024014",
    journal = "Phys. Rev. D",
    volume = "103",
    number = "2",
    pages = "024014",
    year = "2021"
}

@article{Cao:2017ndf,
    author = "Cao, Zhoujian and Han, Wen-Biao",
    title = "{Waveform model for an eccentric binary black hole based on the effective-one-body-numerical-relativity formalism}",
    eprint = "1708.00166",
    archivePrefix = "arXiv",
    primaryClass = "gr-qc",
    doi = "10.1103/PhysRevD.96.044028",
    journal = "Phys. Rev. D",
    volume = "96",
    number = "4",
    pages = "044028",
    year = "2017"
}

@article{Liu:2019jpg,
    author = "Liu, Xiaolin and Cao, Zhoujian and Shao, Lijing",
    title = "{Validating the Effective-One-Body Numerical-Relativity Waveform Models for Spin-aligned Binary Black Holes along Eccentric Orbits}",
    eprint = "1910.00784",
    archivePrefix = "arXiv",
    primaryClass = "gr-qc",
    doi = "10.1103/PhysRevD.101.044049",
    journal = "Phys. Rev. D",
    volume = "101",
    number = "4",
    pages = "044049",
    year = "2020"
}

@article{Liu:2021pkr,
    author = "Liu, Xiaolin and Cao, Zhoujian and Zhu, Zong-Hong",
    title = "{A higher-multipole gravitational waveform model for an eccentric binary black holes based on the effective-one-body-numerical-relativity formalism}",
    eprint = "2102.08614",
    archivePrefix = "arXiv",
    primaryClass = "gr-qc",
    doi = "10.1088/1361-6382/ac4119",
    journal = "Class. Quant. Grav.",
    volume = "39",
    number = "3",
    pages = "035009",
    year = "2022"
}

@article{Riemenschneider:2021ppj,
    author = "Riemenschneider, Gunnar and Rettegno, Piero and Breschi, Matteo and Albertini, Angelica and Gamba, Rossella and Bernuzzi, Sebastiano and Nagar, Alessandro",
    title = "{Assessment of consistent next-to-quasicircular corrections and postadiabatic approximation in effective-one-body multipolar waveforms for binary black hole coalescences}",
    eprint = "2104.07533",
    archivePrefix = "arXiv",
    primaryClass = "gr-qc",
    doi = "10.1103/PhysRevD.104.104045",
    journal = "Phys. Rev. D",
    volume = "104",
    number = "10",
    pages = "104045",
    year = "2021"
}

@article{Chiaramello:2020ehz,
    author = "Chiaramello, Danilo and Nagar, Alessandro",
    title = "{Faithful analytical effective-one-body waveform model for spin-aligned, moderately eccentric, coalescing black hole binaries}",
    eprint = "2001.11736",
    archivePrefix = "arXiv",
    primaryClass = "gr-qc",
    doi = "10.1103/PhysRevD.101.101501",
    journal = "Phys. Rev. D",
    volume = "101",
    number = "10",
    pages = "101501",
    year = "2020"
}

@article{Albanesi:2021rby,
    author = "Albanesi, Simone and Nagar, Alessandro and Bernuzzi, Sebastiano",
    title = "{Effective one-body model for extreme-mass-ratio spinning binaries on eccentric equatorial orbits: Testing radiation reaction and waveform}",
    eprint = "2104.10559",
    archivePrefix = "arXiv",
    primaryClass = "gr-qc",
    doi = "10.1103/PhysRevD.104.024067",
    journal = "Phys. Rev. D",
    volume = "104",
    number = "2",
    pages = "024067",
    year = "2021"
}

@article{Nagar:2021gss,
    author = "Nagar, Alessandro and Bonino, Alice and Rettegno, Piero",
    title = "{Effective one-body multipolar waveform model for spin-aligned, quasicircular, eccentric, hyperbolic black hole binaries}",
    eprint = "2101.08624",
    archivePrefix = "arXiv",
    primaryClass = "gr-qc",
    doi = "10.1103/PhysRevD.103.104021",
    journal = "Phys. Rev. D",
    volume = "103",
    number = "10",
    pages = "104021",
    year = "2021"
}

@article{Nagar:2021xnh,
    author = "Nagar, Alessandro and Rettegno, Piero",
    title = "{Next generation: Impact of high-order analytical information on effective one body waveform models for noncircularized, spin-aligned black hole binaries}",
    eprint = "2108.02043",
    archivePrefix = "arXiv",
    primaryClass = "gr-qc",
    doi = "10.1103/PhysRevD.104.104004",
    journal = "Phys. Rev. D",
    volume = "104",
    number = "10",
    pages = "104004",
    year = "2021"
}

@article{Mora:2002gf,
    author = "Mora, Thierry and Will, Clifford M.",
    title = "{Numerically generated quasiequilibrium orbits of black holes: Circular or eccentric?}",
    eprint = "gr-qc/0208089",
    archivePrefix = "arXiv",
    doi = "10.1103/PhysRevD.66.101501",
    journal = "Phys. Rev. D",
    volume = "66",
    pages = "101501",
    year = "2002"
}

@article{Carullo:2024smg,
    author = "Carullo, Gregorio",
    title = "{Ringdown amplitudes of nonspinning eccentric binaries}",
    eprint = "2406.19442",
    archivePrefix = "arXiv",
    primaryClass = "gr-qc",
    reportNumber = "Virgo document number: VIR-0579A-24",
    doi = "10.1088/1475-7516/2024/10/061",
    journal = "JCAP",
    volume = "10",
    pages = "061",
    year = "2024"
}

@article{Shaikh:2023ypz,
    author = "Shaikh, Md Arif and Varma, Vijay and Pfeiffer, Harald P. and Ramos-Buades, Antoni and van de Meent, Maarten",
    title = "{Defining eccentricity for gravitational wave astronomy}",
    eprint = "2302.11257",
    archivePrefix = "arXiv",
    primaryClass = "gr-qc",
    doi = "10.1103/PhysRevD.108.104007",
    journal = "Phys. Rev. D",
    volume = "108",
    number = "10",
    pages = "104007",
    year = "2023"
}

@article{London:2014cma,
    author = "London, Lionel and Shoemaker, Deirdre and Healy, James",
    title = "{Modeling ringdown: Beyond the fundamental quasinormal modes}",
    eprint = "1404.3197",
    archivePrefix = "arXiv",
    primaryClass = "gr-qc",
    doi = "10.1103/PhysRevD.90.124032",
    journal = "Phys. Rev. D",
    volume = "90",
    number = "12",
    pages = "124032",
    year = "2014",
    note = "[Erratum: Phys.Rev.D 94, 069902 (2016)]"
}

@article{Damour:2014sva,
    author = "Damour, Thibault and Nagar, Alessandro",
    title = "{New effective-one-body description of coalescing nonprecessing spinning black-hole binaries}",
    eprint = "1406.6913",
    archivePrefix = "arXiv",
    primaryClass = "gr-qc",
    doi = "10.1103/PhysRevD.90.044018",
    journal = "Phys. Rev. D",
    volume = "90",
    number = "4",
    pages = "044018",
    year = "2014"
}

@article{Ramos-Buades:2021adz,
    author = "Ramos-Buades, Antoni and Buonanno, Alessandra and Khalil, Mohammed and Ossokine, Serguei",
    title = "{Effective-one-body multipolar waveforms for eccentric binary black holes with nonprecessing spins}",
    eprint = "2112.06952",
    archivePrefix = "arXiv",
    primaryClass = "gr-qc",
    doi = "10.1103/PhysRevD.105.044035",
    journal = "Phys. Rev. D",
    volume = "105",
    number = "4",
    pages = "044035",
    year = "2022"
}

@article{London:2018gaq,
    author = "London, L. T.",
    title = "{Modeling ringdown. II. Aligned-spin binary black holes, implications for data analysis and fundamental theory}",
    eprint = "1801.08208",
    archivePrefix = "arXiv",
    primaryClass = "gr-qc",
    doi = "10.1103/PhysRevD.102.084052",
    journal = "Phys. Rev. D",
    volume = "102",
    number = "8",
    pages = "084052",
    year = "2020"
}

@article{Buonanno:2006ui,
    author = "Buonanno, Alessandra and Cook, Gregory B. and Pretorius, Frans",
    title = "{Inspiral, merger and ring-down of equal-mass black-hole binaries}",
    eprint = "gr-qc/0610122",
    archivePrefix = "arXiv",
    doi = "10.1103/PhysRevD.75.124018",
    journal = "Phys. Rev. D",
    volume = "75",
    pages = "124018",
    year = "2007"
}

@article{Carullo:2019flw,
    author = "Carullo, Gregorio and Del Pozzo, Walter and Veitch, John",
    title = "{Observational Black Hole Spectroscopy: A time-domain multimode analysis of GW150914}",
    eprint = "1902.07527",
    archivePrefix = "arXiv",
    primaryClass = "gr-qc",
    doi = "10.1103/PhysRevD.99.123029",
    journal = "Phys. Rev. D",
    volume = "99",
    number = "12",
    pages = "123029",
    year = "2019",
    note = "[Erratum: Phys.Rev.D 100, 089903 (2019)]"
}

@article{Nee:2025nmh,
    author = "Nee, Peter James and others",
    title = "{Eccentric binary black holes: A new framework for numerical relativity waveform surrogates}",
    eprint = "2510.00106",
    journal = "",
    archivePrefix = "arXiv",
    primaryClass = "gr-qc",
    month = "9",
    year = "2025"
}

@article{Islam:2025bhf,
    author = "Islam, Tousif and Venumadhav, Tejaswi and Mehta, Ajit Kumar and Wadekar, Digvijay and Roulet, Javier and Anantpurkar, Isha and Mushkin, Jonathan and Zackay, Barak and Zaldarriaga, Matias",
    eprint = "2509.20556",
    journal = "",
    archivePrefix = "arXiv",
    primaryClass = "gr-qc",
    month = "9",
    year = "2025"
}

@article{Owen:1995tm,
    author = "Owen, Benjamin J.",
    title = "{Search templates for gravitational waves from inspiraling binaries: Choice of template spacing}",
    eprint = "gr-qc/9511032",
    archivePrefix = "arXiv",
    doi = "10.1103/PhysRevD.53.6749",
    journal = "Phys. Rev. D",
    volume = "53",
    pages = "6749--6761",
    year = "1996"
}

@article{Stevenson:2017tfq,
    author = "Stevenson, Simon and Vigna-G{\'o}mez, Alejandro and Mandel, Ilya and Barrett, Jim W. and Neijssel, Coenraad J. and Perkins, David and de Mink, Selma E.",
    title = "{Formation of the first three gravitational-wave observations through isolated binary evolution}",
    eprint = "1704.01352",
    archivePrefix = "arXiv",
    primaryClass = "astro-ph.HE",
    doi = "10.1038/ncomms14906",
    journal = "Nature Commun.",
    volume = "8",
    pages = "14906",
    year = "2017"
}

@article{Romero-Shaw:2021ual,
    author = "Romero-Shaw, Isobel M. and Lasky, Paul D. and Thrane, Eric",
    title = "{Signs of Eccentricity in Two Gravitational-wave Signals May Indicate a Subpopulation of Dynamically Assembled Binary Black Holes}",
    eprint = "2108.01284",
    archivePrefix = "arXiv",
    primaryClass = "astro-ph.HE",
    doi = "10.3847/2041-8213/ac3138",
    journal = "Astrophys. J. Lett.",
    volume = "921",
    number = "2",
    pages = "L31",
    year = "2021"
}

@article{Favata:2021vhw,
    author = "Favata, Marc and Kim, Chunglee and Arun, K. G. and Kim, JeongCho and Lee, Hyung Won",
    title = "{Constraining the orbital eccentricity of inspiralling compact binary systems with Advanced LIGO}",
    eprint = "2108.05861",
    archivePrefix = "arXiv",
    primaryClass = "gr-qc",
    reportNumber = "LIGO DCC P2100284",
    doi = "10.1103/PhysRevD.105.023003",
    journal = "Phys. Rev. D",
    volume = "105",
    number = "2",
    pages = "023003",
    year = "2022"
}

@article{Romero-Shaw:2022fbf,
    author = "Romero-Shaw, Isobel M. and Gerosa, Davide and Loutrel, Nicholas",
    title = "{Eccentricity or spin precession? Distinguishing subdominant effects in gravitational-wave data}",
    eprint = "2211.07528",
    archivePrefix = "arXiv",
    primaryClass = "astro-ph.HE",
    doi = "10.1093/mnras/stad031",
    journal = "Mon. Not. Roy. Astron. Soc.",
    volume = "519",
    number = "4",
    pages = "5352--5357",
    year = "2023"
}

@article{Islam:2021mha,
    author = "Islam, Tousif and Varma, Vijay and Lodman, Jackie and Field, Scott E. and Khanna, Gaurav and Scheel, Mark A. and Pfeiffer, Harald P. and Gerosa, Davide and Kidder, Lawrence E.",
    title = "{Eccentric binary black hole surrogate models for the gravitational waveform and remnant properties: comparable mass, nonspinning case}",
    eprint = "2101.11798",
    archivePrefix = "arXiv",
    primaryClass = "gr-qc",
    doi = "10.1103/PhysRevD.103.064022",
    journal = "Phys. Rev. D",
    volume = "103",
    number = "6",
    pages = "064022",
    year = "2021"
}

@article{Liu:2023dgl,
    author = "Liu, Xiaolin and Cao, Zhoujian and Shao, Lijing",
    title = "{Upgraded waveform model of eccentric binary black hole based on effective-one-body-numerical-relativity for spin-aligned binary black holes}",
    eprint = "2306.15277",
    archivePrefix = "arXiv",
    primaryClass = "gr-qc",
    doi = "10.1142/S0218271823500153",
    journal = "Int. J. Mod. Phys. D",
    volume = "32",
    number = "04",
    pages = "2350015",
    year = "2023"
}

@article{Gamboa:2024hli,
    author = "Gamboa, Aldo and others",
    title = "{Accurate waveforms for eccentric, aligned-spin binary black holes: The multipolar effective-one-body model seobnrv5ehm}",
    eprint = "2412.12823",
    archivePrefix = "arXiv",
    primaryClass = "gr-qc",
    doi = "10.1103/jxrc-z298",
    journal = "Phys. Rev. D",
    volume = "112",
    number = "4",
    pages = "044038",
    year = "2025"
}

@article{Nagar:2024dzj,
    author = "Nagar, Alessandro and Gamba, Rossella and Rettegno, Piero and Fantini, Veronica and Bernuzzi, Sebastiano",
    title = "{Effective-one-body waveform model for noncircularized, planar, coalescing black hole binaries: The importance of radiation reaction}",
    eprint = "2404.05288",
    archivePrefix = "arXiv",
    primaryClass = "gr-qc",
    doi = "10.1103/PhysRevD.110.084001",
    journal = "Phys. Rev. D",
    volume = "110",
    number = "8",
    pages = "084001",
    year = "2024"
}

@article{Hannam:2013oca,
    author = {Hannam, Mark and Schmidt, Patricia and Boh{\'e}, Alejandro and Haegel, Le{\"\i}la and Husa, Sascha and Ohme, Frank and Pratten, Geraint and P{\"u}rrer, Michael},
    title = "{Simple Model of Complete Precessing Black-Hole-Binary Gravitational Waveforms}",
    eprint = "1308.3271",
    archivePrefix = "arXiv",
    primaryClass = "gr-qc",
    doi = "10.1103/PhysRevLett.113.151101",
    journal = "Phys. Rev. Lett.",
    volume = "113",
    number = "15",
    pages = "151101",
    year = "2014"
}

@article{Crescimbeni:2025ytx,
    author = "Crescimbeni, Francesco and Carullo, Gregorio and Berti, Emanuele and Caneva Santoro, Giada and Cheung, Mark Ho-Yeuk and Pani, Paolo",
    title = "{Accuracy of ringdown models calibrated to numerical relativity simulations}",
    eprint = "2511.02915",
    journal = "",
    archivePrefix = "arXiv",
    primaryClass = "gr-qc",
    month = "11",
    year = "2025"
}

@article{Ajith:2009bn,
    author = "Ajith, P. and others",
    title = "{Inspiral-merger-ringdown waveforms for black-hole binaries with non-precessing spins}",
    eprint = "0909.2867",
    archivePrefix = "arXiv",
    primaryClass = "gr-qc",
    doi = "10.1103/PhysRevLett.106.241101",
    journal = "Phys. Rev. Lett.",
    volume = "106",
    pages = "241101",
    year = "2011"
}

@article{Santamaria:2010yb,
    author = "Santamaria, L. and others",
    title = "{Matching post-Newtonian and numerical relativity waveforms: systematic errors and a new phenomenological model for non-precessing black hole binaries}",
    eprint = "1005.3306",
    archivePrefix = "arXiv",
    primaryClass = "gr-qc",
    reportNumber = "LIGO-P1000048, AEI-2010-122",
    doi = "10.1103/PhysRevD.82.064016",
    journal = "Phys. Rev. D",
    volume = "82",
    pages = "064016",
    year = "2010"
}

@article{Planas:2025feq,
    author = "Planas, Maria de Lluc and Ramos-Buades, Antoni and Garc{\'\i}a-Quir{\'o}s, Cecilio and Estell{\'e}s, H{\'e}ctor and Husa, Sascha and Haney, Maria",
    title = "{Time-domain phenomenological multipolar waveforms for aligned-spin binary black holes in elliptical orbits}",
    eprint = "2503.13062",
    archivePrefix = "arXiv",
    primaryClass = "gr-qc",
    journal = "",
    month = "3",
    year = "2025"
}

@article{Huerta:2013qb,
    author = "Huerta, E. A. and Brown, Duncan A.",
    title = "{Effect of eccentricity on binary neutron star searches in Advanced LIGO}",
    eprint = "1301.1895",
    archivePrefix = "arXiv",
    primaryClass = "gr-qc",
    reportNumber = "LIGO-DCC-1200187, KITP-NUMBER-NSF-KITP-12-194",
    doi = "10.1103/PhysRevD.87.127501",
    journal = "Phys. Rev. D",
    volume = "87",
    number = "12",
    pages = "127501",
    year = "2013"
}

@article{LIGOScientific:2019dag,
    author = "Abbott, B. P. and others",
    collaboration = "LIGO Scientific, Virgo",
    title = "{Search for Eccentric Binary Black Hole Mergers with Advanced LIGO and Advanced Virgo during their First and Second Observing Runs}",
    eprint = "1907.09384",
    archivePrefix = "arXiv",
    primaryClass = "astro-ph.HE",
    reportNumber = "LIGO Document P1900110",
    doi = "10.3847/1538-4357/ab3c2d",
    journal = "Astrophys. J.",
    volume = "883",
    number = "2",
    pages = "149",
    year = "2019"
}

@article{OShea:2021faf,
    author = "O'Shea, Eamonn and Kumar, Prayush",
    title = "{Correlations in gravitational-wave reconstructions from eccentric binaries: A case study with GW151226 and GW170608}",
    eprint = "2107.07981",
    archivePrefix = "arXiv",
    primaryClass = "astro-ph.HE",
    doi = "10.1103/PhysRevD.108.104018",
    journal = "Phys. Rev. D",
    volume = "108",
    number = "10",
    pages = "104018",
    year = "2023"
}

@article{Ramos-Buades:2019uvh,
    author = "Ramos-Buades, Antoni and Husa, Sascha and Pratten, Geraint and Estell{\'e}s, H{\'e}ctor and Garc{\'\i}a-Quir{\'o}s, Cecilio and Mateu-Lucena, Maite and Colleoni, Marta and Jaume, Rafel",
    title = "{First survey of spinning eccentric black hole mergers: Numerical relativity simulations, hybrid waveforms, and parameter estimation}",
    eprint = "1909.11011",
    archivePrefix = "arXiv",
    primaryClass = "gr-qc",
    doi = "10.1103/PhysRevD.101.083015",
    journal = "Phys. Rev. D",
    volume = "101",
    number = "8",
    pages = "083015",
    year = "2020"
}

@article{Stegmann:2025shr,
    author = "Stegmann, Jakob and Gerosa, Davide and Romero-Shaw, Isobel and Fumagalli, Giulia and Tagawa, Hiromichi and Zwick, Lorenz",
    title = "{Distinguishing the origin of eccentric black-hole mergers with gravitational-wave spin measurements}",
    eprint = "2505.13589",
    archivePrefix = "arXiv",
    primaryClass = "astro-ph.HE",
    journal = "",
    month = "5",
    year = "2025"
}

@article{Bhat:2022amc,
    author = "Bhat, Sajad A. and Saini, Pankaj and Favata, Marc and Arun, K. G.",
    title = "{Systematic bias on the inspiral-merger-ringdown consistency test due to neglect of orbital eccentricity}",
    eprint = "2207.13761",
    archivePrefix = "arXiv",
    primaryClass = "gr-qc",
    reportNumber = "LIGO Preprint No. P2200216",
    doi = "10.1103/PhysRevD.107.024009",
    journal = "Phys. Rev. D",
    volume = "107",
    number = "2",
    pages = "024009",
    year = "2023"
}

@article{Bhat:2024hyb,
    author = "Bhat, Sajad A. and Saini, Pankaj and Favata, Marc and Gandevikar, Chinmay and Mishra, Chandra Kant and Arun, K. G.",
    title = "{Parametrized tests of general relativity using eccentric compact binaries}",
    eprint = "2408.14132",
    archivePrefix = "arXiv",
    primaryClass = "gr-qc",
    reportNumber = "LIGO Preprint No. P2400347",
    doi = "10.1103/PhysRevD.110.124062",
    journal = "Phys. Rev. D",
    volume = "110",
    number = "12",
    pages = "124062",
    year = "2024"
}

@article{Gupta:2024gun,
    author = "Gupta, Anuradha and others",
    title = "{Possible causes of false general relativity violations in gravitational wave observations}",
    eprint = "2405.02197",
    archivePrefix = "arXiv",
    primaryClass = "gr-qc",
    doi = "10.21468/SciPostPhysCommRep.5",
    journal = "",
    month = "5",
    year = "2024"
}

@article{Divyajyoti:2023rht,
    author = "Divyajyoti and Kumar, Sumit and Tibrewal, Snehal and Romero-Shaw, Isobel M. and Mishra, Chandra Kant",
    title = "{Blind spots and biases: The dangers of ignoring eccentricity in gravitational-wave signals from binary black holes}",
    eprint = "2309.16638",
    archivePrefix = "arXiv",
    primaryClass = "gr-qc",
    doi = "10.1103/PhysRevD.109.043037",
    journal = "Phys. Rev. D",
    volume = "109",
    number = "4",
    pages = "043037",
    year = "2024"
}

@article{Gonzalez:2025xba,
    author = "Gonzalez, Alejandra and Bernuzzi, Sebastiano and Rashti, Alireza and Brandoli, Francesco and Gamba, Rossella",
    title = "{Black-hole - neutron-star mergers: new numerical-relativity simulations and multipolar effective-one-body model with spin precession and eccentricity}",
    eprint = "2507.00113",
    archivePrefix = "arXiv",
    primaryClass = "gr-qc",
    journal = "",
    month = "6",
    year = "2025"
}

@article{Planas:2025plq,
    author = "Planas, Maria de Lluc and Husa, Sascha and Ramos-Buades, Antoni and Valencia, Jorge",
    title = "{First eccentric inspiral-merger-ringdown analysis of neutron star-black hole mergers}",
    eprint = "2506.01760",
    archivePrefix = "arXiv",
    primaryClass = "astro-ph.HE",
    journal = "",
    month = "6",
    year = "2025"
}

@article{Iglesias:2022xfc,
    author = "Iglesias, H. L. and others",
    title = "{Eccentricity Estimation for Five Binary Black Hole Mergers with Higher-order Gravitational-wave Modes}",
    eprint = "2208.01766",
    archivePrefix = "arXiv",
    primaryClass = "gr-qc",
    reportNumber = "LIGO-P2200208",
    doi = "10.3847/1538-4357/ad5ff6",
    journal = "Astrophys. J.",
    volume = "972",
    number = "1",
    pages = "65",
    year = "2024"
}

@article{Zhu:2023fnf,
    author = "Zhu, Hengrui and others",
    title = "{Black hole spectroscopy for precessing binary black hole coalescences}",
    eprint = "2312.08588",
    archivePrefix = "arXiv",
    primaryClass = "gr-qc",
    doi = "10.1103/PhysRevD.111.064052",
    journal = "Phys. Rev. D",
    volume = "111",
    number = "6",
    pages = "064052",
    year = "2025"
}

@article{Morras:2025xfu,
    author = "Morras, Gonzalo and Pratten, Geraint and Schmidt, Patricia",
    title = "{Orbital eccentricity in a neutron star - black hole binary}",
    eprint = "2503.15393",
    archivePrefix = "arXiv",
    primaryClass = "astro-ph.HE",
    reportNumber = "LIGO-DCC P2500105",
    month = "3",
    year = "2025",
    journal="",
}

@article{Bonino:2024xrv,
    author = "Bonino, Alice and Schmidt, Patricia and Pratten, Geraint",
    title = "{Mapping eccentricity evolutions between numerical relativity and effective-one-body gravitational waveforms}",
    eprint = "2404.18875",
    archivePrefix = "arXiv",
    primaryClass = "gr-qc",
    doi = "10.1103/PhysRevD.110.104002",
    journal = "Phys. Rev. D",
    volume = "110",
    pages = "104002",
    year = "2024"
}

@article{Bonino:2022hkj,
    author = "Bonino, Alice and Gamba, Rossella and Schmidt, Patricia and Nagar, Alessandro and Pratten, Geraint and Breschi, Matteo and Rettegno, Piero and Bernuzzi, Sebastiano",
    title = "{Inferring eccentricity evolution from observations of coalescing binary black holes}",
    eprint = "2207.10474",
    archivePrefix = "arXiv",
    primaryClass = "gr-qc",
    reportNumber = "LIGO DCC P2200219",
    doi = "10.1103/PhysRevD.107.064024",
    journal = "Phys. Rev. D",
    volume = "107",
    number = "6",
    pages = "064024",
    year = "2023"
}

@article{Clarke:2022fma,
    author = "Clarke, Teagan A. and Romero-Shaw, Isobel M. and Lasky, Paul D. and Thrane, Eric",
    title = "{Gravitational-wave inference for eccentric binaries: the argument of periapsis}",
    eprint = "2206.14006",
    archivePrefix = "arXiv",
    primaryClass = "gr-qc",
    doi = "10.1093/mnras/stac2965",
    journal = "Mon. Not. Roy. Astron. Soc.",
    volume = "517",
    number = "3",
    pages = "3778--3784",
    year = "2022"
}

@article{Knee:2022hth,
    author = "Knee, Alan M. and Romero-Shaw, Isobel M. and Lasky, Paul D. and McIver, Jess and Thrane, Eric",
    title = "{A Rosetta Stone for Eccentric Gravitational Waveform Models}",
    eprint = "2207.14346",
    archivePrefix = "arXiv",
    primaryClass = "gr-qc",
    doi = "10.3847/1538-4357/ac8b02",
    journal = "Astrophys. J.",
    volume = "936",
    number = "2",
    pages = "172",
    year = "2022"
}

@article{CalderonBustillo:2020xms,
    author = "Calder{\'o}n Bustillo, Juan and Sanchis-Gual, Nicolas and Torres-Forn{\'e}, Alejandro and Font, Jos{\'e} A.",
    title = "{Confusing Head-On Collisions with Precessing Intermediate-Mass Binary Black Hole Mergers}",
    eprint = "2009.01066",
    archivePrefix = "arXiv",
    primaryClass = "gr-qc",
    reportNumber = "LIGO-P1900363",
    doi = "10.1103/PhysRevLett.126.201101",
    journal = "Phys. Rev. Lett.",
    volume = "126",
    number = "20",
    pages = "201101",
    year = "2021"
}

@article{Ramos-Buades:2022lgf,
    author = {Ramos-Buades, Antoni and van de Meent, Maarten and Pfeiffer, Harald P. and R{\"u}ter, Hannes R. and Scheel, Mark A. and Boyle, Michael and Kidder, Lawrence E.},
    title = "{Eccentric binary black holes: Comparing numerical relativity and small mass-ratio perturbation theory}",
    eprint = "2209.03390",
    archivePrefix = "arXiv",
    primaryClass = "gr-qc",
    doi = "10.1103/PhysRevD.106.124040",
    journal = "Phys. Rev. D",
    volume = "106",
    number = "12",
    pages = "124040",
    year = "2022"
}

@article{Ramos-Buades:2023yhy,
    author = "Ramos-Buades, Antoni and Buonanno, Alessandra and Gair, Jonathan",
    title = "{Bayesian inference of binary black holes with inspiral-merger-ringdown waveforms using two eccentric parameters}",
    eprint = "2309.15528",
    archivePrefix = "arXiv",
    primaryClass = "gr-qc",
    doi = "10.1103/PhysRevD.108.124063",
    journal = "Phys. Rev. D",
    volume = "108",
    number = "12",
    pages = "124063",
    year = "2023"
}

@article{Gupte:2024jfe,
    author = "Gupte, Nihar and others",
    title = "{Evidence for eccentricity in the population of binary black holes observed by LIGO-Virgo-KAGRA}",
    eprint = "2404.14286",
    archivePrefix = "arXiv",
    journal = "",
    primaryClass = "gr-qc",
    month = "4",
    year = "2024"
}

@article{Gamba:2024cvy,
    author = "Gamba, Rossella and Chiaramello, Danilo and Neogi, Sayan",
    title = "{Toward efficient effective-one-body models for generic, nonplanar orbits}",
    eprint = "2404.15408",
    archivePrefix = "arXiv",
    primaryClass = "gr-qc",
    doi = "10.1103/PhysRevD.110.024031",
    journal = "Phys. Rev. D",
    volume = "110",
    number = "2",
    pages = "024031",
    year = "2024"
}

@article{Albanesi:2024xus,
    author = "Albanesi, Simone and Rashti, Alireza and Zappa, Francesco and Gamba, Rossella and Cook, William and Daszuta, Boris and Bernuzzi, Sebastiano and Nagar, Alessandro and Radice, David",
    title = "{Scattering and dynamical capture of two black holes: Synergies between numerical and analytical methods}",
    eprint = "2405.20398",
    archivePrefix = "arXiv",
    primaryClass = "gr-qc",
    doi = "10.1103/PhysRevD.111.024069",
    journal = "Phys. Rev. D",
    volume = "111",
    number = "2",
    pages = "024069",
    year = "2025"
}

@article{Kankani:2024may,
    author = "Kankani, Anuj and McWilliams, Sean T.",
    title = "{Testing the boundary-to-bound correspondence with numerical relativity}",
    eprint = "2404.03607",
    archivePrefix = "arXiv",
    primaryClass = "gr-qc",
    doi = "10.1103/PhysRevD.110.064033",
    journal = "Phys. Rev. D",
    volume = "110",
    number = "6",
    pages = "064033",
    year = "2024"
}

@article{Planas:2025jny,
    author = "Planas, Maria de Lluc and Ramos-Buades, Antoni and Garc{\'\i}a-Quir{\'o}s, Cecilio and Estell{\'e}s, H{\'e}ctor and Husa, Sascha and Haney, Maria",
    title = "{Eccentric or circular? A reanalysis of binary black hole gravitational wave events for orbital eccentricity signatures}",
    eprint = "2504.15833",
    journal = "",
    archivePrefix = "arXiv",
    primaryClass = "gr-qc",
    month = "4",
    year = "2025"
}

@article{Ajith:2007kx,
    author = "Ajith, P. and others",
    title = "{A Template bank for gravitational waveforms from coalescing binary black holes. I. Non-spinning binaries}",
    eprint = "0710.2335",
    archivePrefix = "arXiv",
    primaryClass = "gr-qc",
    reportNumber = "LIGO-P070111-00-Z, AEI-2007-143",
    doi = "10.1103/PhysRevD.77.104017",
    journal = "Phys. Rev. D",
    volume = "77",
    pages = "104017",
    year = "2008",
    note = "[Erratum: Phys.Rev.D 79, 129901 (2009)]"
}

@article{Isi:2021iql,
    author = "Isi, Maximiliano and Farr, Will M.",
    title = "{Analyzing black-hole ringdowns}",
    eprint = "2107.05609",
    archivePrefix = "arXiv",
    primaryClass = "gr-qc",
    reportNumber = "LIGO-P2100227",
    journal = "",
    month = "7",
    year = "2021"
}

@article{Stein:2019mop,
    author = "Stein, Leo C.",
    title = "{qnm: A Python package for calculating Kerr quasinormal modes, separation constants, and spherical-spheroidal mixing coefficients}",
    eprint = "1908.10377",
    archivePrefix = "arXiv",
    primaryClass = "gr-qc",
    doi = "10.21105/joss.01683",
    journal = "J. Open Source Softw.",
    volume = "4",
    number = "42",
    pages = "1683",
    year = "2019"
}

@article{Cheung:2023vki,
    author = "Cheung, Mark Ho-Yeuk and Berti, Emanuele and Baibhav, Vishal and Cotesta, Roberto",
    title = "{Extracting linear and nonlinear quasinormal modes from black hole merger simulations}",
    eprint = "2310.04489",
    archivePrefix = "arXiv",
    primaryClass = "gr-qc",
    doi = "10.1103/PhysRevD.109.044069",
    journal = "Phys. Rev. D",
    volume = "109",
    number = "4",
    pages = "044069",
    year = "2024",
    note = "[Erratum: Phys.Rev.D 110, 049902 (2024)]"
}

@article{Zhu:2024dyl,
    author = "Zhu, Hengrui and others",
    title = "{Imprints of changing mass and spin on black hole ringdown}",
    eprint = "2404.12424",
    archivePrefix = "arXiv",
    primaryClass = "gr-qc",
    doi = "10.1103/PhysRevD.110.124028",
    journal = "Phys. Rev. D",
    volume = "110",
    number = "12",
    pages = "124028",
    year = "2024"
}

@article{Radia:2021hjs,
    author = "Radia, Miren and Sperhake, Ulrich and Berti, Emanuele and Croft, Robin",
    title = "{Anomalies in the gravitational recoil of eccentric black-hole mergers with unequal mass ratios}",
    eprint = "2101.11015",
    archivePrefix = "arXiv",
    primaryClass = "gr-qc",
    doi = "10.1103/PhysRevD.103.104006",
    journal = "Phys. Rev. D",
    volume = "103",
    number = "10",
    pages = "104006",
    year = "2021"
}

@article{Nagar:2020xsk,
    author = "Nagar, Alessandro and Rettegno, Piero and Gamba, Rossella and Bernuzzi, Sebastiano",
    title = "{Effective-one-body waveforms from dynamical captures in black hole binaries}",
    eprint = "2009.12857",
    archivePrefix = "arXiv",
    primaryClass = "gr-qc",
    doi = "10.1103/PhysRevD.103.064013",
    journal = "Phys. Rev. D",
    volume = "103",
    number = "6",
    pages = "064013",
    year = "2021"
}

@article{Faggioli:2026alx,
    author = "Faggioli, Guglielmo and Buonanno, Alessandra and van de Meent, Maarten and Khanna, Gaurav",
    title = "{Modeling the merger-ringdown of an eccentric test-mass inspiral into a Kerr black hole using the effective-one-body framework}",
    eprint = "2603.19913",
    journal = "",
    archivePrefix = "arXiv",
    primaryClass = "gr-qc",
    month = "3",
    year = "2026"
}

@article{Albanesi:2026qtx,
    author = "Albanesi, Simone and Bernuzzi, Sebastiano and Nagar, Alessandro",
    title = "{Ringdown modeling for effective-one-body waveforms in the test-mass limit for eccentric equatorial orbits around a Kerr black hole}",
    eprint = "2603.19413",
    journal = "",
    archivePrefix = "arXiv",
    primaryClass = "gr-qc",
    month = "3",
    year = "2026"
}

@article{Gamba:2021gap,
    author = "Gamba, Rossella and Breschi, Matteo and Carullo, Gregorio and Albanesi, Simone and Rettegno, Piero and Bernuzzi, Sebastiano and Nagar, Alessandro",
    title = "{GW190521 as a dynamical capture of two nonspinning black holes}",
    eprint = "2106.05575",
    archivePrefix = "arXiv",
    primaryClass = "gr-qc",
    doi = "10.1038/s41550-022-01813-w",
    journal = "Nature Astron.",
    volume = "7",
    number = "1",
    pages = "11--17",
    year = "2023"
}

@article{Berti:2006wq,
    author = "Berti, Emanuele and Cardoso, Vitor",
    title = "{Quasinormal ringing of Kerr black holes. I. The Excitation factors}",
    eprint = "gr-qc/0605118",
    archivePrefix = "arXiv",
    doi = "10.1103/PhysRevD.74.104020",
    journal = "Phys. Rev. D",
    volume = "74",
    pages = "104020",
    year = "2006"
}

@article{Lagos:2022otp,
    author = "Lagos, Macarena and Hui, Lam",
    title = "{Generation and propagation of nonlinear quasinormal modes of a Schwarzschild black hole}",
    eprint = "2208.07379",
    archivePrefix = "arXiv",
    primaryClass = "gr-qc",
    doi = "10.1103/PhysRevD.107.044040",
    journal = "Phys. Rev. D",
    volume = "107",
    number = "4",
    pages = "044040",
    year = "2023"
}

@article{Albanesi:2023bgi,
    author = "Albanesi, Simone and Bernuzzi, Sebastiano and Damour, Thibault and Nagar, Alessandro and Placidi, Andrea",
    title = "{Faithful effective-one-body waveform of small-mass-ratio coalescing black hole binaries: The eccentric, nonspinning case}",
    eprint = "2305.19336",
    archivePrefix = "arXiv",
    primaryClass = "gr-qc",
    doi = "10.1103/PhysRevD.108.084037",
    journal = "Phys. Rev. D",
    volume = "108",
    number = "8",
    pages = "084037",
    year = "2023"
}

@article{Lovelace:2016uwp,
    author = "Lovelace, Geoffrey and others",
    title = "{Modeling the source of GW150914 with targeted numerical-relativity simulations}",
    eprint = "1607.05377",
    archivePrefix = "arXiv",
    primaryClass = "gr-qc",
    doi = "10.1088/0264-9381/33/24/244002",
    journal = "Class. Quant. Grav.",
    volume = "33",
    number = "24",
    pages = "244002",
    year = "2016"
}

@article{Gleiser:1996yc,
    author = "Gleiser, Reinaldo J. and Nicasio, Carlos O. and Price, Richard H. and Pullin, Jorge",
    title = "{Colliding black holes: How far can the close approximation go?}",
    eprint = "gr-qc/9609022",
    archivePrefix = "arXiv",
    reportNumber = "CGPG-96-9-1",
    doi = "10.1103/PhysRevLett.77.4483",
    journal = "Phys. Rev. Lett.",
    volume = "77",
    pages = "4483--4486",
    year = "1996"
}

@article{Sberna:2021eui,
    author = "Sberna, Laura and Bosch, Pablo and East, William E. and Green, Stephen R. and Lehner, Luis",
    title = "{Nonlinear effects in the black hole ringdown: Absorption-induced mode excitation}",
    eprint = "2112.11168",
    archivePrefix = "arXiv",
    primaryClass = "gr-qc",
    doi = "10.1103/PhysRevD.105.064046",
    journal = "Phys. Rev. D",
    volume = "105",
    number = "6",
    pages = "064046",
    year = "2022"
}

@article{Cheung:2022rbm,
    author = "Cheung, Mark Ho-Yeuk and others",
    title = "{Nonlinear Effects in Black Hole Ringdown}",
    eprint = "2208.07374",
    archivePrefix = "arXiv",
    primaryClass = "gr-qc",
    doi = "10.1103/PhysRevLett.130.081401",
    journal = "Phys. Rev. Lett.",
    volume = "130",
    number = "8",
    pages = "081401",
    year = "2023"
}

@article{Mitman:2022qdl,
    author = "Mitman, Keefe and others",
    title = "{Nonlinearities in Black Hole Ringdowns}",
    eprint = "2208.07380",
    archivePrefix = "arXiv",
    primaryClass = "gr-qc",
    doi = "10.1103/PhysRevLett.130.081402",
    journal = "Phys. Rev. Lett.",
    volume = "130",
    number = "8",
    pages = "081402",
    year = "2023"
}

@article{Baibhav:2023clw,
    author = "Baibhav, Vishal and Cheung, Mark Ho-Yeuk and Berti, Emanuele and Cardoso, Vitor and Carullo, Gregorio and Cotesta, Roberto and Del Pozzo, Walter and Duque, Francisco",
    title = "{Agnostic black hole spectroscopy: Quasinormal mode content of numerical relativity waveforms and limits of validity of linear perturbation theory}",
    eprint = "2302.03050",
    archivePrefix = "arXiv",
    primaryClass = "gr-qc",
    doi = "10.1103/PhysRevD.108.104020",
    journal = "Phys. Rev. D",
    volume = "108",
    number = "10",
    pages = "104020",
    year = "2023"
}

@article{Bucciotti:2023ets,
    author = "Bucciotti, Bruno and Kuntz, Adrien and Serra, Francesco and Trincherini, Enrico",
    title = "{Nonlinear quasi-normal modes: uniform approximation}",
    eprint = "2309.08501",
    archivePrefix = "arXiv",
    primaryClass = "hep-th",
    doi = "10.1007/JHEP12(2023)048",
    journal = "JHEP",
    volume = "12",
    pages = "048",
    year = "2023"
}

@article{Perrone:2023jzq,
    author = "Perrone, Davide and Barreira, Thomas and Kehagias, Alex and Riotto, Antonio",
    title = "{Non-linear black hole ringdowns: An analytical approach}",
    eprint = "2308.15886",
    archivePrefix = "arXiv",
    primaryClass = "gr-qc",
    doi = "10.1016/j.nuclphysb.2023.116432",
    journal = "Nucl. Phys. B",
    volume = "999",
    pages = "116432",
    year = "2024"
}

@article{Redondo-Yuste:2023seq,
    author = "Redondo-Yuste, Jaime and Carullo, Gregorio and Ripley, Justin L. and Berti, Emanuele and Cardoso, Vitor",
    title = "{Spin dependence of black hole ringdown nonlinearities}",
    eprint = "2308.14796",
    archivePrefix = "arXiv",
    primaryClass = "gr-qc",
    doi = "10.1103/PhysRevD.109.L101503",
    journal = "Phys. Rev. D",
    volume = "109",
    number = "10",
    pages = "L101503",
    year = "2024"
}

@article{Ma:2024qcv,
    author = "Ma, Sizheng and Yang, Huan",
    title = "{Excitation of quadratic quasinormal modes for Kerr black holes}",
    eprint = "2401.15516",
    archivePrefix = "arXiv",
    primaryClass = "gr-qc",
    doi = "10.1103/PhysRevD.109.104070",
    journal = "Phys. Rev. D",
    volume = "109",
    number = "10",
    pages = "104070",
    year = "2024"
}

@article{Bucciotti:2024zyp,
    author = "Bucciotti, Bruno and Juliano, Leonardo and Kuntz, Adrien and Trincherini, Enrico",
    title = "{Quadratic quasinormal modes of a Schwarzschild black hole}",
    eprint = "2405.06012",
    archivePrefix = "arXiv",
    primaryClass = "gr-qc",
    doi = "10.1103/PhysRevD.110.104048",
    journal = "Phys. Rev. D",
    volume = "110",
    number = "10",
    pages = "104048",
    year = "2024"
}

@article{Bourg:2024jme,
    author = "Bourg, Patrick and Panosso Macedo, Rodrigo and Spiers, Andrew and Leather, Benjamin and Bonga, B{\'e}atrice and Pound, Adam",
    title = "{Quadratic Quasinormal Mode Dependence on Linear Mode Parity}",
    eprint = "2405.10270",
    archivePrefix = "arXiv",
    primaryClass = "gr-qc",
    doi = "10.1103/PhysRevLett.134.061401",
    journal = "Phys. Rev. Lett.",
    volume = "134",
    number = "6",
    pages = "061401",
    year = "2025"
}

@article{Zhu:2024rej,
    author = "Zhu, Hengrui and others",
    title = "{Nonlinear effects in black hole ringdown from scattering experiments: Spin and initial data dependence of quadratic mode coupling}",
    eprint = "2401.00805",
    archivePrefix = "arXiv",
    primaryClass = "gr-qc",
    doi = "10.1103/PhysRevD.109.104050",
    journal = "Phys. Rev. D",
    volume = "109",
    number = "10",
    pages = "104050",
    year = "2024"
}

@article{Redondo-Yuste:2023ipg,
    author = "Redondo-Yuste, Jaime and Pere{\~n}iguez, David and Cardoso, Vitor",
    title = "{Ringdown of a dynamical spacetime}",
    eprint = "2312.04633",
    archivePrefix = "arXiv",
    primaryClass = "gr-qc",
    doi = "10.1103/PhysRevD.109.044048",
    journal = "Phys. Rev. D",
    volume = "109",
    number = "4",
    pages = "044048",
    year = "2024"
}

@article{May:2024rrg,
    author = "May, Taillte and Ma, Sizheng and Ripley, Justin L. and East, William E.",
    title = "{Nonlinear effect of absorption on the ringdown of a spinning black hole}",
    eprint = "2405.18303",
    archivePrefix = "arXiv",
    primaryClass = "gr-qc",
    doi = "10.1103/PhysRevD.110.084034",
    journal = "Phys. Rev. D",
    volume = "110",
    number = "8",
    pages = "084034",
    year = "2024"
}

@article{Capuano:2024qhv,
    author = "Capuano, Lodovico and Santoni, Luca and Barausse, Enrico",
    title = "{Perturbations of the Vaidya metric in the frequency domain: Quasinormal modes and tidal response}",
    eprint = "2407.06009",
    archivePrefix = "arXiv",
    primaryClass = "gr-qc",
    doi = "10.1103/PhysRevD.110.084081",
    journal = "Phys. Rev. D",
    volume = "110",
    number = "8",
    pages = "084081",
    year = "2024"
}

@article{Damour:2014yha,
    author = "Damour, Thibault and Nagar, Alessandro",
    title = "{A new analytic representation of the ringdown waveform of coalescing spinning black hole binaries}",
    eprint = "1406.0401",
    archivePrefix = "arXiv",
    primaryClass = "gr-qc",
    doi = "10.1103/PhysRevD.90.024054",
    journal = "Phys. Rev. D",
    volume = "90",
    number = "2",
    pages = "024054",
    year = "2014"
}

@article{Bohe:2016gbl,
    author = "Boh{\'e}, Alejandro and others",
    title = "{Improved effective-one-body model of spinning, nonprecessing binary black holes for the era of gravitational-wave astrophysics with advanced detectors}",
    eprint = "1611.03703",
    archivePrefix = "arXiv",
    primaryClass = "gr-qc",
    reportNumber = "LIGO-P1600315",
    doi = "10.1103/PhysRevD.95.044028",
    journal = "Phys. Rev. D",
    volume = "95",
    number = "4",
    pages = "044028",
    year = "2017"
}

@article{Cotesta:2018fcv,
    author = "Cotesta, Roberto and Buonanno, Alessandra and Boh{\'e}, Alejandro and Taracchini, Andrea and Hinder, Ian and Ossokine, Serguei",
    title = "{Enriching the Symphony of Gravitational Waves from Binary Black Holes by Tuning Higher Harmonics}",
    eprint = "1803.10701",
    archivePrefix = "arXiv",
    primaryClass = "gr-qc",
    doi = "10.1103/PhysRevD.98.084028",
    journal = "Phys. Rev. D",
    volume = "98",
    number = "8",
    pages = "084028",
    year = "2018"
}

@article{Scheel:2025jct,
    author = "Scheel, Mark A. and others",
    title = "{The SXS Collaboration's third catalog of binary black hole simulations}",
    eprint = "2505.13378",
    archivePrefix = "arXiv",
    primaryClass = "gr-qc",
    doi = "10.1088/1361-6382/adfd34",
    journal = "Class. Quant. Grav.",
    year = "2025"
}

\end{document}